\documentclass[12pt]{article}

\usepackage{latexsym}
\usepackage{feynmp}

\newbox\SlashedBox  
\def\fs#1{\setbox\SlashedBox=\hbox{#1} 
\hbox to 0pt{\hbox to 1\wd\SlashedBox{\hfil/\hfil}\hss}{#1}} 
\def\hboxtosizeof#1#2{\setbox\SlashedBox=\hbox{#1} 
\hbox to 1\wd\SlashedBox{#2}} 

\def\ms#1{\setbox\SlashedBox=\hbox{$#1$}
\hbox to 0pt{\hbox to 1\wd\SlashedBox{\hfil/\hfil}\hss}#1}

\newcommand{\db}{\raisebox{10pt}
           {\tiny$\scriptscriptstyle\longleftarrow$}\hspace{-11pt}}
\newcommand{\dbo}{\raisebox{11.5pt}
           {\tiny$\scriptscriptstyle\longleftarrow$}\hspace{-11pt}}           
\newcommand{\tr}{{\rm tr}}
\newcommand{\ie}{{\em i.e.~}}
\begin{document}

\thispagestyle{empty}

\begin{flushright}
ROM2F/99/01
\end{flushright}

\vspace{1.5cm}

\begin{center}

{\LARGE {\bf A perturbative re-analysis of ${\cal N}$=4 
supersymmetric Yang--Mills theory\rule{0pt}{25pt} }} \\ \vspace{1cm}
{\large Stefano Kovacs} \\ \vspace{0.6cm} 
{\large {\it Dipartimento di Fisica, \ Universit{\`a} di Roma \  
``Tor Vergata''}} \\  {\large {\it I.N.F.N.\ -\ Sezione di Roma \ 
``Tor Vergata''}} \\ {\large {\it Via della Ricerca  Scientifica, 1}} 
\\ {\large {\it 00173 \ Roma, \ ITALY}} \\
%\vspace{0.6cm}
%\date{\today}	

\end{center}

\vspace{1cm}

\begin{abstract}
The finiteness properties of the ${\cal N}$=4 supersymmetric 
Yang--Mills theory are reanalyzed both in the component formulation 
and using ${\cal N}$=1 superfields, in order to discuss some 
subtleties that emerge in the computation of gauge dependent 
quantities.	The one-loop corrections to various Green functions of 
elementary fields are calculated. In the component formulation 
it is shown that the choice of the Wess--Zumino gauge, that is standard 
in supersymmetric gauge theories, introduces ultraviolet divergences in 
the propagators at the one-loop level. Such divergences are exactly 
cancelled when the contributions of the fields that are put to zero in 
the Wess--Zumino gauge are taken into account. 

In the description in terms of ${\cal N}$=1 superfields infrared 
divergences are found for every choice of gauge different from the 
supersymmetric generalization of the Fermi--Feynman gauge. Two-, 
three- and four-point functions of ${\cal N}$=1 superfields are 
computed and some general features of the infrared problem are discussed.

We also examine the effect of the introduction of mass terms for 
the (anti) chiral superfields in the theory, which break supersymmetry 
from ${\cal N}$=4 to ${\cal N}$=1. It is shown that in the mass 
deformed model no ultraviolet divergences appear in two-point 
functions. It argued that this result can be generalized to 
$n$-point functions, supporting the proposal of a possible of use of 
this modified model as a supersymmetry-preserving regularization scheme 
for ${\cal N}$=1 theories.
\end{abstract}
	
\newpage 

\setcounter{page}{1}

\section{Introduction}

The action of ${\cal N}$=4 four dimensional supersymmetric Yang--Mills 
theory was given for the first time in \cite{brinkscherkschwarz,gso} 
in the context of toroidal compactifications of the type I 
superstring. The theory has the maximal amount of supersymmetry 
allowed for a rigid supersymmetric theory in four dimensions, 
namely the symmetry generated by sixteen real supercharges, 
and has been proved to be finite, 
possessing a vanishing $\beta$-function. For this reason it has long 
been considered to be a rather trivial theory. However recent 
developments in the study of the correspondence with type IIB 
superstring theory on anti de Sitter space have shown that it is 
actually a very interesting quantum field theory, displaying very 
peculiar properties.

The field content of the theory, which is unique apart from the choice 
of the gauge group G, consists of six real scalars, four Weyl spinors 
and one vector, which are all in the adjoint representation of the 
gauge group G. In the Abelian case the theory is free, whereas in the 
non Abelian case it has a moduli space of vacua parameterized by the 
vacuum expectation values (vev's) of the six real scalars. In the 
Coulomb phase, reached giving non vanishing vev's to the scalars in 
the Cartan subalgebra of G, the theory is again free, while the origin 
of the moduli space corresponds to a highly non-trivial superconformal 
field theory.

The model is classically invariant under the ${\cal N}$=4 superconformal 
group and this property is supposed to be preserved at the quantum 
level \cite{sohniuswest,white}. Furthermore it has a global 
SU(4)$\sim$SO(6) symmetry, which naturally emerges in the 
compactification from ten dimensions on a six-torus and is identified, 
in a four dimensional perspective, with the R-symmetry group of 
automorphisms of the ${\cal N}$=4 superconformal algebra \cite{hagsohlop}.

The ${\cal N}$=4 super Yang-Mills theory is supposed to exactly 
realize a generalization of the electric-magnetic duality of Montonen 
and Olive \cite{olivemont} (S-duality), thanks to the presence in the 
spectrum of an infinite tower of stable BPS dyonic states 
\cite{osborn,n4qm,sen}.

At the perturbative level the theory is finite up 
to three loops. The $\beta$-function has been shown to be vanishing 
both in the component formulation \cite{sym1,2loop,tarasov} and using 
superspace techniques \cite{grirocsieg2,caswzanon}. In particular the 
formulation in terms of ${\cal N}$=1 superfields has proved to be an 
extremely powerful tool in the perturbative analysis. Strong arguments 
have been proposed in order to extend the proof of the finiteness to 
all orders in perturbation theory 
\cite{sohniuswest,white,n2finite,mandelstam} and moreover general 
considerations from instanton calculus and duality arguments suggest 
that the same should hold at the non-perturbative level \cite{dinesei}.

Following the proposal of Maldacena \cite{maldacena} of a correspondence 
relating (super) conformal field theories in $d$ dimensions and type 
IIB superstring theory on $d$+1 anti de Sitter space there has 
been recently a renewal of interest in ${\cal N}$=4 supersymmetric 
Yang--Mills theory. According to this conjectured duality both 
perturbative \cite{adspert} and non-perturbative \cite{adsnonpert} 
contributions to correlation functions of gauge invariant composite 
operators in ${\cal N}$=4 super Yang--Mills theory should be related 
to type IIB superstring amplitudes in AdS$_{5}\times S^{5}$. 
Work in this context is leading to very 
interesting results in the study of peculiar properties of 
${\cal N}$=4 Yang--Mills theory as a superconformal field theory. 

The aim of this paper is to present a careful re-analysis of perturbation 
theory, which allows to point out various problems related 
to the choice of gauge. Both in the component field formulation and 
using ${\cal N}$=1 superfields the gauge fixing procedure appears very 
subtle. As we will discuss in detail later, in both cases one finds 
ultraviolet and/or infrared divergences in off-shell Green functions. 

Throughout the paper calculations will be carried out in Euclidean 
space. Two different formulations of the ${\cal N}$=4 
supersymmetric Yang--Mills theory will be employed, one in terms 
of ${\cal N}$=1 superfields and the other in component (`physical') 
fields. Notations and conventions that will be used are 
those of \cite{wessbagger}. The ${\cal N}$=4 ``on-shell'' multiplet can 
be obtained by combining three ${\cal N}$=1 chiral superfields and one 
${\cal N}$=1 vector superfield, so that the six real scalars of the 
model are assembled into three complex scalars which, together with 
three of the Weyl spinors, form the chiral multiplets, the fourth 
spinor and the vector gives rise to the vector multiplet. In this 
description a SU(3)$\times$U(1) subgroup of the SU(4) R-symmetry group 
is manifest. The component-field formulation that will be used is 
directly related to that in ${\cal N}$=1 superspace, actually it is 
obtained from the former by integrating over the Grassmannian 
coordinates of ${\cal N}$=1 superspace.

In the component formulation the propagators of the 
elementary fields are ultraviolet divergent in the Wess--Zumino (WZ) 
gauge and these infinities are exactly cancelled when the contributions 
of the `gauge-dependent' fields, that are put to zero in the WZ gauge, 
are taken into account. The choice of the WZ gauge, that is almost 
unavoidable in explicit computations, introduces divergences that 
require a wave function renormalization. 

Different problems emerge in the formulation of the theory in terms 
of ${\cal N}$=1 superfields. Almost all of the calculations showing 
the vanishing of the quantum corrections to two- and three-point 
functions, that are presented in the literature, were performed in the 
supersymmetric generalization of the Fermi--Feynman gauge. With a 
different choice of gauge two- and three-point functions develop infrared 
singularities, leading to the result that 
the choice of the Fermi--Feynman gauge is somehow privileged. 
This conclusion was proposed for the first time in \cite{piguetrouet} 
and then it was discussed in the case of the ${\cal N}$=4 theory in 
\cite{storey}; however no possible explanation was proposed. Unlike 
those of the elementary fields, correlation functions of gauge 
invariant composite operators, that play a crucial r\^ole in the 
correspondence with AdS type IIB supergravity/superstring theory, 
should not suffer from problems related to the gauge fixing.

We will also discuss, in the superfield formulation, 
the effect of introducing of a mass term for the (anti) chiral 
superfields, which breaks supersymmetry from ${\cal N}$=4 down to 
${\cal N}$=1. It is shown that this deformation of the model does not 
modify the ultraviolet properties of the original theory. This result 
was first proposed in \cite{parkeswest}, where it was proved that the 
inclusion of  mass terms for the (anti) chiral superfields does not 
generate divergent corrections to the effective action. 
We will argue that this statement can be reinforced, showing that, at 
least at one loop, no new divergences, not even corresponding to wave  
function renormalizations, appear as a consequence of the addition of 
the mass terms. This result supports the claim put forward in 
\cite{japan,yoshida}, where the `mass deformed' ${\cal N}$=4 theory 
was proposed as a supersymmetry-preserving regularization scheme for 
a class of ${\cal N}$=1 theories.

The paper is organized as follows. Section \ref{wzproblems} deals with 
the problems introduced by the choice of the Wess--Zumino gauge 
when the component formalism is used. The following sections report 
calculations performed in the ${\cal N}$=1 superfield formalism of 
two-, three- and four-point functions. A discussion of the results is 
presented in the concluding section.

\section{Perturbation theory in components: problems with the 
Wess--Zumino gauge}
\label{wzproblems}

As already remarked the field content of ${\cal N}$=4 super 
Yang--Mills theory can be obtained by coupling in a gauge invariant 
way one ${\cal N}$=1 vector superfield 
$V_{a}(x,\theta,{\overline \theta})$ and three chiral superfields 
$\Phi_{a}^{I}(x,\theta,{\overline \theta})$, $I=1,2,3$, all in 
the adjoint representation of the gauge group G, so that the colour 
index $a$ takes the values $a=1,2,\ldots,{\rm dim~G}$. A 
SU(3)$\times$U(1) subgroup of the SU(4) R-symmetry group is manifest 
and under this global symmetry the chiral superfields $\Phi^{I}$ 
transform in the {\bf 3}, while the vector $V$ is a singlet. 

The complete expression for $V$ is 
\begin{eqnarray}
	&& \hspace{-1cm}V(x,\theta,{\overline \theta}) =C(x) + 
	i \theta \chi(x) - i {\overline \theta} {\overline \chi}(x) + 
	\frac{i}{2}\theta \theta S(x) 
	-\frac{i}{2} {\overline \theta}{\overline \theta} 
	S^{\dagger}(x) - \theta \sigma^{\mu}{\overline \theta} A_{\mu}(x) + 
	\nonumber \\
	&& \hspace{-1cm} + i\theta \theta {\overline \theta} 
	\left[ {\overline \lambda}(x) + 
	\frac{i}{2}{\overline \sigma}^{\mu}\partial_{\mu}\chi(x) \right] 
	-i{\overline \theta}{\overline \theta} \theta \left[ 
	\lambda(x) +\frac{i}{2}\sigma^{\mu}\partial_{\mu}{\overline \chi}(x) 
	\right]+\frac{1}{2} \theta \theta {\overline \theta}{\overline \theta} 
	\left[ D(x) + \frac{1}{2} \Box C(x) \right] .
\label{vector}
\end{eqnarray}
The gauge transformations of the superfields take the form 
\begin{displaymath}
	\Phi \longrightarrow \Phi^{\prime} = e^{-i \Lambda} \Phi \, , 
	\hspace{1.5cm} \Phi^{\dagger} \longrightarrow 
	\Phi^{\dagger \, \prime} = \Phi^{\dagger} e^{i \Lambda^{\dagger}}
\end{displaymath}
and 
\begin{displaymath}
	V \longrightarrow V^{\prime}   \hspace{1cm} {\rm where} 
	\hspace{1cm} e^{V^{\prime}} = 
	e^{-i\Lambda^{\dagger}} e^{V} e^{i\Lambda} \, ,
\end{displaymath}
where $\Lambda$ is a matrix-valued chiral superfield.
For infinitesimal gauge transformations use of the Becker--Hausdorff's 
formula allows to write 
\begin{equation}
	\delta V = V^{\prime} - V = i {\cal L}_{V/2} \left[ (\Lambda + 
	\Lambda^{\dagger} + \coth({\cal L}_{V/2}) (\Lambda - \Lambda^{\dagger})
	\right] \, ,
	\label{vvariaz}
\end{equation}
where ${\cal L}_{A}(B) = [A,B]$ is the Lie derivative and 
(\ref{vvariaz}) is a compact form to be understood in terms of the 
power expansion of $\coth({\cal L})$.
By explicitly writing the component expansion of these relations one 
can show that it is always possible to put to zero the lower 
components, $C$, $\chi$ and $S$, of $V$ by a suitable gauge 
transformation. As a result for the superfield $V$ one obtains
\begin{eqnarray*}
	V(x,\theta,{\overline \theta}) & = & -\theta\sigma^{\mu}
	{\overline \theta} A_{\mu}(x) +i \theta \theta 
	{\overline \theta} {\overline \lambda}(x) - i 
	{\overline \theta}{\overline \theta}  \theta \lambda(x) +\frac{1}{2}
	\theta\theta {\overline \theta}{\overline \theta} D(x) \\
	V^{2}(x,\theta,{\overline \theta}) & = & -\frac{1}{2} 
	\theta\theta {\overline \theta}{\overline \theta} A_{\mu}(x)
	A^{\mu}(x) \\
	V^{n}(x,\theta,{\overline \theta}) & = & 0 \, , \hspace{1cm}
	\forall n \geq 3  \: .
\end{eqnarray*}
This choice is known as the Wess--Zumino gauge. Fixing the 
Wess--Zumino gauge still leaves with the ordinary non Abelian gauge 
freedom on the remaining fields.

In this section we will discuss perturbation theory in components. 
The one-loop  correction to the propagators of elementary fields will 
be computed first in the Wess-Zumino gauge and then taking into 
account the contribution of the fields $C$, $\chi$ and $S$ that are 
absent in this gauge. We will show that the choice of 
the WZ gauge introduces ultraviolet divergences in the propagators. 

The standard formulation of the model is obtained by eliminating the 
auxiliary fields, $F^{I}$ from the chiral multiplet and $D$ from the 
vector multiplet. This procedure in the WZ gauge results in a 
polynomial action with a scalar potential containing a quadrilinear 
term for the scalars. However in order to correctly 
deal with the lower components of the vector superfield $V$, that are 
put to zero in the Wess--Zumino gauge, it is more convenient not to 
eliminate the auxiliary fields through their equations of 
motion, but rather keep them in the action: in the computation of 
Green functions the corresponding $x$-space propagators are 
$\delta$-functions. 
 
The action is written in terms of the non Abelian field strength 
superfield $W_{\alpha}$ 
\begin{equation}
	W_{\alpha}=-\frac{1}{4} {\overline D}{\overline D} e^{-V}
	D_{\alpha}e^{V}=\sum_{k=1}^{\infty} W_{\alpha}^{(k)} \, ,
	\label{wexpansion}
\end{equation}
where 
\begin{equation}
\begin{array}{ccl}
	\begin{displaystyle}W_{\alpha}^{(1)} \end{displaystyle}
	&=& \begin{displaystyle} -\frac{1}{4}{\overline D}{\overline D}
	D_{\alpha}V \end{displaystyle} \\
    \begin{displaystyle}W_{\alpha}^{(2)} \end{displaystyle} 
    &=& \begin{displaystyle} \frac{1}{8}{\overline D}{\overline D}
	[V,D_{\alpha}V] \rule{0pt}{24pt}\end{displaystyle}
	\label{w1-2}
\end{array}
\end{equation}
and the terms $W_{\alpha}^{(k)}$, with $k\geq 3$, contain $k$ factors of 
$V$ and vanish in the WZ gauge. 
In Euclidean space the action in the ${\cal N}$=1 superfield 
formalism takes the form
\begin{eqnarray}
	S^{({\rm E})} & = & \int d^{4}x \, d^{2}\theta d^{2}{\overline\theta}
	\, \left\{ \left[ 
	\frac{1}{4} W^{(1)\alpha} W^{(1)}_{\alpha} \delta({\overline \theta})
	+\frac{1}{4} {\overline W}^{(1)}_{\dot \alpha} 
	{\overline W}^{(1)\dot \alpha}\delta(\theta) - \frac{1}{8\alpha} 
	{\overline D}^{2} V D^{2} V \right] + \right. \nonumber \\
	&+& \left. \Phi_{I}^{\dagger} V \Phi^{I} + \left[ \left( \frac{1}{2} 
	W^{(1)\alpha} W^{(2)}_{\alpha} 
	+\frac{1}{4} W^{(2)\alpha} W^{(2)}_{\alpha} \right) \delta(
	{\overline \theta}) + \nonumber \right. \right. \\ 
	&+&  \left. \left. \left( \frac{1}{2} {\overline W}^{(1)}_{\dot 
	\alpha} {\overline W}^{(2)\dot \alpha} + 
	\frac{1}{4} {\overline W}^{(2)}_{\dot \alpha} 
	{\overline W}^{(2)\dot \alpha} \right) \delta(\theta) \right] + 
	\frac{1}{2} \Phi_{I}^{\dagger} V^{2} \Phi^{I}  + \ldots \right\}
	\label{relevantaction} \, ,
\end{eqnarray}
where a standard gauge fixing term has been included, whereas no ghost 
term is displayed since it will not be relevant for the computations 
to be discussed in this section. 
In equation (\ref{relevantaction}) dots denote terms 
of higher order in $V$, which do not contribute to the Green functions 
that will be considered and thus will be suppressed from now on. 
In the following calculations the 
Fermi--Feynman gauge, $\alpha$=1, will be used. With this choice one 
obtains\footnote{From now on the superscript E will be suppressed. 
Euclidean signature is to be understood unless otherwise stated.}
\begin{eqnarray}
	S & = & \int d^{4}x \, d^{2}\theta d^{2}{\overline\theta}
	\, \left\{ \left[ 
	\frac{1}{2} V \Box V + \Phi_{I}^{\dagger} \Phi^{I} \right] + 
	\left[ \left( \frac{1}{2} W^{(1)\alpha} W^{(2)}_{\alpha} 
	+\frac{1}{4} W^{(2)\alpha} W^{(2)}_{\alpha} \right) \delta(
	{\overline \theta}) + \right. \right. \nonumber \\ 
	&+& \hspace{-0.2cm} \left. \left. \left( \frac{1}{2} 
	{\overline W}^{(1)}_{\dot \alpha} {\overline W}^{(2)\dot \alpha} + 
	\frac{1}{4} {\overline W}^{(2)}_{\dot \alpha} 
	{\overline W}^{(2)\dot \alpha} \right) \delta(\theta) \right] + 
	\left[ \Phi_{I}^{\dagger} V \Phi^{I} + \frac{1}{2} 
	\Phi_{I}^{\dagger} V^{2} \Phi^{I} \right] \right\} \, ,
\label{ffaction}
\end{eqnarray}
where from the definition (\ref{w1-2}) one gets
\begin{eqnarray}
	W^{(1)}_{\alpha} & = & -i\lambda_{\alpha} + \left[ 
	\delta_{\alpha}{}^{\beta} D -\frac{1}{2} 
	\delta_{\alpha}{}^{\beta} \Box C 
	-\frac{i}{2} (\sigma^{\mu}{\overline \sigma}^{\nu})_{\alpha}{}^{\beta}
	\left( \partial_{\mu} A_{\nu}-\partial_{\nu} A_{\mu} \right) \right] 
	\theta_{\beta} + \nonumber \\  
	 && + \theta \theta \sigma^{\mu}_{\alpha{\dot \alpha}} 
	\partial_{\mu}{\overline \lambda}^{\dot \alpha} \, .
	\label{w1}
\end{eqnarray}
The explicit form of $W^{(2)}_{\alpha}$ is not necessary for 
the moment (see however equation (\ref{w2nonwz}) at the end of this 
section). Expansion of the action (\ref{ffaction}) in components 
using the complete expression of $V$ gives
\begin{displaymath}
	S = S_{0}+S_{{\rm int}} \, .
\end{displaymath}
The free action $S_{0}$ comes from the terms $\frac{1}{2} V \Box V$ 
and $\Phi_{I}^{\dagger} \Phi^{I}$ and reads
\begin{eqnarray}
	S_{0} & = & \int d^{4}x \, \left\{ \left[ 
	(\partial_{\mu}\varphi^{a\dagger}_{I})(\partial^{\mu}
	\varphi_{a}^{I}) + {\overline \psi}_{I}^{a}{\overline \sigma}^{\mu} 
	(\partial_{\mu}\psi^{I}_{a}) + F^{a\dagger}_{I}F^{I}_{a} \right] + 
	\left[ \frac{1}{2} S^{\dagger}_{a}\Box S^{a} +
	\right. \right. \nonumber \\
	&& \hspace{-1.6cm} \left. \left. -\frac{1}{2} 
	C^{a} \Box D_{a} -\frac{1}{2} D^{a}\Box C_{a} - \frac{1}{2} C^{a} 
	\Box^{2} C_{a} + \frac{1}{2} \chi^{a} \Box \lambda_{a} + \frac{1}{2} 
	{\overline \chi}^{a}\Box {\overline \lambda}_{a} + 
	\frac{1}{2} \lambda^{a}\Box \chi_{a} + 
	\right. \right. \nonumber \\ 
	&& \hspace{-1.6cm} \left. \left. + \frac{1}{2} 
	{\overline \lambda}^{a}\Box {\overline \chi}_{a} + \frac{1}{2} 
	\chi^{a} \Box \sigma^{\mu} (\partial_{\mu}{\overline \chi}_{a}) + 
	\frac{1}{2} {\overline \chi}^{a} \Box {\overline \sigma}^{\mu} 
	(\partial_{\mu}\chi_{a}) +\frac{1}{2} (\partial_{\mu}A^{a}_{\nu})
	(\partial_{\nu}A_{a\mu}) \right] \right\} .
	\label{freeact}
\end{eqnarray}
Note that the scalar field $C$ and the spinor $\chi$ have ``wrong'' 
physical dimension, resulting in non standard free propagators, as will 
be shown below. 
The interaction part $S_{{\rm int}}$ contains an infinite number of 
terms. The propagators of the fermions $\psi^{I}$ and of the scalars 
$\varphi^{I}$ belonging to the ${\cal N}$=1 chiral multiplets will 
now be computed at one loop. The terms that are relevant for these 
calculations come from the expansion of $\Phi^{\dagger}V\Phi$ and 
$\Phi^{\dagger}V^{2}\Phi$ in (\ref{ffaction}). The latter generates 
tadpole type diagrams in the propagator $\langle \varphi^{\dagger}
\varphi \rangle$, but not in $\langle {\overline \psi} \psi \rangle$. 
The interaction part of the action to be considered is 
\begin{eqnarray}
	S_{{\rm int}} & = & \int d^{4}x \,\left\{ ig f_{abc} \left( \left[ 
	-\frac{1}{4} \varphi_{I}^{a\dagger} C^{b}(\Box \varphi^{cI}) - 
	\frac{1}{4} \varphi_{I}^{a\dagger} (\Box C^{b}) \varphi^{cI} - 
	\frac{1}{4} (\Box \varphi_{I}^{a\dagger}) C^{b} \varphi^{cI} + 
	\right. \right. \right. \nonumber \\
	&&\hspace{-1cm} \left. \left. \left. + \frac{1}{2} (\partial_{\mu}
	\varphi^{a\dagger}_{I}) C^{b}(\partial^{\mu}
	\varphi^{cI}) + \frac{i}{2} \left( \varphi^{a\dagger}
	S^{b\dagger}F^{cI} - F^{a\dagger}_{I} S^{b}\varphi^{cI} \right) - 
	\frac{1}{2} \varphi^{a\dagger}D^{b\dagger}\varphi^{cI} + 
	\right. \right. \right. \nonumber \\
	&&\hspace{-1cm} \left. \left. \left. - \frac{i}{2} 
	\left( \varphi^{a\dagger}_{I}A_{\mu}^{b}(\partial^{\mu}
	\varphi^{cI}) - (\partial^{\mu}\varphi^{a\dagger}_{I}) 
	A_{\mu}^{b}\varphi^{cI} \right) - F^{a\dagger}_{I}C^{b}F^{cI} + 
	\frac{i}{\sqrt{2}} \left( {\overline \psi}^{a}_{I}
	{\overline \lambda}^{b}\varphi^{cI} + 
	\nonumber \right. \right. \right. \right. \\
	&&\hspace{-1cm} \left. \left. \left. \left. - 
	\varphi^{a\dagger}\lambda^{b}\psi^{cI} \right) + \frac{i}{\sqrt{2}}
	\left( F^{a\dagger}_{I}\chi^{b}\psi^{cI} - {\overline \psi}^{a}_{I} 
	{\overline \chi}^{b}F^{cI} \right) + \frac{i}{\sqrt{2}} \left( 
	\varphi^{a\dagger}(\partial_{\mu}{\overline 
	\chi}^{b})\sigma^{\mu}\psi^{cI} + 
	\nonumber \right. \right. \right. \right. \\ 
	&&\hspace{-1cm} \left. \left. \left. \left. 
	+ {\overline \psi}^{a}_{I}{\overline 
	\sigma}^{\mu}(\partial_{\mu}\chi^{b}) \varphi^{cI} \right) +
	\frac{1}{2} \left( C^{b}{\overline \psi}^{a}_{I}{\overline 
	\sigma}^{\mu}(\partial_{\mu}\psi^{cI}) - 
	C^{b}(\partial_{\mu}{\overline \psi}^{a}_{I}){\overline \sigma}^{\mu} 
	\psi^{cI} \right) +
	\right. \right. \right. \nonumber \\  
	&&\hspace{-1cm} \left. \left. \left. 
	-\frac{i}{2} {\overline \psi}^{a}_{I}{\overline 
	\sigma}^{\mu}\psi^{cI}A_{\mu}^{b} \right]
	- \frac{1}{2} \left[ \varepsilon^{IJK} \left( \varphi^{a\dagger}_{I} 
	\varphi^{b\dagger}_{J}F^{c\dagger}_{K} - 
	\varphi_{I}^{a\dagger}{\overline\psi}^{b}_{J}{\overline\psi}^{c}_{K}
	\right) + \right. \right. \right. \nonumber \\
	&&\hspace{-1cm} + \left. \left. \left.
	\varepsilon_{IJK} \left( \varphi^{aI}\varphi^{bJ} F^{cK} - 
	\varphi^{aI} \psi^{bJ}\psi^{cK} \right) \rule{0pt}{14pt} \right]
	\right) -\frac{g^{2}}{2} f_{abe}f^{e}{}_{cd}\left( \left[ \rule{0pt}
	{18pt} -C^{b}\psi^{d} \sigma^{\mu}{\overline \psi}^{a} A^{c}_{\mu} + 
	\right. \right. \right. \nonumber \\
	&&\hspace{-1cm} \left. \left. \left. + 
	({\overline \chi}^{c}{\overline \psi}^{d})(\chi^{b} \psi^{a}) +
	\frac{i}{2} \left( \partial_{\mu} 
	\psi^{d} \sigma^{\mu} {\overline \psi}^{a} - \psi^{d} \sigma^{\mu}  
	\partial_{\mu}{\overline \psi}^{a} \right)
	C^{b}C^{c} \right] + \varphi^{a\dagger} \left[ C^{b} \left( D^{c}+
	\rule{0pt}{17pt} 
	\right. \right. \right. \right. \nonumber \\ 
	&&\hspace{-1cm} \left. \left. \left. \left. +
	\frac{1}{2} \Box C^{c} \right)  - 
	\chi^{b} \left( \lambda^{c} + \frac{i}{2} 
	\sigma^{\mu}\partial_{\mu}{\overline \chi}^{c} \right) - 
	{\overline \chi}^{b} \left( {\overline \lambda}^{c} + \frac{i}{2} 
	\sigma^{\mu} \partial_{\mu} \chi^{c}\right) + \frac{1}{2} 
	S^{b\dagger}S^{c} + \rule{0pt}{17pt} 
	\right. \right. \right. \nonumber \\
	&&\hspace{-1cm} \left.\left.\left. 
	-\frac{1}{2} A^{bc}_{\mu}A^{c\mu}\right] 
	\varphi^{d} + \frac{i}{2} C^{b} A^{c\mu} 
	\left( \varphi^{a\dagger} (\partial_{\mu} \varphi^{d}) - 
	(\partial_{\mu}\varphi^{a\dagger}) \varphi^{d} \right) +
	\frac{1}{4} C^{b}C^{c}\left[ \varphi^{a\dagger}(\Box \varphi^{d})+ 
	 \right. \right. \right. \nonumber \\ 
	&&\hspace{-1cm} \left. \left. \left. + (\Box \varphi^{a\dagger}) 
	\varphi^{d} \right] +\frac{i}{2} \chi^{b}\sigma^{\mu}
	{\overline \chi}^{c} \left[ \varphi^{a\dagger}
	(\partial_{\mu} \varphi^{d}) - 
	(\partial_{\mu}\varphi^{a\dagger}) \varphi^{d} \right] 
	\rule{0pt}{18pt}\right) \right\} \, .
	\label{intaction}
\end{eqnarray}
The  quadratic part of the action, $S_{0}$, can be written in a more 
compact form introducing the notation 
\begin{equation}
	{\cal B}^{a}(x) = \left( \begin{array}{c} C^{a}(x) \\ 
	\rule{0pt}{16pt} D^{a}(x) \end{array} \right) \, , 
	\hspace{1cm} {\cal F}^{a}(x) = \left( \begin{array}{c} \chi^{a}(x) \\
	{\overline \chi}^{a}(x) \rule{0pt}{16pt} \\ \lambda^{a}(x) 
	\rule{0pt}{16pt} \\ {\overline \lambda}^{a}(x) \rule{0pt}{16pt} 
	\end{array} \right) \, , 
	\label{fieldvector}
\end{equation}
so that 
\begin{eqnarray}
	S_{0} &=& \int d^{4}x \, \left\{ \left[ 
	(\partial_{\mu}\varphi^{a\dagger}_{I})(\partial^{\mu}
	\varphi_{a}^{I}) + {\overline \psi}_{I}^{a}{\overline \sigma}^{\mu} 
	(\partial_{\mu}\psi^{I}_{a}) + F^{a\dagger}_{I}F^{I}_{a} \right] + 
	\right. \nonumber \\ 
	&& + \left. \left[ \frac{1}{2} S^{\dagger}_{a}\Box S^{a} + 
	\frac{1}{2} A^{a}_{\mu}A_{a}^{\mu} + {\cal B}_{a}^{T} M {\cal B}^{a} 
	+ {\cal F}_{a}^{T} N {\cal F}^{a} \right] \right\} \, , 
	\label{compactfree}
\end{eqnarray}
where 
\begin{equation}
	M = \frac{1}{2} \left( \begin{array}{cc} \Box^{2} & \Box \\ 
	\Box & 0 \end{array} \right) \, , \hspace{0.5cm} 
	N = \frac{1}{2} \left( \begin{array}{cccc} 0 & \Box 
	\sigma^{\mu}\partial_{\mu} & -\Box & 0 \\ \Box 
	{\overline \sigma}^{\mu}\partial_{\mu} & 0 & 0 & -\Box \\
	-\Box & 0 & 0 & 0 \\ 0 & -\Box & 0 & 0 \end{array} \right) \, .
	\label{kineticmatrices}
\end{equation}
Inverting the kinetic matrices $M$ and $N$ one gets the free 
propagators. From (\ref{kineticmatrices}) it follows 
\begin{equation}
	M^{-1} = \left( \begin{array}{cc} 0 & \begin{displaystyle}
	\frac{1}{\Box} \end{displaystyle} \\ \begin{displaystyle}
	\frac{1}{\Box} \end{displaystyle} & 1 \end{array} \right) \, ,
	\hspace{0.5cm} N^{-1} = \left( \begin{array}{cccc} 0 & 0 & 
	\begin{displaystyle} -\frac{1}{\Box} \end{displaystyle} & 0 \\
	0 & 0 & 0 & \begin{displaystyle} -\frac{1}{\Box} \end{displaystyle} 
	\\ \begin{displaystyle} -\frac{1}{\Box} \end{displaystyle} & 0 & 0 
	& \begin{displaystyle} \frac{\rule{0pt}{14pt}\sigma^{\mu}
	\partial_{\mu}}{\Box} \end{displaystyle} \\ 
	0 & \begin{displaystyle} -\frac{1}{\Box} 
    \end{displaystyle} & \begin{displaystyle} \frac{\rule{0pt}{14pt}
	{\overline \sigma}^{\mu} \partial_{\mu}}{\Box} \end{displaystyle} 
	& 0 \end{array} \right) \, ,
	\label{inverskinmatr}
\end{equation}
so that the free propagators are \\
\hspace*{0.5cm}
\begin{fmffile}{phi+phi}
\begin{fmfgraph*}(90,50) 
  \fmfleft{i1} \fmfright{o1}
  \fmf{scalar}{i1,o1}
  \fmflabel{\raisebox{22pt}{\hspace{-20pt} $I,a$ \hspace{20pt}} 
  \raisebox{12pt}{$\longrightarrow \hspace{0.2cm} \langle 
  \varphi^{b\dagger}_{J}(x) \varphi^{I}_{a}(y) \rangle_{{\rm free}} = 
  - \begin{displaystyle} \frac{\delta^{b}_{a}\delta^{I}_{J}}{\Box} 
  \end{displaystyle}\delta(x-y)=\Delta^{bI}_{aJ}(x-y)$}}{o1}
  \fmflabel{\raisebox{22pt}{\hspace{18pt} $J,b$ 
  \hspace{-18pt}}}{i1}
\end{fmfgraph*}   
\end{fmffile} \\
\hspace*{0.5cm}
\begin{fmffile}{S+S}
\begin{fmfgraph*}(90,50) 
  \fmfleft{i1} \fmfright{o1}
  \fmf{scalar}{i1,o1}
  \fmflabel{\raisebox{22pt}{\hspace{-20pt} $a$ \hspace{20pt}} 
  \raisebox{12pt}{$\longrightarrow \hspace{0.2cm} 
  \langle S^{b\dagger}(x) S_{a}(y) \rangle_{{\rm free}} = 
  \begin{displaystyle} \frac{2\delta^{b}_{a}}{\Box}\end{displaystyle}
  \delta(x-y)=2\Delta^{b}_{a}(x-y)$}}{o1}
  \fmflabel{\raisebox{22pt}{\hspace{18pt} $b$ 
  \hspace{-18pt}}}{i1}
\end{fmfgraph*}   
\end{fmffile} \\
\hspace*{0.5cm} 
\begin{fmffile}{CD}
\begin{fmfgraph*}(90,50) 
  \fmfleft{i1} \fmfright{o1}
  \fmf{dashes}{i1,v}
  \fmf{dashes}{v,o1}
  \fmfv{d.sh=cross,d.size=3mm}{v}
  \fmflabel{\raisebox{22pt}{\hspace{-20pt} $a$ \hspace{20pt}} 
  \raisebox{12pt}{$\longrightarrow \hspace{0.2cm} 
  \langle C^{b}(x) D_{a}(y) \rangle_{{\rm free}} = 
  \begin{displaystyle} \frac{\delta^{b}_{a}}{\Box}\end{displaystyle}
  \delta(x-y)=\Delta^{b}_{a}(x-y)$}}{o1}
  \fmflabel{\raisebox{22pt}{\hspace{18pt} $b$ 
  \hspace{-18pt}}}{i1}
\end{fmfgraph*}   
\end{fmffile} \\
\hspace*{0.5cm}
\begin{fmffile}{DD}
\begin{fmfgraph*}(90,50) 
  \fmfleft{i1} \fmfright{o1}
  \fmf{dashes}{i1,o1}
  \fmflabel{\raisebox{22pt}{\hspace{-20pt} $a$ \hspace{20pt}} 
  \raisebox{12pt}{$\longrightarrow \hspace{0.2cm} \langle D^{b}(x) 
  D_{a}(y) \rangle_{{\rm free}} = \delta^{b}_{a}\delta(x-y)$}}{o1}
  \fmflabel{\raisebox{22pt}{\hspace{18pt} $b$ 
  \hspace{-18pt}}}{i1}
\end{fmfgraph*}   
\end{fmffile} \\
\hspace*{0.5cm}
\begin{fmffile}{F+F}
\begin{fmfgraph*}(90,50) 
  \fmfleft{i1} \fmfright{o1}
  \fmf{scalar}{i1,o1}
  \fmflabel{\raisebox{22pt}{\hspace{-20pt} $I,a$ \hspace{20pt}} 
  \raisebox{12pt}{$\longrightarrow \hspace{0.2cm} 
  \langle F^{b\dagger}_{J}(x) F^{I}_{a}(y) \rangle_{{\rm free}} = 
  \delta^{I}_{J}\delta^{b}_{a}\delta(x-y)$}}{o1}
  \fmflabel{\raisebox{22pt}{\hspace{18pt} $J,b$ 
  \hspace{-18pt}}}{i1}
\end{fmfgraph*}   
\end{fmffile} \\
\hspace*{0.5cm}
\begin{fmffile}{AA}
\begin{fmfgraph*}(90,50) 
  \fmfleft{i1} \fmfright{o1}
  \fmf{boson}{i1,o1}
  \fmflabel{\raisebox{22pt}{\hspace{-20pt} $\nu,a$ \hspace{10pt}} 
  \raisebox{12pt}{$\longrightarrow \hspace{0.2cm} 
  \langle A^{b}_{\mu}(x)A_{a\nu}(y)\rangle_{{\rm free}}=\begin{displaystyle} 
  -\frac{\delta_{\mu\nu}\delta^{b}_{a}}{\Box} \end{displaystyle} 
  \delta(x-y)=\Delta^{b}_{a\mu\nu}(x-y)$}}{o1}
  \fmflabel{\raisebox{22pt}{\hspace{18pt} $\mu,b$ 
  \hspace{-18pt}}}{i1}
\end{fmfgraph*}   
\end{fmffile} \\
\hspace*{0.5cm}
\begin{fmffile}{chilam}
\begin{fmfgraph*}(90,50) 
  \fmfleft{i1} \fmfright{o1}
  \fmf{fermion}{v,i1}
  \fmf{fermion}{v,o1}
  \fmfv{d.sh=cross,d.size=3mm}{v}
  \fmflabel{\raisebox{22pt}{\hspace{-20pt} $\beta,a$ \hspace{10pt}} 
  \raisebox{12pt}{$\longrightarrow \hspace{0.2cm} \langle 
  \chi^{b\alpha}(x) \lambda_{a}^{\beta}(y) \rangle_{{\rm free}} = 
  \begin{displaystyle} -\frac{\varepsilon^{\alpha\beta}\delta^{b}_{a}}
  {\Box} \end{displaystyle}
  \delta(x-y)=R_{a}^{b\alpha\beta}(x-y)$}}{o1}  
  \fmflabel{\raisebox{22pt}{\hspace{18pt} $\alpha,b$ 
  \hspace{-18pt}}}{i1}
\end{fmfgraph*}
\end{fmffile} \\
\hspace*{0.5cm}
\begin{fmffile}{bchiblam}
\begin{fmfgraph*}(90,50) 
  \fmfleft{i1} \fmfright{o1}
  \fmf{fermion}{i1,v}
  \fmf{fermion}{o1,v}
  \fmfv{d.sh=cross,d.size=3mm}{v}
  \fmflabel{\raisebox{22pt}{\hspace{-20pt} ${\dot\beta},a$\hspace{10pt}} 
  \raisebox{12pt}{$\longrightarrow \hspace{0.2cm} 
  \langle {\overline \chi}^{b\dot\alpha}
  (x) {\overline \lambda}_{a}^{\dot\beta}(y) \rangle_{{\rm free}} = 
  \begin{displaystyle} - \frac{\varepsilon^{{\dot\alpha}{\dot\beta}}
  \delta^{b}_{a}}{\Box} \end{displaystyle}\delta(x-y)=
  {\overline R}_{a}^{b{\dot\alpha}{\dot\beta}}(x-y)$}}{o1}
  \fmflabel{\raisebox{22pt}{\hspace{18pt} ${\dot\alpha},b$ 
  \hspace{-18pt}}}{i1}  
\end{fmfgraph*} 
\end{fmffile} \\
\hspace*{0.5cm}
\begin{fmffile}{blamlam}
\begin{fmfgraph*}(90,50) 
  \fmfleft{i1} \fmfright{o1}
  \fmf{fermion}{i1,o1}
  \fmflabel{\raisebox{-20pt}{\hspace{-20pt} $\alpha,a$ \hspace{20pt}} 
  \raisebox{-30pt}{$\longrightarrow \hspace{0.2cm} 
  \langle {\overline \lambda}^{b\dot\alpha}(x) \lambda_{a}^{\alpha}(y) 
  \rangle_{{\rm free}} = \begin{displaystyle} 
  \frac{\delta^{b}_{a}{\overline \sigma}_{{\dot\alpha}\alpha}^{\mu}
  \partial_{\mu}}{\Box} 
  \end{displaystyle} \delta(x-y)=$
  \raisebox{-1cm}{\hspace{-4cm}=${\overline S}^{b}_{a{\dot\alpha}\alpha}
  (x-y)$}}}{o1}
  \fmflabel{\raisebox{22pt}{\hspace{18pt} ${\dot\alpha},b$ 
  \hspace{-18pt}}}{i1}  
\end{fmfgraph*}   
\end{fmffile} 

\vspace{0.3cm}
\noindent
\hspace*{0.5cm}
\begin{fmffile}{bpsipsi}
\begin{fmfgraph*}(90,50) 
  \fmfleft{i1} \fmfright{o1}
  \fmf{fermion}{i1,o1}
  \fmflabel{\raisebox{-20pt}{\hspace{-25pt} $\alpha,J,a$ \hspace{20pt}} 
  \raisebox{-30pt}{$\longrightarrow \hspace{0.2cm} 
  \langle {\overline \psi}_{I}^{b\dot\alpha}(x)\psi_{a}^{\alpha J}(y) 
  \rangle_{{\rm free}} = \begin{displaystyle} 
  \frac{\delta^{b}_{a}\delta^{J}_{I}{\overline \sigma}_{{\dot\alpha}
  \alpha}^{\mu}\partial_{\mu}}{\Box} 
  \end{displaystyle} \delta(x-y)=$
  \raisebox{-1cm}{\hspace{-4.35cm}=${\overline S}^{bJ}_{aI{\dot\alpha}
  \alpha}(x-y)$}}}{o1}
  \fmflabel{\raisebox{22pt}{\hspace{23pt} ${\dot\alpha},I,b$ 
  \hspace{-18pt}}}{i1}
\end{fmfgraph*}   
\end{fmffile}

\vspace{0.3cm}
\noindent
To summarize beyond the ordinary propagators for the physical 
fields, $\varphi$, $\psi$, $\lambda$ and $A_{\mu}$, and those for the 
auxiliary fields, $F$ and $D$, one further obtains the propagators 
$\langle S^{\dagger}S\rangle$, $\langle CD\rangle$, $\langle 
\chi\lambda \rangle$ and $\langle {\overline \chi} {\overline \lambda}
\rangle$. The latter are absent in the Wess--Zumino gauge.

\subsection{One loop corrections to the propagator of the fermions 
belonging to the chiral multiplet}

The one-loop correction to the propagator of the fermions $\psi^{I}$ 
in the chiral multiplet is the simplest to compute. 
In the WZ gauge there are three 
contributions at the one loop level, that will be shown to lead to a 
logarithmically divergent result. 

From the action (\ref{freeact}), (\ref{intaction}), with 
$C$=$\chi$=$S$=0 
one obtains the following three diagrams 

\vspace{0.3cm}
\noindent
\hspace*{1cm}
\begin{fmffile}{psipsiA}
 \begin{fmfgraph*}(150,80) 
  \fmfleft{i1} \fmfright{o1}
  \fmf{fermion,tension=1.2}{i1,v1}
  \fmf{boson,label=$A_{\mu}A_{\nu}$,left=0.6,tension=0.5}{v1,v2}
  \fmf{fermion,label=${\overline\psi}\psi$,right=0.6,tension=0.5}{v1,v2}
  \fmf{fermion,tension=1.2}{v2,o1}
  \fmfv{l=${\overline \psi}^{a}_{J}(x)$,l.a=180}{i1}
  \fmfv{l=$\psi^{I}_{b}(y)$ \hspace{0.5cm} $\longrightarrow ~~~ 
        A^{aI}_{bJ}(x;y)$,l.a=0}{o1}
  \fmfdot{v1,v2}
 \end{fmfgraph*}   
\end{fmffile} \\
\hspace*{1cm}
\begin{fmffile}{psipsiB}
 \begin{fmfgraph*}(150,80) 
  \fmfleft{i1} \fmfright{o1}
  \fmf{fermion,tension=1.2}{i1,v1}
  \fmf{scalar,label=$\varphi\varphi^{\dagger}$,left=0.6,tension=0.5}
      {v2,v1}
  \fmf{fermion,label=$\psi{\overline\psi}$,right=0.6,tension=0.5}{v2,v1}
  \fmf{fermion,tension=1.2}{v2,o1}
  \fmfv{l=${\overline \psi}^{a}_{J}(x)$,l.a=180}{i1}
  \fmfv{l=$\psi^{I}_{b}(y)$ \hspace{0.5cm} $\longrightarrow ~~~ 
        B^{aI}_{bJ}(x;y)$,l.a=0}{o1}
  \fmfdot{v1,v2}
 \end{fmfgraph*}   
\end{fmffile} \\
\hspace*{1cm}
\begin{fmffile}{psipsiC}
 \begin{fmfgraph*}(150,80) 
  \fmfleft{i1} \fmfright{o1}
  \fmf{fermion,tension=1.2}{i1,v1}
  \fmf{scalar,label=$\varphi^{\dagger}\varphi$,right=0.6,tension=0.5}
      {v1,v2}
  \fmf{fermion,label=$\lambda{\overline\lambda}$,right=0.6,tension=0.5}
      {v2,v1}
  \fmf{fermion,tension=1.2}{v2,o1}
  \fmfv{l=${\overline \psi}^{a}_{J}(x)$,l.a=180}{i1}
  \fmfv{l=$\psi^{I}_{b}(y)$ \hspace{0.5cm} $\longrightarrow ~~~ 
        C^{aI}_{bJ}(x;y)$,l.a=0}{o1}
  \fmfdot{v1,v2}
 \end{fmfgraph*}   
\end{fmffile} 

Notice that the insertion of tadpoles such as 

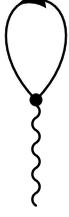
\vspace*{2cm}
\noindent
\hspace*{0.7cm}
\begin{fmffile}{tadpole}
 \begin{fmfgraph*}(100,40) 
  \fmfleft{i1,i2} \fmfright{o1,o2}
  \fmf{phantom,tension=1}{i1,v1}
  \fmf{phantom,tension=1}{v1,o1}
  \fmf{boson,tension=0}{v1,v2}
  \fmf{phantom,tension=1}{i2,v2}
  \fmf{phantom,tension=1}{v2,o2}
  \fmffreeze  
  \fmf{fermion,tension=0.7}{v2,v2}
  \fmfdot{v2}
 \end{fmfgraph*}   
\end{fmffile} 

\vspace*{0.3cm}
\noindent 
in a diagram gives a vanishing result because all the propagators are 
diagonal in colour space, so that the tadpole contains a factor 
$\delta_{ab}f^{abc}\equiv 0$. The same is true also for diagrams in 
${\cal N}$=1 superspace that will be discussed in subsequent sections. 

The three contributions depicted above can be easily evaluated and 
give the results
\begin{eqnarray}
	A^{{\dot\alpha}\alpha aI}_{bJ}(x;y) &=& -\frac{1}{4}g^{2}
	f^{d}{}_{ef}f^{l}{}_{mn}\int d^{4}x_{1}
	d^{4}x_{2} \, \left\{ \Delta^{me}_{\nu\mu}(x_{2}-x_{1}) \left[
	{\overline S}^{{\dot\alpha}\beta 
	aI}_{lL}(x-x_{2})\sigma^{\nu}_{\beta{\dot \gamma}} \cdot
	\right. \right.\nonumber \\
	&& \left. \left. \hspace{2cm}  \cdot 
	{\overline S}^{{\dot\gamma}\gamma nL}_{dK}(x_{2}-x_{1})  
	\sigma^{\mu}_{\gamma{\dot\beta}} {\overline S}^{{\dot\beta}\alpha 
	fK}_{bJ}(x_{1}-y)\right] \right\} \, ,
	\label{bpsipsi-a}
\end{eqnarray}
\begin{eqnarray}
    B^{{\dot\alpha}\alpha aI}_{bJ}(x;y) &=& -\frac{1}{4}g^{2}
    \varepsilon^{LMN}\varepsilon^{PQ}{}_{R}f_{de}{}^{f}f_{lmn}\int 
    d^{4}x_{1}d^{4}x_{2} \, \left\{ \Delta^{ld}_{PL}(x_{2}-x_{1}) 
    \cdot \right. \nonumber \\ 
    && \left. \cdot
    \left[ {\overline S}^{{\dot\alpha}\beta 
	aI}_{fN}(x-x_{2}) S_{\beta{\dot\beta}MQ}^{em}(x_{2}-x_{1})
	{\overline S}^{{\dot\beta}\alpha 
	nR}_{bJ}(x_{1}-y)\right] \right\} \, ,
\end{eqnarray}
\begin{eqnarray}
    C^{{\dot\alpha}\alpha aI}_{bJ}(x;y) &=& -\frac{1}{2}g^{2}
    f_{de}{}^{f}f_{lm}{}^{n} \int 
    d^{4}x_{1}d^{4}x_{2} \, \left\{ \Delta^{lK}_{fL}(x_{2}-x_{1}) 
    \cdot \right. \nonumber \\ 
    && \left.  \cdot
    \left[ {\overline S}^{{\dot\alpha}\beta 
	aI}_{nK}(x-x_{2}) S_{\beta{\dot\beta}}^{em}(x_{2}-x_{1})
	{\overline S}^{{\dot\beta}\alpha 
	dL}_{bJ}(x_{1}-y)\right] \right\} \, .
\end{eqnarray}
Taking the Fourier transform one obtains 
\begin{displaymath}
	\tilde{A}^{{\dot\alpha}\alpha aI}_{bJ}(p) = \frac{i}{4} g^{2}
	\delta^{a}_{b} \delta^{I}_{J} 
	\tilde{\overline{S}}^{{\dot\alpha}\beta}(p) 
	\sigma^{\lambda}_{\beta{\dot\beta}}p_{\lambda} 
	\tilde{\overline{S}}^{{\dot\beta}\alpha}(p) 
	\left[ \int \frac{d^{4}k}{(2\pi)^{4}} 
	\frac{1}{\left(k+\frac{p}{2}\right)^{2}\left(k-\frac{p}{2}
	\right)^{2}} \right] \, ,
\end{displaymath}
\begin{displaymath}
    \tilde{B}^{{\dot\alpha}\alpha aI}_{bJ}(p) = 
    -\tilde{C}^{{\dot\alpha}\alpha aI}_{bJ}(p) = 
    \tilde{A}^{{\dot\alpha}\alpha aI}_{bJ}(p) \, ,
\end{displaymath}
where 
\begin{displaymath}
    \tilde{\overline{S}}^{{\dot\alpha}\alpha}(p) = -i 
    \frac{{\overline \sigma}_{\mu}^{{\dot\alpha}\alpha} 
    p^{\mu}}{p^{2}} \, .
\end{displaymath}
As anticipated above, the one-loop correction to the fermion propagator 
in the Wess--Zumino gauge turns out to be logarithmically 
ultraviolet-divergent
\begin{eqnarray}
	&& \langle \left({\overline \psi}\psi\right)^{{\dot\alpha}
	\alpha aI}_{bJ} \rangle^{{\rm 1-loop,WZ}}_{{\rm FT}} = 
	\tilde{A}^{{\dot\alpha}\alpha aI}_{bJ}(p) + 
    \tilde{B}^{{\dot\alpha}\alpha aI}_{bJ}(p) + 
    \tilde{C}^{{\dot\alpha}\alpha aI}_{bJ}(p) = \nonumber \\ 
    && = \frac{i}{4} g^{2}
	\delta^{a}_{b} \delta^{I}_{J} 
	\tilde{\overline{S}}^{{\dot\alpha}\beta}(p) 
	\sigma^{\lambda}_{\beta{\dot\beta}}p_{\lambda} 
	\tilde{\overline{S}}^{{\dot\beta}\alpha}(p) 
	\left[ \int \frac{d^{4}k}{(2\pi)^{4}} 
	\frac{1}{\left(k+\frac{p}{2}\right)^{2}\left(k-\frac{p}{2}
	\right)^{2}} \right] \, . ~~~
	\label{onelooppsiwz}
\end{eqnarray}
Equation (\ref{onelooppsiwz}) shows that a logarithmically divergent wave 
function renormalization is required in the WZ gauge.

This divergence will now be shown to be a gauge artifact due to the 
choice of the WZ gauge. In fact if the fields $C$, $\chi$ and $S$ are 
included two further contributions must be added, corresponding 
to the diagrams 

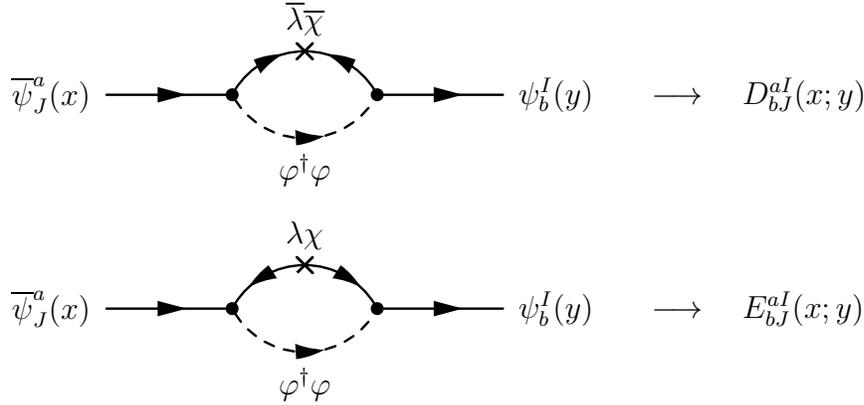
\vspace{0.3cm}
\noindent
\hspace*{1cm}
\begin{fmffile}{psipsiD}
 \begin{fmfgraph*}(150,80) 
  \fmfleft{i1} \fmfright{o1}
  \fmf{fermion,tension=1.5}{i1,v1}
  \fmf{fermion,tension=1.5}{v2,o1}
  \fmf{phantom,right=0.6,tension=0.4,tag=1}{v1,v2}
  \fmf{phantom,right=0.6,tension=0.4,tag=2}{v2,v1}
  \fmf{phantom,left=0.6,tension=0.5,tag=3}{v1,v2} 
  \fmfv{l=${\overline \psi}^{a}_{J}(x)$,l.a=180}{i1}
  \fmfv{l=$\psi^{I}_{b}(y)$ \hspace{0.5cm} $\longrightarrow ~~~ 
        D^{aI}_{bJ}(x;y)$,l.a=0}{o1}
  \fmfdot{v1,v2}
  \fmffreeze
  \fmfipath{p[]}
  \fmfiset{p1}{vpath1(__v1,__v2)}
  \fmfiset{p2}{vpath2(__v2,__v1)}
  \fmfiset{p3}{vpath3(__v1,__v2)}
  \fmfi{scalar,l=$\varphi^{\dagger}\varphi$}{subpath (0,length(p1)) of p1}
  \fmfi{fermion}{subpath (0,length(p3)/2) of p3}
  \fmfi{fermion}{subpath (0,length(p2)/2) of p2}  
  \fmfiv{d.sh=cross,d.si=3mm,l=${\overline\lambda}{\overline\chi}$,
       l.a=90}{point length(p2)/2 of p2}      
 \end{fmfgraph*}   
\end{fmffile} \\
\hspace*{1cm}
\begin{fmffile}{psipsiE}
 \begin{fmfgraph*}(150,80) 
  \fmfleft{i1} \fmfright{o1}
  \fmf{fermion,tension=1.5}{i1,v1}
  \fmf{fermion,tension=1.5}{v2,o1}
  \fmf{phantom,right=0.6,tension=0.4,tag=1}{v1,v2}
  \fmf{phantom,right=0.6,tension=0.4,tag=2}{v2,v1}
  \fmf{phantom,left=0.6,tension=0.5,tag=3}{v1,v2} 
  \fmfv{l=${\overline \psi}^{a}_{J}(x)$,l.a=180}{i1}
  \fmfv{l=$\psi^{I}_{b}(y)$ \hspace{0.5cm} $\longrightarrow ~~~ 
        E^{aI}_{bJ}(x;y)$,l.a=0}{o1}
  \fmfdot{v1,v2}
  \fmffreeze
  \fmfipath{p[]}
  \fmfiset{p1}{vpath1(__v1,__v2)}
  \fmfiset{p2}{vpath2(__v2,__v1)}
  \fmfiset{p3}{vpath3(__v1,__v2)}
  \fmfi{scalar,l=$\varphi^{\dagger}\varphi$}{subpath (0,length(p1)) of p1}
  \fmfi{fermion}{subpath (length(p3)/2,length(p3)) of p3}
  \fmfi{fermion}{subpath (length(p2)/2,length(p2)) of p2}  
  \fmfiv{d.sh=cross,d.si=3mm,l=$\lambda\chi$,
       l.a=90}{point length(p2)/2 of p2}      
 \end{fmfgraph*}   
\end{fmffile} 

The contribution of these two diagrams exactly cancels the divergence 
in equation (\ref{onelooppsiwz}) giving a net vanishing one-loop result 
for the propagator. In fact the computation of $D$ and $E$ gives
\begin{eqnarray}
    && D^{{\dot\alpha}\alpha aI}_{bJ}(x;y) = -\frac{1}{4}g^{2}
    f_{de}{}^{f}f_{lm}{}^{n} \int 
    d^{4}x_{1}d^{4}x_{2} \, \left\{ \Delta^{lK}_{fL}(x_{2}-x_{1}) 
    \rule{0pt}{18pt}\cdot \right. \nonumber \\ 
    && \left.  \cdot
    \left[ {\overline S}^{{\dot\alpha}\gamma aI}_{mK}
	(x-x_{2}) \sigma^{\mu}_{\gamma{\dot\gamma}} 
	\left(\partial_{\mu}^{(2)} R^{{\dot\gamma}ne}_{\dot\beta}
	(x_{2}-x_{1}) \right)
	{\overline S}^{{\dot\beta}\alpha 
	dL}_{bJ}(x_{1}-y)\right] \right\} \, ,
\end{eqnarray}
\begin{displaymath}
	E^{{\dot\alpha}\alpha aI}_{bJ}(x;y) = 
	D^{{\dot\alpha}\alpha aI}_{bJ}(x;y) \, .
\end{displaymath}
In momentum space one finds 
\begin{displaymath}
	\tilde{D}^{{\dot\alpha}\alpha aI}_{bJ}(p) = -\frac{i}{8} g^{2}
	\delta^{a}_{b} \delta^{I}_{J} 
	\tilde{\overline{S}}^{{\dot\alpha}\beta}(p) 
	\sigma^{\lambda}_{\beta{\dot\beta}}p_{\lambda} 
	\tilde{\overline{S}}^{{\dot\beta}\alpha}(p) 
	\left[ \int \frac{d^{4}k}{(2\pi)^{4}} 
	\frac{1}{\left(k+\frac{p}{2}\right)^{2}\left(k-\frac{p}{2}
	\right)^{2}} \right] \, ,
\end{displaymath}
\begin{displaymath}
	\tilde{E}^{{\dot\alpha}\alpha aI}_{bJ}(p) =
	\tilde{D}^{{\dot\alpha}\alpha aI}_{bJ}(p) \, .
\end{displaymath}
In conclusion summing up all the terms gives
\begin{eqnarray}
	\langle \left({\overline \psi}\psi\right)^{{\dot\alpha}
	\alpha aI}_{bJ} \rangle^{{\rm 1-loop,WZ}}_{{\rm FT}} &=& 
	\tilde{A}^{{\dot\alpha}\alpha aI}_{bJ}(p) + 
    \tilde{B}^{{\dot\alpha}\alpha aI}_{bJ}(p) + 
    \tilde{C}^{{\dot\alpha}\alpha aI}_{bJ}(p) +
    \tilde{D}^{{\dot\alpha}\alpha aI}_{bJ}(p) +
    \tilde{E}^{{\dot\alpha}\alpha aI}_{bJ}(p) = \nonumber \\
    &=& 0 \, ,
	\label{onelooppsi}
\end{eqnarray}
so that the one-loop correction to the $\langle{\overline 
\psi}\psi\rangle$ propagator without fixing the Wess--Zumino 
gauge is zero, as expected in ${\cal N}$=4 super Yang--Mills 
theory.

\subsection{One loop corrections to the propagator of the scalars 
belonging to the chiral multiplet}

The calculation of the propagator $\langle\varphi^{\dagger}\varphi
\rangle$ is more complicated because more diagrams are involved. In 
particular tadpole type graphs, coming from the term 
$\Phi^{\dagger}V^{2}\Phi$ in the action, are present. However the 
result is analogous to what was found for the 
$\langle{\overline\psi}\psi\rangle$ propagator: in the WZ gauge the 
one-loop correction is logarithmically ultraviolet-divergent, requiring 
a wave function renormalization, but when the contributions neglected 
in the WZ gauge are taken into account the total one-loop result vanishes.

The diagrams contributing to $\langle\varphi^{\dagger}\varphi\rangle$ 
at the one-loop level in the Wess--Zumino gauge are the following

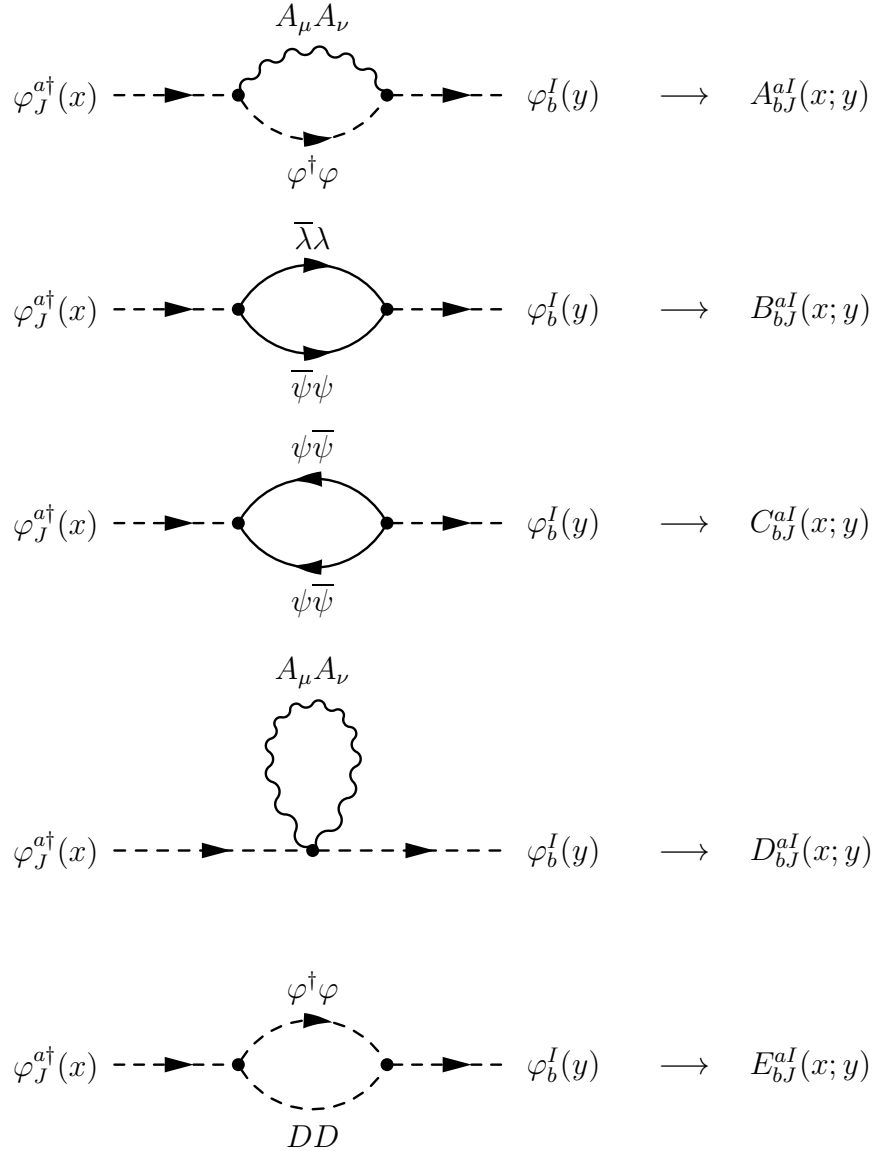
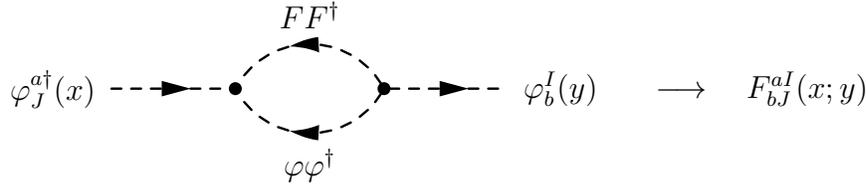
\vspace{0.3cm}
\noindent
\hspace*{1cm}
\begin{fmffile}{phiphiA}
 \begin{fmfgraph*}(150,80) 
  \fmfleft{i1} \fmfright{o1}
  \fmf{scalar,tension=1.2}{i1,v1}
  \fmf{boson,label=$A_{\mu}A_{\nu}$,left=0.6,tension=0.5}{v1,v2}
  \fmf{scalar,label=$\varphi^{\dagger}\varphi$,right=0.6,tension=0.5}
      {v1,v2}
  \fmf{scalar,tension=1.2}{v2,o1}
  \fmfv{l=${\varphi}^{a\dagger}_{J}(x)$,l.a=180}{i1}
  \fmfv{l=$\varphi^{I}_{b}(y)$ \hspace{0.5cm} $\longrightarrow ~~~ 
        A^{aI}_{bJ}(x;y)$,l.a=0}{o1}
  \fmfdot{v1,v2}
 \end{fmfgraph*}   
\end{fmffile} \\
\hspace*{1cm}
\begin{fmffile}{phiphiB}
 \begin{fmfgraph*}(150,80) 
  \fmfleft{i1} \fmfright{o1}
  \fmf{scalar,tension=1.2}{i1,v1}
  \fmf{fermion,label=${\overline\lambda}\lambda$,left=0.6,tension=0.5}
      {v1,v2}
  \fmf{fermion,label=${\overline\psi}\psi$,right=0.6,tension=0.5}{v1,v2}
  \fmf{scalar,tension=1.2}{v2,o1}
  \fmfv{l=${\varphi}^{a\dagger}_{J}(x)$,l.a=180}{i1}
  \fmfv{l=$\varphi^{I}_{b}(y)$ \hspace{0.5cm} $\longrightarrow ~~~ 
        B^{aI}_{bJ}(x;y)$,l.a=0}{o1}
  \fmfdot{v1,v2}
 \end{fmfgraph*}   
\end{fmffile} \\
\hspace*{1cm}
\begin{fmffile}{phiphiC}
 \begin{fmfgraph*}(150,80) 
  \fmfleft{i1} \fmfright{o1}
  \fmf{scalar,tension=1.2}{i1,v1}
  \fmf{fermion,label=$\psi{\overline\psi}$,left=0.6,tension=0.5}{v2,v1}
  \fmf{fermion,label=$\psi{\overline\psi}$,right=0.6,tension=0.5}{v2,v1}
  \fmf{scalar,tension=1.2}{v2,o1}
  \fmfv{l=${\varphi}^{a\dagger}_{J}(x)$,l.a=180}{i1}
  \fmfv{l=$\varphi^{I}_{b}(y)$ \hspace{0.5cm} $\longrightarrow ~~~ 
        C^{aI}_{bJ}(x;y)$,l.a=0}{o1}
  \fmfdot{v1,v2}
 \end{fmfgraph*}   
\end{fmffile} \\

\vspace*{1cm}
\noindent
\hspace*{1cm}
\begin{fmffile}{phiphiD}
 \begin{fmfgraph*}(150,80) 
  \fmfleft{i1} \fmfright{o1}
  \fmf{scalar,tension=1.4}{i1,v1}
  \fmf{scalar,tension=1.4}{v1,o1}
  \fmf{phantom,tension=0.9}{v1,v1}
  \fmfv{l=$\varphi^{a\dagger}_{J}(x)$,l.a=180}{i1}
  \fmfv{l=$\varphi^{I}_{b}(y)$ \hspace{0.5cm} $\longrightarrow ~~~ 
        D^{aI}_{bJ}(x;y)$,l.a=0}{o1}
  \fmfdot{v1}
  \fmffreeze
  \fmfipath{p[]}
  \fmfiset{p1}{vpath(__v1,__v1)}
  \fmfi{boson}{subpath (0,length(p1)) of p1}
  \fmfiv{l=$A_{\mu}A_{\nu}$,l.a=90}{point length(p1)/2 of p1}      
 \end{fmfgraph*}   
\end{fmffile} \\
\hspace*{1cm}
\begin{fmffile}{phiphiE}
 \begin{fmfgraph*}(150,80) 
  \fmfleft{i1} \fmfright{o1}
  \fmf{scalar,tension=1.2}{i1,v1}
  \fmf{scalar,label=$\varphi^{\dagger}\varphi$,left=0.6,tension=0.5}
      {v1,v2}
  \fmf{dashes,label=$DD$,right=0.6,tension=0.5}{v1,v2}
  \fmf{scalar,tension=1.2}{v2,o1}
  \fmfv{l=${\varphi}^{a\dagger}_{J}(x)$,l.a=180}{i1}
  \fmfv{l=$\varphi^{I}_{b}(y)$ \hspace{0.5cm} $\longrightarrow ~~~ 
        E^{aI}_{bJ}(x;y)$,l.a=0}{o1}
  \fmfdot{v1,v2}
 \end{fmfgraph*}   
\end{fmffile} \\
\hspace*{1cm}
\begin{fmffile}{phiphiF}
 \begin{fmfgraph*}(150,80) 
  \fmfleft{i1} \fmfright{o1}
  \fmf{scalar,tension=1.2}{i1,v1}
  \fmf{scalar,label=$\varphi\varphi^{\dagger}$,left=0.6,tension=0.5}
      {v2,v1}
  \fmf{scalar,label=$FF^{\dagger}$,right=0.6,tension=0.5}{v2,v1}
  \fmf{scalar,tension=1.2}{v2,o1}
  \fmfv{l=${\varphi}^{a\dagger}_{J}(x)$,l.a=180}{i1}
  \fmfv{l=$\varphi^{I}_{b}(y)$ \hspace{0.5cm} $\longrightarrow ~~~ 
        F^{aI}_{bJ}(x;y)$,l.a=0}{o1}
  \fmfdot{v1,v2}
 \end{fmfgraph*}   
\end{fmffile} 

The last three diagrams are all tadpole corrections since the free 
propagators for the auxiliary fields $F$ and $D$ are 
$\delta$-functions, so that the corresponding lines shrink to a point. 
Each one of these diagrams is quadratically divergent. There is also a 
quadratically divergent contribution coming from the two diagrams 
$B(x,y)$ and $C(x,y)$ and the sum of all these terms exactly vanishes.
The total sum gives a final result that is 
logarithmically divergent. In momentum space it is schematically of 
the form
\begin{equation}
	\langle \varphi^{a\dagger}_{J}\varphi^{I}_{b} 
	\rangle^{{\rm 1-loop,WZ}}_{{\rm FT}} \sim g^{2} \delta^{I}_{J}
	\delta^{a}_{b} p^{2} \int \frac{d^{4}k}{(2\pi)^{4}} 
	\frac{1}{\left(k+\frac{p}{2}\right)^{2}
	\left(k-\frac{p}{2}\right)^{2}} \, .
	\label{phioneloopwz}
\end{equation}
Just like in the case of $\langle {\overline \psi}\psi\rangle$ this 
divergence can be reabsorbed by a wave function renormalization for the 
field $\varphi$. 

Taking into account the contribution of the fields $C$, $\chi$ and $S$ 
there are eight more diagrams to be calculated 

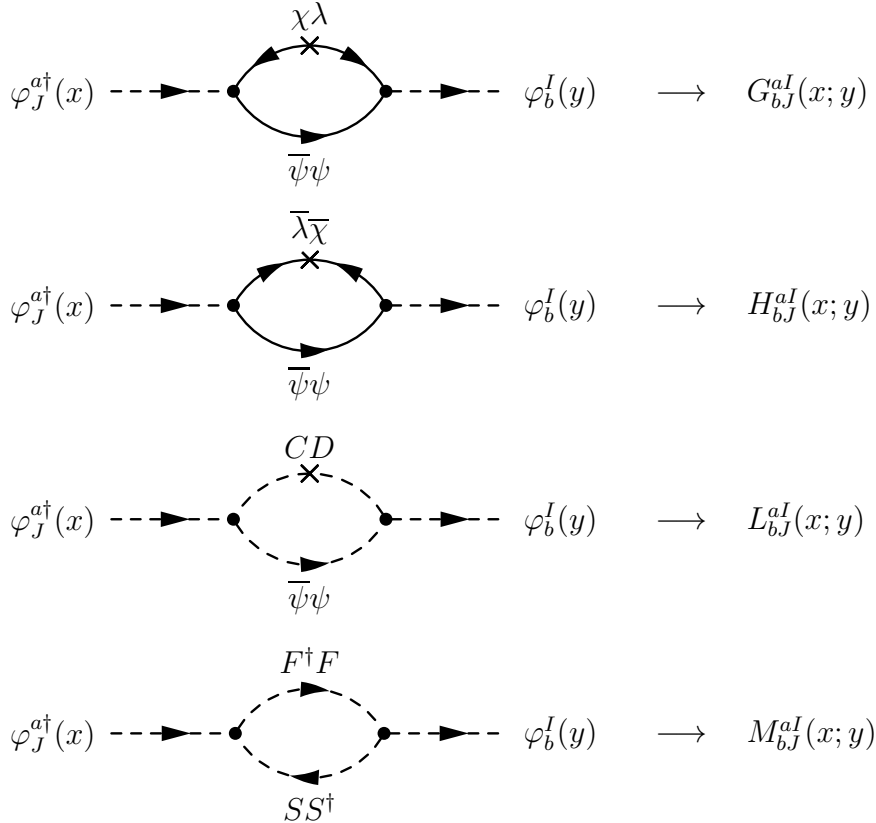
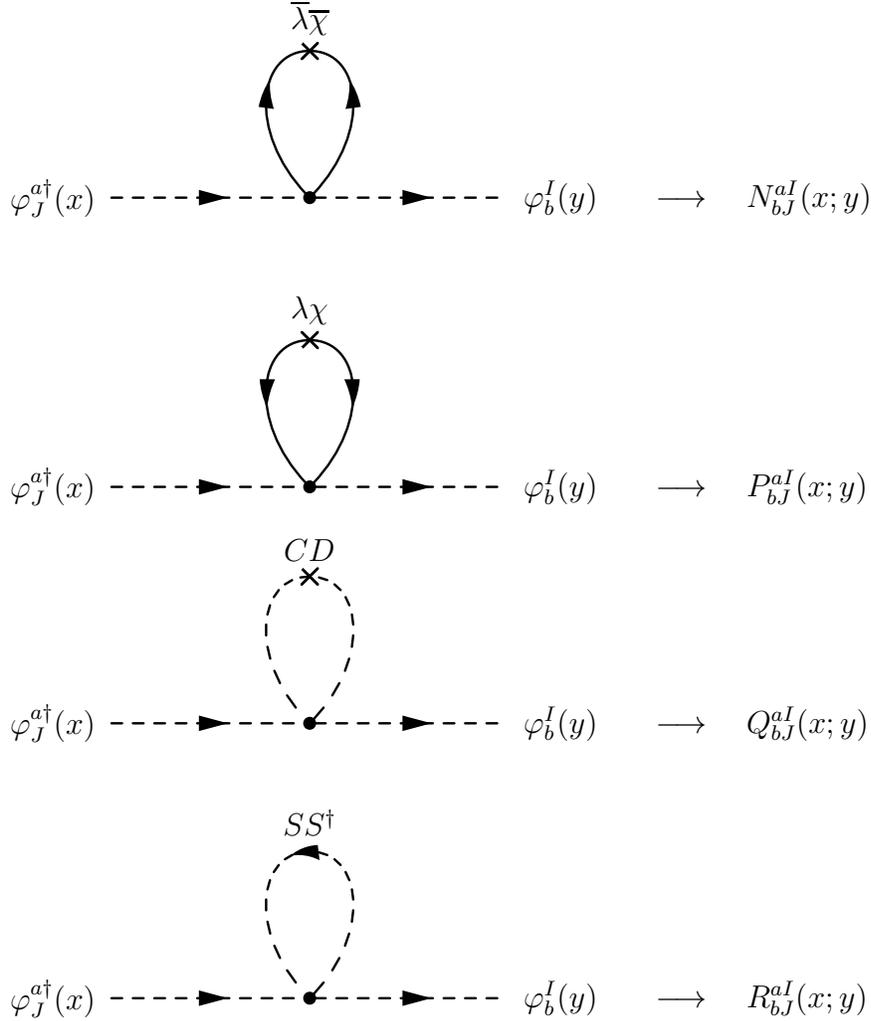
\vspace{0.3cm}
\noindent
\hspace*{1cm}
\begin{fmffile}{phiphiG}
 \begin{fmfgraph*}(150,80) 
  \fmfleft{i1} \fmfright{o1}
  \fmf{scalar,tension=1.5}{i1,v1}
  \fmf{scalar,tension=1.5}{v2,o1}
  \fmf{phantom,right=0.6,tension=0.4,tag=1}{v1,v2}
  \fmf{phantom,right=0.6,tension=0.4,tag=2}{v2,v1}
  \fmf{phantom,left=0.6,tension=0.4,tag=3}{v1,v2} 
  \fmfv{l=$\varphi^{a\dagger}_{J}(x)$,l.a=180}{i1}
  \fmfv{l=$\varphi^{I}_{b}(y)$ \hspace{0.5cm} $\longrightarrow ~~~ 
        G^{aI}_{bJ}(x;y)$,l.a=0}{o1}
  \fmfdot{v1,v2}
  \fmffreeze
  \fmfipath{p[]}
  \fmfiset{p1}{vpath1(__v1,__v2)}
  \fmfiset{p2}{vpath2(__v2,__v1)}
  \fmfiset{p3}{vpath3(__v1,__v2)}
  \fmfi{fermion,l=${\overline\psi}\psi$}{subpath (0,length(p1)) of p1}
  \fmfi{fermion}{subpath (length(p3)/2,length(p3)) of p3}
  \fmfi{fermion}{subpath (length(p2)/2,length(p2)) of p2}  
  \fmfiv{d.sh=cross,d.si=3mm,l=$\chi\lambda$,
       l.a=90}{point length(p2)/2 of p2}      
 \end{fmfgraph*}   
\end{fmffile} \\
\hspace*{1cm}
\begin{fmffile}{phiphiH}
 \begin{fmfgraph*}(150,80) 
  \fmfleft{i1} \fmfright{o1}
  \fmf{scalar,tension=1.5}{i1,v1}
  \fmf{scalar,tension=1.5}{v2,o1}
  \fmf{phantom,right=0.6,tension=0.4,tag=1}{v1,v2}
  \fmf{phantom,right=0.6,tension=0.4,tag=2}{v2,v1}
  \fmf{phantom,left=0.6,tension=0.4,tag=3}{v1,v2} 
  \fmfv{l=$\varphi^{a\dagger}_{J}(x)$,l.a=180}{i1}
  \fmfv{l=$\varphi^{I}_{b}(y)$ \hspace{0.5cm} $\longrightarrow ~~~ 
        H^{aI}_{bJ}(x;y)$,l.a=0}{o1}
  \fmfdot{v1,v2}
  \fmffreeze
  \fmfipath{p[]}
  \fmfiset{p1}{vpath1(__v1,__v2)}
  \fmfiset{p2}{vpath2(__v2,__v1)}
  \fmfiset{p3}{vpath3(__v1,__v2)}
  \fmfi{fermion,l=${\overline\psi}\psi$}{subpath (0,length(p1)) of p1}
  \fmfi{fermion}{subpath (0,length(p3)/2) of p3}
  \fmfi{fermion}{subpath (0,length(p2)/2) of p2}  
  \fmfiv{d.sh=cross,d.si=3mm,l=${\overline\lambda}{\overline\chi}$,
       l.a=90}{point length(p2)/2 of p2}      
 \end{fmfgraph*}   
\end{fmffile} \\
\hspace*{1cm}
\begin{fmffile}{phiphiL}
 \begin{fmfgraph*}(150,80) 
  \fmfleft{i1} \fmfright{o1}
  \fmf{scalar,tension=1.5}{i1,v1}
  \fmf{scalar,tension=1.5}{v2,o1}
  \fmf{phantom,right=0.6,tension=0.4,tag=1}{v1,v2}
  \fmf{phantom,right=0.6,tension=0.4,tag=2}{v2,v1}
  \fmf{phantom,left=0.6,tension=0.4,tag=3}{v1,v2} 
  \fmfv{l=$\varphi^{a\dagger}_{J}(x)$,l.a=180}{i1}
  \fmfv{l=$\varphi^{I}_{b}(y)$ \hspace{0.5cm} $\longrightarrow ~~~ 
        L^{aI}_{bJ}(x;y)$,l.a=0}{o1}
  \fmfdot{v1,v2}
  \fmffreeze
  \fmfipath{p[]}
  \fmfiset{p1}{vpath1(__v1,__v2)}
  \fmfiset{p2}{vpath2(__v2,__v1)}
  \fmfiset{p3}{vpath3(__v1,__v2)}
  \fmfi{scalar,l=${\overline\psi}\psi$}{subpath (0,length(p1)) of p1}
  \fmfi{dashes}{subpath (0,length(p2)) of p2}
  \fmfiv{d.sh=cross,d.si=3mm,l=$CD$,
       l.a=90}{point length(p2)/2 of p2}      
 \end{fmfgraph*}   
\end{fmffile} \\
\hspace*{1cm}
\begin{fmffile}{phiphiM}
 \begin{fmfgraph*}(150,80) 
  \fmfleft{i1} \fmfright{o1}
  \fmf{scalar,tension=1.2}{i1,v1}
  \fmf{scalar,label=$SS^{\dagger}$,left=0.6,tension=0.5}
      {v2,v1}
  \fmf{scalar,label=$F^{\dagger}F$,left=0.6,tension=0.5}{v1,v2}
  \fmf{scalar,tension=1.2}{v2,o1}
  \fmfv{l=${\varphi}^{a\dagger}_{J}(x)$,l.a=180}{i1}
  \fmfv{l=$\varphi^{I}_{b}(y)$ \hspace{0.5cm} $\longrightarrow ~~~ 
        M^{aI}_{bJ}(x;y)$,l.a=0}{o1}
  \fmfdot{v1,v2}
 \end{fmfgraph*}   
\end{fmffile} 

\vspace{1.3cm}
\noindent
\hspace*{1cm}
\begin{fmffile}{phiphiN}
 \begin{fmfgraph*}(150,80) 
  \fmfleft{i1} \fmfright{o1}
  \fmf{scalar,tension=1.4}{i1,v1}
  \fmf{scalar,tension=1.4}{v1,o1}
  \fmf{phantom,tension=0.9}{v1,v1}
  \fmfv{l=$\varphi^{a\dagger}_{J}(x)$,l.a=180}{i1}
  \fmfv{l=$\varphi^{I}_{b}(y)$ \hspace{0.5cm} $\longrightarrow ~~~ 
        N^{aI}_{bJ}(x;y)$,l.a=0}{o1}
  \fmfdot{v1}
  \fmffreeze
  \fmfipath{p[]}
  \fmfiset{p1}{vpath(__v1,__v1)}
  \fmfi{fermion}{subpath (0,length(p1)/2) of p1}
  \fmfi{fermion}{subpath (length(p1),length(p1)/2) of p1}  
  \fmfiv{d.sh=cross,d.si=3mm,l=${\overline\lambda}{\overline\chi}$,
       l.a=90}{point length(p1)/2 of p1}      
 \end{fmfgraph*}   
\end{fmffile} 

\vspace{1cm}
\noindent
\hspace*{1cm}
\begin{fmffile}{phiphiP}
 \begin{fmfgraph*}(150,80) 
  \fmfleft{i1} \fmfright{o1}
  \fmf{scalar,tension=1.4}{i1,v1}
  \fmf{scalar,tension=1.4}{v1,o1}
  \fmf{phantom,tension=0.9}{v1,v1}
  \fmfv{l=$\varphi^{a\dagger}_{J}(x)$,l.a=180}{i1}
  \fmfv{l=$\varphi^{I}_{b}(y)$ \hspace{0.5cm} $\longrightarrow ~~~ 
        P^{aI}_{bJ}(x;y)$,l.a=0}{o1}
  \fmfdot{v1}
  \fmffreeze
  \fmfipath{p[]}
  \fmfiset{p1}{vpath(__v1,__v1)}
  \fmfi{fermion}{subpath (length(p1)/2,0) of p1}
  \fmfi{fermion}{subpath (length(p1)/2,length(p1)) of p1}  
  \fmfiv{d.sh=cross,d.si=3mm,l=$\lambda\chi$,
       l.a=90}{point length(p1)/2 of p1}      
 \end{fmfgraph*}   
\end{fmffile} 

\vspace*{0.3cm}
\noindent
\hspace*{1cm}
\begin{fmffile}{phiphiQ}
 \begin{fmfgraph*}(150,80) 
  \fmfleft{i1} \fmfright{o1}
  \fmf{scalar,tension=1.4}{i1,v1}
  \fmf{scalar,tension=1.4}{v1,o1}
  \fmf{phantom,tension=0.9}{v1,v1}
  \fmfv{l=$\varphi^{a\dagger}_{J}(x)$,l.a=180}{i1}
  \fmfv{l=$\varphi^{I}_{b}(y)$ \hspace{0.5cm} $\longrightarrow ~~~ 
        Q^{aI}_{bJ}(x;y)$,l.a=0}{o1}
  \fmfdot{v1}
  \fmffreeze
  \fmfipath{p[]}
  \fmfiset{p1}{vpath(__v1,__v1)}
  \fmfi{dashes}{subpath (0,length(p1)) of p1}
  \fmfiv{d.sh=cross,d.si=3mm,l=$CD$,
       l.a=90}{point length(p1)/2 of p1}      
 \end{fmfgraph*}   
\end{fmffile} \\

\vspace*{0.3cm}
\noindent
\hspace*{1cm}
\begin{fmffile}{phiphiR}
 \begin{fmfgraph*}(150,80) 
  \fmfleft{i1} \fmfright{o1}
  \fmf{scalar,tension=1.4}{i1,v1}
  \fmf{scalar,tension=1.4}{v1,o1}
  \fmf{phantom,tension=0.9}{v1,v1}
  \fmfv{l=$\varphi^{a\dagger}_{J}(x)$,l.a=180}{i1}
  \fmfv{l=$\varphi^{I}_{b}(y)$ \hspace{0.5cm} $\longrightarrow ~~~ 
        R^{aI}_{bJ}(x;y)$,l.a=0}{o1}
  \fmfdot{v1}
  \fmffreeze
  \fmfipath{p[]}
  \fmfiset{p1}{vpath(__v1,__v1)}
  \fmfi{scalar}{subpath (0,length(p1)) of p1}
  \fmfiv{l=$SS^{\dagger}$,l.a=90}{point length(p1)/2 of p1}      
 \end{fmfgraph*}   
\end{fmffile} 

Single diagrams contain quadratically divergent terms that cancel in 
the sum leaving a total contribution that is again logarithmically 
divergent. This correction is exactly what is needed to 
cancel the divergence in (\ref{phioneloopwz}). As a result the complete 
one-loop correction to the $\langle\varphi^{\dagger}\varphi\rangle$ 
propagator is actually zero.

Notice that the situation induced by the choice of the WZ gauge is a 
feature common to every supersymmetric gauge theory, since the 
contribution of the gauge-dependent fields $C$, $\chi$ and $S$ is in 
general logarithmically divergent. However in theories with less 
supersymmetry a logarithmic wave function renormalization is in any 
case unavoidable, so that this effect is completely irrelevant. 
On the contrary it becomes important in the ${\cal N}$=4 Yang--Mills 
theory and in general in finite theories. Of course the divergences 
encountered here are gauge artifacts and disappear in gauge invariant 
correlation functions. Examples of computations of correlators of gauge 
invariant operators, in which the choice of the WZ gauge does not lead 
to this kind of problems, have been considered within the discussion of 
the correspondence with type IIB superstring theory on AdS space, 
\cite{adspert,adsnonpert}.

\vspace{0.7cm}
The calculations we described seem to suggest the possibility of 
constructing improved Feynman rules in which the effect of the gauge 
dependent fields, set to zero in the WZ gauge, is dealt with by a suitable 
redefinition of the free propagators. However this program proves 
extremely complicated when the propagators of fields in the vector 
multiplet or three- and more-point functions are considered. The 
calculation of these  correlation functions, already at the 
one-loop level, requires many other terms to be included in the 
action $S_{{\rm int}}$. In particular terms coming from $W^{(1)}W^{(2)}$ 
and $W^{(2)}W^{(2)}$ must be considered. The calculation of $W^{(2)}$ 
without the simplifications introduced by the Wess--Zumino gauge is 
rather lengthy and gives
\begin{eqnarray}
	W^{a(2)}_{\alpha} &=& -\frac{i}{2} f^{a}{}_{bc} \left\{ 
	-iC^{b}\lambda^{c}_{\alpha}-\frac{1}{2} 
	\sigma^{\mu}_{\alpha{\dot\alpha}}{\overline\chi}^{{\dot\alpha}b} 
	\partial_{\mu}C^{c}-\frac{i}{2}\sigma^{\mu}_{\alpha{\dot\alpha}}
	{\overline\chi}^{{\dot\alpha}b}A^{c}_{\mu} + 
	\frac{1}{2} S^{b\dagger}\chi^{c}_{\alpha} \right. \nonumber \\
	&&\left. + 
	\left[ \frac{1}{2} S^{b\dagger}
	S^{c}\delta_{\alpha}{}^{\beta} + \delta_{\alpha}{}^{\beta} 
	C^{b}D^{c}-\frac{i}{2}(\sigma^{\mu}
	{\overline\sigma}^{\nu})_{\alpha}{}^{\beta} C^{b} 
	(\partial_{\mu}A^{c}_{\nu}-\partial_{\nu}
	A^{c}_{\mu}) + \right. \right.\nonumber \\ 
	&& \left. \left. -\frac{i}{2}(\sigma^{\mu}
	{\overline\sigma}^{\nu})_{\alpha}{}^{\beta}
	(A^{b}_{\nu}\partial_{\mu}C^{c}+A_{\mu}^{b}\partial_{\nu}C^{c})-
	\frac{1}{2} \delta_{\alpha}{}^{\beta}C^{b}\Box C^{c} - 
	\delta_{\alpha}{}^{\beta}{\overline\chi}^{b}{\overline\lambda}^{c}+
	\right. \right. \nonumber \\
	&& \left. \left. 
	+\frac{1}{2}(\sigma^{\mu}\sigma^{\nu})_{\alpha}{}^{\beta}
	\partial_{\mu}C^{b}\partial_{\nu}C^{c}-
	\chi^{b\beta}\lambda^{c}_{\alpha} -
	\lambda^{b\beta}\chi^{c}_{\alpha}-i\sigma^{\mu}_{\alpha{\dot\alpha}}
	{\overline\chi}^{b{\dot\alpha}}\partial_{\mu}\chi^{c\beta} + 
	\right. \right. \nonumber \\
	&& \left. \left. +i\varepsilon ^{\gamma\beta}
	\sigma^{\mu}_{\gamma{\dot\beta}}\partial_{\mu}{\overline\chi}^{b} 
	\chi^{c}_{\alpha} + \frac{1}{2} 
	(\sigma^{\mu}\sigma^{\nu})_{\alpha}{}^{\beta}A_{\nu}^{b}A_{\mu}^{c} 
	\right] \theta_{\beta} + \left[ 
	C^{b}\sigma^{\mu}_{\alpha{\dot\alpha}} 
	\partial_{\mu}{\overline\lambda}^{c{\dot\alpha}} + 
	\right. \right. \nonumber \\
	&& \left. \left. -i\chi_{\alpha}^{b}D^{c}-\frac{1}{4} (\sigma^{\mu}
	{\overline\sigma}^{\nu})_{\alpha}{}^{\beta} \chi_{\beta}^{b} 
	(\partial_{\mu}A^{c}_{\nu}-\partial_{\nu} A^{c}_{\mu}) - 
	\frac{1}{2} A_{\nu}^{b} (\sigma^{\mu}
	{\overline\sigma}^{\nu})_{\alpha}{}^{\beta} \partial_{\mu}
	\chi_{\beta}^{c} + \right. \right. \nonumber \\
	&& \left. \left. 
	+ \frac{1}{2}\partial_{\mu}A^{b\mu}\chi_{\alpha}^{c} - 
	\frac{i}{2} \partial_{\nu}C^{b} (\sigma^{\mu}
	{\overline\sigma}^{\nu})_{\alpha}{}^{\beta} \partial_{\mu}
	\chi_{\beta}^{c} - \frac{i}{2} \sigma^{\mu}_{\alpha{\dot\alpha}}
	\partial_{\mu}{\overline\chi}^{{\dot\alpha}b}S^{c} + 
	S^{b}\lambda^{c}_{\alpha}+ \right. \right. \nonumber \\
	&& \left. \left. -i A_{\mu}^{b} \sigma^{\mu}_{\alpha{\dot\alpha}}
	{\overline\lambda}^{c{\dot\alpha}} \right] \theta\theta
	\rule{0pt}{18pt}\right\} \label{w2nonwz}
\end{eqnarray}
Substituting (\ref{w2nonwz}) into $S_{{\rm int}}$ results in a number 
of relevant interaction terms of the order of 100, making this formulation 
totally impractical in explicit calculations. In conclusion perturbation 
theory in components almost unavoidably requires the WZ gauge, but the latter 
introduces divergences in gauge dependent quantities. In the following 
sections the superfield formalism, that allows to avoid these problems, 
will be employed. However we will show that new difficulties related to 
``ordinary'' gauge fixing emerge.

\section{Perturbation theory in ${\cal N}$=1 superspace: propagators}
\label{n1propagators}

The difficulties encountered in the previous section in the 
calculation of gauge dependent quantities, because of the divergences 
present in the Wess--Zumino gauge, can be overcome using the 
superfield formulation.

Because of the lack of a completely consistent ${\cal N}$=4 
formulation, we will employ the ${\cal N}$=1 superfield formalism 
which has proved to be a powerful tool 
in the proof of the finiteness of the theory up to three loops. 
In this approach there is no particular difficulty in working without 
fixing the WZ gauge. If one does not choose to work in  the WZ gauge 
the action is non polynomial, however a finite number of terms is 
relevant at each order in perturbation theory, so that only at very 
high order the choice of the Wess--Zumino gauge introduces significant 
simplifications. The aim of this section is to show that 
there are other subtleties related to further fixing the gauge for 
the vector superfield, even if one does not work in the WZ gauge.

To be more general and for the purpose of studying the possibility of 
finding a supersymmetric regularization of a class of ${\cal N}$=1 
theories, following the proposal of \cite{japan,yoshida}, we will 
consider a formulation in ${\cal N}$=1 superspace of ${\cal N}$=4 
super Yang--Mills theory deformed with the addition of mass terms 
for the (anti) chiral superfields
\begin{equation}
	S_{{\rm m}} = -\int d^{4}x d^{2}\theta d^{2}{\overline\theta}
	\, \left[ \frac{1}{2} 
	m \delta_{IJ} \Phi_{a}^{I} \Phi^{Ja} \delta({\overline \theta}) + 
	\frac{1}{2} m^{*} \delta^{IJ}{\Phi^{\dagger}}^{a}_{I} 
	{\Phi^{\dagger}}_{Ja} \delta(\theta) \right] .
	\label{massterm} 
\end{equation}
The inclusion of this terms to the action breaks ${\cal N}$=4 
supersymmetry down to ${\cal N}$=1. In \cite{parkeswest} 
it was argued, by means of 
dimensional arguments, that this term should not affect the ultraviolet 
properties of the ${\cal N}$=4 theory. More precisely the statement of 
\cite{parkeswest} is that no divergences appear in gauge invariant 
quantities, so that no divergent contribution to the quantum 
effective action is generated perturbatively. 
As a result the model obtained in this way would be an 
example of a finite ${\cal N}$=1 theory. The inverse 
construction, in which one deforms a ${\cal N}$=1 model to ${\cal N}$=4 
super Yang--Mills plus a mass term, has been proposed in 
\cite{japan,yoshida} as a regularization procedure preserving 
supersymmetry. The calculations presented in this section will show 
that in the presence of the term (\ref {massterm}) no divergence is 
generated, at the one-loop level, in the two-, three- and four-point 
Green functions that are computed. The results presented suggest the  
possibility of reinforcing the conclusions of \cite{parkeswest}; 
namely the ${\cal N}$=4 theory augmented with (\ref{massterm}) appears 
to be finite in the sense that no divergences appear in the complete 
$n$-point irreducible Green functions, at least at one loop. In 
particular no wave function renormalization is required. Notice that 
this result is what is actually necessary for the consistency of 
the approach advocated in \cite{japan,yoshida}.

The complete action we will be using for our perturbative 
calculations is thus\footnote{In the following the mass parameter $m$ 
will be taken to be real for simplicity of notation.}
\begin{displaymath}
	S = S_{{\cal N}=4} + S_{{\rm m}}
\end{displaymath}
and reads
\begin{eqnarray}
    S &=& \int d^{4}x\,d^{2}\theta d^{2}{\overline\theta}\: 
	\left\{ \frac{1}{2} 
	V^{a} \left[ - \Box P_{T} - \xi (P_{1}+P_{2}) \Box \right] V_{a} +  
	{\Phi^{\dagger}}^{a}_{I} \Phi_{a}^{I} - \frac{1}{2} 
	m \delta_{IJ} \Phi_{a}^{I} \Phi^{Ja}
	\delta({\overline \theta}) + \right. \nonumber \\
	&&  \hspace{-1cm} -\frac{1}{2} m 
	\delta^{IJ}{\Phi^{\dagger}}^{a}_{I}{\Phi^{\dagger}}_{Ja}
	\delta(\theta)+\frac{i}{\sqrt{2}} g f_{abc}{\Phi^{\dagger}}^{a}_{I}
	V^{b} \Phi^{Ic}- \frac{g^{2}}{2} {f_{ab}}^{e}f_{ecd} 
	{\Phi^{\dagger}}^{a}_{I} V^{b} V^{c} \Phi^{Id} + \nonumber \\
	&& \hspace{-1cm} - \frac{i}{16 \sqrt{2}} g f_{abc} 
	\left[ {\overline D}^{2} 
	\left( D^{\alpha}V^{a} \right) \right] V^{b} 
	\left( D_{\alpha}V^{c} \right) -\frac{1}{128} g^{2} f_{ab}{}^{e}
	f_{ecd} V^{a} \left( D^{\alpha}
	V^{b} \right) \left[ \left({\overline D}^{2} V^{c} 
	\right) \left( D_{\alpha}V^{d} \right) 
	\right] \hspace{-0.15cm} + \nonumber \\
	&& \hspace{-1cm} +\ldots - \frac{1}{3!} g f^{abc} 
	\left[ \varepsilon_{IJK} \Phi_{a}^{I}
	\Phi_{b}^{J}\Phi_{c}^{K} \delta({\overline \theta}) 
	+ \varepsilon^{IJK} {\Phi^{\dagger}}_{Ia}{\Phi^{\dagger}}_{Jb} 
	{\Phi^{\dagger}}_{Kc} \delta(\theta) \right] + 
	\left( {{\overline C}^{\prime}}_{a} C^{a} - {C^{\prime}}_{a}
	{\overline C}^{a} \right)  + \\
	&& \hspace{-1cm} \left. 
	+\frac{i}{2\sqrt{2}} g f_{abc} \left( {C^{\prime}}^{a} +
	{{\overline C}^{\prime}}^{a} \right)  V^{b} 
	\left( C^{c} + {\overline C}^{c} \right) - 
	\frac{1}{8} g^{2} {f_{ab}}^{e}f_{ecd}  \left( 
	{C^{\prime}}^{a} +{{\overline C}^{\prime}}^{a} 
	\right) V^{b} V^{c} \left( C^{d} +{\overline C}^{d} \right)  
	+ \ldots \rule{0pt}{18pt} \right\} \, , \nonumber 
\label{n1sfaction}	
\end{eqnarray}
where dots stand for terms that are not relevant for the 
considerations of this paper.

Notice that in the action (\ref{n1sfaction}) a gauge fixing 
term corresponding to a family of gauges parameterized by 
$\alpha=\frac{1}{\xi}$ has been introduced. It will now be shown, by 
explicitly computing the propagators of both the chiral and the vector 
superfields, that the supersymmetric generalization of the 
Fermi--Feynman gauge, corresponding to $\alpha=1$, is somehow 
privileged (see also \cite{storey,juerstorey2}), since any other 
choice of the parameter $\alpha$ leads to infrared divergences in 
Green functions.

\subsection{Propagator of the chiral superfield} 

\noindent 
The propagator of the chiral superfield is the simplest Green 
function  to compute. The calculation will be reported in detail in 
order to illustrate the superfield technique. 

It follows from the form of the action (\ref{n1sfaction}) that there 
are three diagrams contributing to the propagator $\langle \Phi
\Phi^{\dagger} \rangle$ at the one loop level. The one-particle 
irreducible parts of these diagrams will be directly evaluated in 
momentum space using the improved super Feynman rules of 
\cite{grisaruroceksiegel}. The convention employed is that all momenta 
are taken to be incoming. The diagrams are the following

\noindent
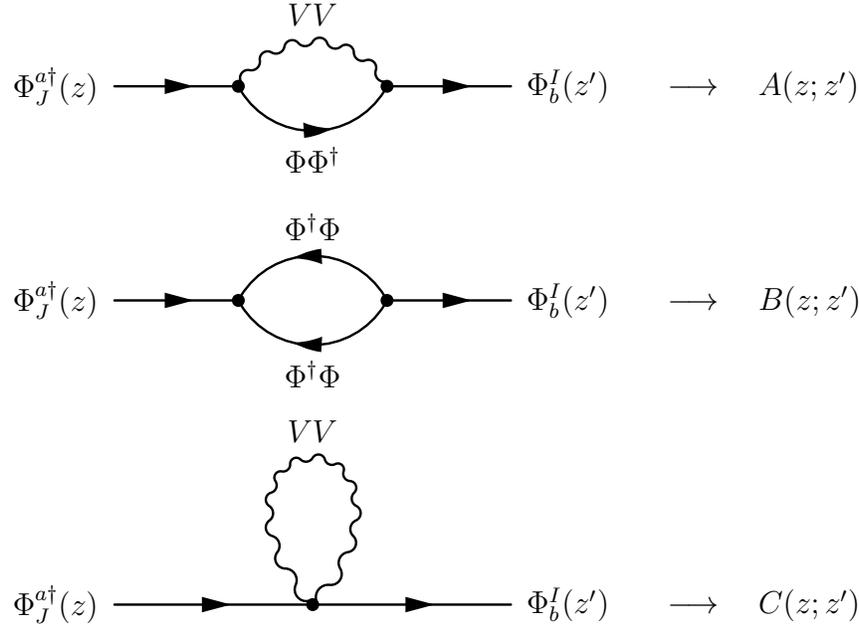
\hspace*{1cm}
\begin{fmffile}{sphiphiA}
 \begin{fmfgraph*}(150,80) 
  \fmfleft{i1} \fmfright{o1}
  \fmf{fermion,tension=1.2}{i1,v1}
  \fmf{boson,label=$VV$,left=0.6,tension=0.5}{v1,v2}
  \fmf{fermion,label=$\Phi\Phi^{\dagger}$,right=0.6,tension=0.5}{v1,v2}
  \fmf{fermion,tension=1.2}{v2,o1}
  \fmfv{l=$\Phi^{a\dagger}_{J}(z)$,l.a=180}{i1}
  \fmfv{l=$\Phi^{I}_{b}(z^{\prime})$ \hspace{0.5cm} $\longrightarrow ~~~ 
        A(z;z^{\prime})$,l.a=0}{o1}
  \fmfdot{v1,v2}
 \end{fmfgraph*}   
\end{fmffile} \\
\hspace*{1cm}
\begin{fmffile}{sphiphiB}
 \begin{fmfgraph*}(150,80) 
  \fmfleft{i1} \fmfright{o1}
  \fmf{fermion,tension=1.2}{i1,v1}
  \fmf{fermion,label=$\Phi^{\dagger}\Phi$,left=0.6,tension=0.5}
      {v2,v1}
  \fmf{fermion,label=$\Phi^{\dagger}\Phi$,right=0.6,tension=0.5}{v2,v1}
  \fmf{fermion,tension=1.2}{v2,o1}
  \fmfv{l=$\Phi^{a\dagger}_{J}(z)$,l.a=180}{i1}
  \fmfv{l=$\Phi^{I}_{b}(z^{\prime})$ \hspace{0.5cm} $\longrightarrow ~~~ 
        B(z;z^{\prime})$,l.a=0}{o1}
  \fmfdot{v1,v2}
 \end{fmfgraph*}   
\end{fmffile} 

\vspace*{1.2cm}
\noindent
\hspace*{1cm}
\begin{fmffile}{sphiphiC}
 \begin{fmfgraph*}(150,80) 
  \fmfleft{i1} \fmfright{o1}
  \fmf{fermion,tension=1.4}{i1,v1}
  \fmf{fermion,tension=1.4}{v1,o1}
  \fmf{phantom,tension=0.9}{v1,v1}
  \fmfv{l=$\Phi^{a\dagger}_{J}(z)$,l.a=180}{i1}
  \fmfv{l=$\Phi^{I}_{b}(z^{\prime})$ \hspace{0.5cm} $\longrightarrow ~~~ 
        C(z;z^{\prime})$,l.a=0}{o1}
  \fmfdot{v1}
  \fmffreeze
  \fmfipath{p[]}
  \fmfiset{p1}{vpath(__v1,__v1)}
  \fmfi{boson}{subpath (0,length(p1)) of p1}
  \fmfiv{l=$VV$,l.a=90}{point length(p1)/2 of p1}      
 \end{fmfgraph*}   
\end{fmffile} \\
where $z=(x,\theta,{\overline\theta})$. The notations for the 
internal propagators and the corresponding $x$-space expressions are 

\noindent
\begin{fmffile}{sphi+phi}
\begin{fmfgraph*}(90,50) 
  \fmfleft{i1} \fmfright{o1}
  \fmf{fermion}{i1,o1}
  \fmflabel{\raisebox{22pt}{\hspace{-20pt} $J,b$ \hspace{20pt}} 
  \raisebox{12pt}{$\longrightarrow \hspace{0.2cm} 
  \langle \Phi^{aI}(z)\Phi^{b\dagger}_{J}(z^{\prime}) 
  \rangle_{{\rm free}} = \begin{displaystyle}
  \delta^{I}_{J}\delta^{a}_{b}\frac{1}{\Box + m^{2}}
  \end{displaystyle} \delta_{8}(z-z^{\prime})$}}{o1}
  \fmflabel{\raisebox{22pt}{\hspace{18pt} $I,a$ 
  \hspace{-18pt}}}{i1}
\end{fmfgraph*}   
\end{fmffile} \\
\begin{fmffile}{svv}
\begin{fmfgraph*}(90,50) 
  \fmfleft{i1} \fmfright{o1}
  \fmf{boson}{i1,o1}
  \fmflabel{\raisebox{22pt}{\hspace{-20pt} $b$ \hspace{20pt}} 
  \raisebox{12pt}{$\longrightarrow \hspace{0.2cm} 
  \langle V^{a}(z)V_{b}(z^{\prime})\rangle_{{\rm free}}=
  \begin{displaystyle} -\frac{\delta^{a}_{b}}{\Box}
  [1+(\alpha -1)(P_{1}+P_{2})] \end{displaystyle} 
  \delta_{8}(z-z^{\prime})$}}{o1}
  \fmflabel{\raisebox{22pt}{\hspace{18pt} $a$ 
  \hspace{-18pt}}}{i1}
\end{fmfgraph*}   
\end{fmffile} \\
$P_{1}$ and $P_{2}$ in the $\langle VV \rangle$ propagator are the 
projectors \cite{wessbagger} 
\begin{eqnarray*}
	& P_{1} = \displaystyle{\frac{1}{16} 
	\frac{D^{2}{\overline D}^{2}}{\Box}} \, , 
	\hspace{1.5cm} & P_{1}\Phi^{\dagger} = \Phi^{\dagger} \, , \hspace{1cm}
	P_{1}\Phi = 0 \\
	& P_{2} = \displaystyle{\frac{1}{16} 
	\frac{{\overline D}^{2} D^{2}}{\Box}} \, ,
	\hspace{1.5cm} & P_{2}\Phi = \Phi \, , \hspace{1cm}
	P_{2}\Phi^{\dagger} = 0 \, .
\end{eqnarray*}
Moreover in the presence of mass terms 
for $\Phi$ and $\Phi^{\dagger}$ there are extra $\langle\Phi\Phi
\rangle$ and $\langle\Phi^{\dagger}\Phi^{\dagger}\rangle$ propagators, 
that will enter the calculation of the vector superfield propagator at 
one loop

\noindent
\begin{fmffile}{sphiphi}
\begin{fmfgraph*}(90,50) 
  \fmfleft{i1} \fmfright{o1}
  \fmf{fermion}{v1,i1}
  \fmf{fermion}{v1,o1}
  \fmflabel{\raisebox{22pt}{\hspace{-20pt} $J,b$ \hspace{20pt}} 
  \raisebox{12pt}{$\longrightarrow \hspace{0.2cm} 
  \langle \Phi^{aI}(z)\Phi_{b}^{J}(z^{\prime}) 
  \rangle_{{\rm free}} = \begin{displaystyle}
  -\delta^{IJ}\delta^{a}_{b}\,\frac{m}{4}
  \frac{D^{2}}{\Box(\Box + m^{2})} \end{displaystyle}
  \delta_{8}(z-z^{\prime})$}}{o1}
  \fmflabel{\raisebox{22pt}{\hspace{18pt} $I,a$ 
  \hspace{-18pt}}}{i1}
  \fmfv{d.sh=cross,d.si=3mm}{v1}
\end{fmfgraph*}   
\end{fmffile} \\
\begin{fmffile}{sphi+phi+}
\begin{fmfgraph*}(90,50) 
  \fmfleft{i1} \fmfright{o1}
  \fmf{fermion}{i1,v1}
  \fmf{fermion}{o1,v1}
  \fmflabel{\raisebox{22pt}{\hspace{-20pt} $J,b$ \hspace{20pt}} 
  \raisebox{12pt}{$\longrightarrow \hspace{0.2cm} 
  \langle \Phi^{a\dagger}_{I}(z)\Phi_{bJ}^{\dagger}(z^{\prime}) 
  \rangle_{{\rm free}} = \begin{displaystyle}
  -\delta_{IJ}\delta^{a}_{b}\,\frac{m}{4}
  \frac{{\overline D}^{2}}{\Box(\Box + m^{2})}
  \end{displaystyle} \delta_{8}(z-z^{\prime})$}}{o1}
  \fmflabel{\raisebox{22pt}{\hspace{18pt} $I,a$ 
  \hspace{-18pt}}}{i1}
  \fmfv{d.sh=cross,d.si=3mm}{v1}
\end{fmfgraph*}   
\end{fmffile} 

Using the above expressions for the free propagators and the rules of 
\cite{grisaruroceksiegel}, with the vertices read from the action 
(\ref{n1sfaction}), the three contributions can be evaluated without 
too much effort. For the Fourier transform of $A(z,z^{\prime})$ one 
finds
\begin{eqnarray}
	\tilde{A}(p) & = & \int \frac{d^{4}k}{(2\pi)^{4}} 
	d^{2}\theta_{1}d^{2}{\overline\theta}_{1}
	d^{2}\theta_{2}d^{2}{\overline\theta}_{2} \, \left\{ \frac{i}{\sqrt{2}} 
	g f_{acd}\Phi^{a\dagger}_{I}(p,\theta_{1},{\overline \theta}_{1}) 
	\left[\left(-\frac{1}{4}{\overline D}_{1}^{2}\right) 
	\frac{\delta^{I}_{J}\delta^{cf}\delta(1,2)}{(p-k)^{2}+m^{2}} 
	\right. \cdot \right. \nonumber  \\
	&& \hspace{-1.7cm}\cdot \left. \left.\left(-\frac{1}{4}
	\db{D}_{1}^{2} \right) \right] \left[ \left(1+\gamma(D^{2}_{1}
	{\overline D}^{2}_{1}+ {\overline D}^{2}_{1}D^{2}_{1})\right) 
	\left(-\frac{\delta^{de}\delta(1,2)}{k^{2}}\right)\right]
	\frac{i}{\sqrt{2}} g f_{efb}
	\Phi^{bJ}(-p,\theta_{2},{\overline \theta}_{2}) 
	\rule{0pt}{20pt}\right\} ,
	\label{atildephiphi}
\end{eqnarray}
where $\gamma=(\alpha-1)$ and the compact notation 
$\delta(1,2)=\delta_{2}(\theta_{1}-\theta_{2})\delta_{2}({\overline 
\theta_{1}}-{\overline \theta_{2}})$ has been introduced. The 
computation uses the properties of the Grassmannian $\delta$-function, 
which imply for example \cite{wessbagger,grisaruroceksiegel}
\begin{eqnarray*}
	&& D_{1\alpha}\delta(1,2) = - \delta(1,2) \db{D}_{2\alpha} \, , 
	\qquad {\overline D}_{1{\dot \alpha}} \delta(1,2) = - \delta(1,2) 
	\dbo{{\overline D}}_{2{\dot\alpha}} \, , \\ 
	&& {\overline D}_{1{\dot\alpha}} D_{1\alpha}\delta(1,2)=\delta(1,2)
	\dbo{{\overline D}}_{2{\dot\alpha}} \db{D}_{2\alpha} \, , \qquad
	{\overline D}^{2}_{1}D^{2}_{1} \delta(1,2) = \delta(1,2) 
	\dbo{{\overline D}}^{2}_{2} \db{D}^{2}_{2} \, ,
\end{eqnarray*}
and integrations by parts on the Grassmannian variables in order to 
remove the $D$ and ${\overline D}$ derivatives from one $\delta$, so 
that one $\theta$ integration can be performed immediately. In this 
way one obtains an expression that is local in $\theta$ as is expected 
from the ${\cal N}$=1 non-renormalization theorem \cite{n1nonrenormal}. 
From (\ref{atildephiphi}) one obtains
\begin{eqnarray*}
	\tilde{A}(p) & = & -g^{2}\delta^{a}_{b} \delta_{J}^{I}\left( 
	\frac{1}{4} \right)^{2} \int \frac{d^{4}k}{(2\pi)^{4}} 
	d^{2}\theta_{1}d^{2}{\overline\theta}_{1}
	d^{2}\theta_{2}d^{2}{\overline\theta}_{2} 
	\, \frac{1}{k^{2}[(p-k)^{2}+m^{2}]}
	\left\{ \Phi^{\dagger}_{aI}(p,1) \cdot \right. \\
	&& \hspace{-1.7cm} \left. \cdot \Phi^{bJ}(-p,2) \left[
	{\overline D}^{2}_{1} D^{2}_{1} \delta(1,2) \right]\left[ 1+\gamma
	(D^{2}_{1}{\overline D}^{2}_{1}+ {\overline D}^{2}_{1}D^{2}_{1}) 
	\delta(1,2) \right] \right\} = \tilde{A}_{1}(p) + 
	\tilde{A}_{2}(p) \rule{0pt}{20pt} \, .
\end{eqnarray*}
The first term is trivially calculated using 
\begin{eqnarray}
	&& \hspace{-1cm} \int d^{2}\theta d^{2}{\overline\theta} \, \left[ 
	{\overline D}^{2} D^{2} \delta(\theta-\theta^{\prime}) \right] 
	\delta(\theta-\theta^{\prime}) = 16 \nonumber \\ 
	&& \hspace{-1cm} \int d^{2}\theta d^{2}{\overline\theta}
	\, \left[ {\overline D}^{m} 
	D^{n} \delta(\theta-\theta^{\prime}) \right] 
	\delta(\theta-\theta^{\prime}) = 0 \quad {\rm if} \quad
	(m,n) \neq (2,2) 
	\label{thetaint}
\end{eqnarray}
and gives
\begin{equation}
    \tilde{A}_{1}(p) = -2 g^{2}\delta^{a}_{b} \delta_{J}^{I} 
    \int \frac{d^{4}k}{(2\pi)^{4}} d^{2}\theta d^{2}{\overline\theta}\, 
    \frac{1}{k^{2}[(p-k)^{2}+m^{2}]} \left\{ 
	\Phi^{\dagger}_{aI}(p,\theta,{\overline \theta}) 
	\Phi^{bJ}(-p,\theta,{\overline \theta}) \right\} \, .
	\label{a1tildephiphi} 
\end{equation}
In the computation of the second term one must use the (anti) 
commutators of covariant derivatives, which in particular imply 
\cite{wessbagger,grisaruroceksiegel}
\begin{equation}
	{\overline D}^{2}D^{2}{\overline D}^{2} = 16 \Box {\overline D}^{2} 
	\qquad D^{2}{\overline D}^{2}D^{2} = 16 \Box D^{2} \, .
	\label{dprop1}
\end{equation}
Then integration by parts gives 
\begin{equation}
    \tilde{A}_{2}(p) =  2 \, \gamma \, g^{2}\delta^{a}_{b}\delta_{J}^{I}
     \int \frac{d^{4}k}{(2\pi)^{4}} d^{2}\theta d^{2}{\overline\theta} \, 
	\frac{[(p-k)^{2}+p^{2}]}{k^{4}[(p-k)^{2}+m^{2}]} 
	\left\{ \Phi^{\dagger}_{aI}(p,\theta,{\overline \theta}) 
	\Phi^{bJ}(-p,\theta,{\overline \theta}) \right\} \, . 
\end{equation}
In conclusion from the first diagram one obtains two contributions, 
the first proportional to $\gamma$ and the second independent of it. 
The second diagram gives one single $\gamma$-independent contribution. 
The Feynman rules give
\begin{eqnarray*}
	\tilde{B}(p) &=& \int \frac{d^{4}k}{(2 \pi)^{4}} 
	d^{2}{\theta}_{1}d^{2}{\overline\theta}_{1} 
	d^{2}{\theta}_{2}d^{2}{\overline\theta}_{2} 
	\left\{ \left( -\frac{1}{3!} \right) 
	g \varepsilon^{I}{}_{KL} f_{acd} {{\Phi}^{\dagger}}_{I}^{a}
	(p,1) \,\left[ \left( -\frac{1}{4}{D_{1}}^{2}\right)
	\cdot \right. \right. \\
	&& \cdot \left. \left. \frac{\delta^{ce}\delta^{K}_{M}
	\delta(1,2)}{(p-k)^{2} + m^{2}} 
	\left( -\frac{1}{4} \dbo{{\overline D}_{2}}^{2} 
	\right) \right] \left[ \frac{\delta^{d}_{f}\delta^{L}_{N}
	\delta(1,2)}{k^{2} + m^{2}} \right] 
	\left(- \frac{1}{3!} \right) \varepsilon_{J}{}^{MN} f_{be}{}^{f} 
	{\Phi}^{bJ}(-p,2) \right\} \, .
\end{eqnarray*}
Proceeding exactly as for $\tilde{A}_{1}$ one obtains
\begin{equation}
    \tilde{B}(p) = 2 g^{2}\delta^{a}_{b} \delta_{J}^{I} 
    \int \frac{d^{4}k}{(2\pi)^{4}} d^{2}\theta d^{2}{\overline\theta}\, 
    \frac{1}{(k^{2}+m^{2})[(p-k)^{2}+m^{2}]} \left\{ 
	\Phi^{\dagger}_{aI}(p,\theta,{\overline \theta}) 
	\Phi^{bJ}(-p,\theta,{\overline \theta}) \right\} \, .
	\label{btildephiphi} 
\end{equation}
From the last diagram one gets
\begin{eqnarray*}
	&& \tilde{C}(p) = \int \frac{d^{4}k}{(2\pi)^{4}} 
	d^{2}\theta d^{2}{\overline\theta} \, 
	\left\{ \left(- \frac{g^{2}}{2} \right) {f_{ac}}^{d} 
	f_{deb} {\Phi}^{a\dagger}_{I}(p,1) \cdot \right. \\
	&& \cdot \left. \left[ - \left(1+ \gamma 
	({\overline D}_{1}^{2}D_{1}^{2}+D_{1}^{2}{\overline D}_{1}^{2}) 
	\right) \frac{\delta^{ce}\delta(1,1)}{k^{2}} \right] 
	\Phi^{bJ}(-p,1) \right\} \, .
\end{eqnarray*}
The only non-vanishing contribution comes from the term in which the 
projection operators act on the $\delta$-function, since 
$\delta_{4}(\theta-\theta)=0$, so that 
\begin{equation}
	\tilde{C}(p) =-2 \, \gamma \, g^{2}\delta_{a}^{b}\delta_{J}^{I} 
	\int \frac{d^{4}k}{(2\pi)^{4}} d^{2}\theta d^{2}{\overline\theta}
	\frac{1}{k^{4}} \left\{ \Phi^{a\dagger}_{I}(p,\theta,
	{\overline \theta}){\Phi}_{b}^{J}(-p,\theta,{\overline \theta}) 
	\right\} \, ,
	\label{ctildephiphi}
\end{equation}
from which one sees that the $\tilde{C}$ contribution is absent in 
the gauge $\alpha=1$, \ie $\gamma=0$. 

Putting the various corrections together gives the following result. 
The sum of the terms $\tilde{A}_{1}(p)$ and $\tilde{B}(p)$, which does 
not depend on $\gamma$, is 
\begin{eqnarray*}
	{\tilde{A}}_{1}(p) + \tilde{B}(p) &=& 2 g^{2} \delta_{a}^{b} 
	\delta_{J}^{I} \int \frac{d^{4}k}{(2\pi)^{4}} 
	d^{2}\theta d^{2}{\overline\theta}
	\left[ \Phi^{a\dagger}_{I}(p,\theta,{\overline \theta})
	{\Phi}_{b}^{J}(-p,\theta,{\overline \theta}) \right] \cdot \\
	&& \cdot \left\{ \frac{1}{k^{2} \left[ (p-k)^{2} + m^{2} \right]} - 
	\frac{1}{(k^{2} + m^{2}) \left[ (p-k)^{2} + m^{2} \right]} \right\} 
	\, .
\end{eqnarray*}
This is finite and exactly vanishes for $m=0$, \ie in the limit in 
which the ${\cal N}$=4 theory is recovered. 

The sum of the terms $\tilde{A}_{2}(p)$ and $\tilde{C}(p)$ is 
proportional to $\gamma$ and reads
\begin{eqnarray*}
	{\tilde{A}}_{2}(p)+\tilde{C}(p) &=& 2\gamma \, g^{2}\delta_{a}^{b} 
	\delta_{J}^{I} \int \frac{d^{4}k}{(2\pi)^{4}} 
	d^{2}\theta d^{2}{\overline\theta}
	\left[ \Phi^{a\dagger}_{I}(p,\theta,{\overline \theta})
	{\Phi}_{b}^{J}(-p,\theta,{\overline \theta}) \right] \cdot \\
	&& \cdot \left\{ \left( \frac{(p-k)^{2}}{k^{4} \left[ (p-k)^{2} + 
	m^{2} \right]} - \frac{1}{k^{4}} \right) + \left( 
	\frac{p^{2}}{k^{4} \left[ (p-k)^{2} + m^{2} \right]} 
	\right) \right\} \, .
\end{eqnarray*}
Both terms in the last integral are infrared divergent. The 
first one is zero in the limit $m \to 0$, while the second one gives an 
infrared divergence that survives in the ${\cal N}$=4 
theory, \ie in the limit $m\to 0$. More explicitly putting 
\begin{equation}
	{\tilde{A}}_{2}(p)+\tilde{C}(p)=2\gamma \, g^{2}\delta_{a}^{b} 
	\delta_{J}^{I}\left[ \Phi^{a\dagger}_{I}(p,\theta,{\overline 
	\theta}){\Phi}_{b}^{J}(-p,\theta,{\overline \theta}) \right]
	\left[ I_{1}(p) + I_{2}(p) \right] \, ,
	\label{totphiphi}
\end{equation}
one has
\begin{eqnarray*}
	&& I_{1}(p) = \int  \frac{d^{4}k}{(2\pi)^{4}}
	\left\{ \frac{(p-k)^{2}}{k^{4} \left[ (p-k)^{2} + m^{2} \right]} - 
	\frac{1}{k^{4}} \right\} = \\
	&& = - \int  \frac{d^{4}k}{(2\pi)^{4}} \left\{ 
	\frac{m^{2}}{k^{4} \left[ (p-k)^{2} + m^{2} \right]} \right\} = \\
	&& = - 2 m^{2}  \int_{0}^{1}  d\zeta \, \zeta \int 
	\frac{d^{4}k}{(2\pi)^{4}}\frac{1}{{\left[ k^{2} + 
	(p^{2} \zeta + m^{2})(1-\zeta) \right]}^{3}} = \\
	&& = - \frac{2 m^{2}}{2(4\pi)^{2}}  \int_{0}^{1}  d\zeta \, \zeta
	\frac{1}{(p^{2}\zeta + m^{2})(1-\zeta)} = \\
	&& = \frac{m^{2}}{(4\pi)^{2}} \left[ \frac{1}{(p^{2}+m^{2})} \log 
	\epsilon \; + \; \frac{m^{2}}{p^{2}(p^{2}+m^{2})} \log \left(
	\frac{p^{2}+m^{2}}{m^{2}} \right) \right] \, ,
\end{eqnarray*}
where a standard Feynman parameterization has been used. Analogously
\begin{eqnarray*}
	&& I_{2}(p) = \int \frac{d^{4}k}{(2\pi)^{4}} 
	\left\{ \frac{p^{2}}{k^{4}\left[(p-k)^{2}+m^{2}\right]}\right\} = \\ 
	&& = 2p^{2}  \int_{0}^{1}  d\zeta \, \zeta 
	\int \frac{d^{4}k}{(2\pi)^{4}}
	\frac{1}{{\left[ k^{2} + (p^{2} \zeta + m^{2})(1-\zeta) 
	\right]}^{3}} = \\ 
	&& = \frac{p^{2}}{(4\pi)^{2}}  \int_{0}^{1}  d\zeta \, \zeta
	\frac{1}{(p^{2}\zeta + m^{2})(1-\zeta)} = \\
	&& = -\frac{1}{(4\pi)^{2}} \left[ \frac{p^{2}}{(p^{2}+m^{2})} \log 
	\epsilon \; + \; \frac{m^{2}}{(p^{2}+m^{2})} \log \left(
	\frac{p^{2}+m^{2}}{m^{2}} \right) \right] \, .
\end{eqnarray*}
In the above expressions an infrared regulator $\epsilon$ has been 
introduced. Notice that the total correction (\ref{totphiphi}) is 
exactly zero on-shell, \ie for $p^{2}$=$m^{2}$. 

To summarize the results, the propagator of the chiral superfields of 
the ${\cal N}$=4 super Yang--Mills theory in ${\cal N}$=1 superspace 
is logarithmically infrared-divergent for any choice of the 
gauge parameter $\alpha \neq 1$. This divergence corresponds to a 
wave function renormalization for the superfields $\Phi^{I}$. 
In the Fermi--Feynman gauge $\alpha$=1 the one-loop correction 
exactly vanishes.

\subsection{Propagator of the vector superfield}

The one-loop calculation of the propagator of the vector superfield 
is much more complicated, because many more diagrams are involved 
producing a large number of contributions. The final result is however 
completely analogous: in the Fermi--Feynman gauge the one-loop 
correction is zero, whereas off-shell infrared divergences arise for 
$\alpha \neq 1$. In the presence of a mass term for the (anti) chiral 
superfields no new divergences are generated.

From a calculational viewpoint the new feature 
with respect to the $\langle \Phi \Phi^{\dagger} \rangle$ 
case is that there are also diagrams involving the 
ghosts. There are two multiplets of ghosts, described by the chiral 
superfields $C$ and $C^{\prime}$. The free propagators for these 
superfields will be denoted by

\noindent
\begin{fmffile}{scprime+c}
\begin{fmfgraph*}(90,50) 
  \fmfleft{i1} \fmfright{o1}
  \fmf{scalar}{i1,o1}
  \fmflabel{\raisebox{22pt}{\hspace{-20pt} $b$ \hspace{20pt}} 
  \raisebox{12pt}{$\longrightarrow \hspace{0.2cm} 
  \langle {\overline C}^{a\prime}(z) C^{b}(z^{\prime}) 
  \rangle_{{\rm free}} = \begin{displaystyle}
  \delta^{a}_{b}\frac{1}{\Box}
  \end{displaystyle} \delta_{8}(z-z^{\prime})$}}{o1}
  \fmflabel{\raisebox{22pt}{\hspace{18pt} $a$ 
  \hspace{-18pt}}}{i1}
\end{fmfgraph*}   
\end{fmffile} 

\noindent
\begin{fmffile}{sc+cprime}
\begin{fmfgraph*}(90,50) 
  \fmfleft{i1} \fmfright{o1}
  \fmf{scalar}{i1,o1}
  \fmflabel{\raisebox{22pt}{\hspace{-20pt} $b$ \hspace{20pt}} 
  \raisebox{12pt}{$\longrightarrow \hspace{0.2cm} 
  \langle {\overline C}^{a}(z) C^{b\prime}(z^{\prime}) 
  \rangle_{{\rm free}} = \begin{displaystyle}
  \delta^{a}_{b}\frac{1}{\Box}
  \end{displaystyle} \delta_{8}(z-z^{\prime})$}}{o1}
  \fmflabel{\raisebox{22pt}{\hspace{18pt} $a$ 
  \hspace{-18pt}}}{i1}
\end{fmfgraph*}   
\end{fmffile} 

\noindent
The ghosts are treated exactly like ordinary chiral superfields 
with the only difference that there is a minus sign associated with 
loops, because $C$ and $C^{\prime}$ are anticommuting fields 
\cite{grisaruroceksiegel} 

The corrections to the $\langle VV \rangle$ propagator at the 
one-loop level are given by the following diagrams

\vspace{0.3cm}
\noindent
\hspace*{1cm}
\begin{fmffile}{vvA}
 \begin{fmfgraph*}(150,80) 
  \fmfleft{i1} \fmfright{o1}
  \fmf{boson,tension=1.6}{i1,v1}
  \fmf{boson,tension=1.6}{v2,o1}
  \fmf{phantom,left=0.6,tension=0.3,tag=1}{v1,v2}
  \fmf{phantom,right=0.6,tension=0.3,tag=2}{v2,v1}
  \fmf{phantom,right=0.6,tension=0.3,tag=3}{v1,v2}
  \fmf{phantom,left=0.6,tension=0.3,tag=4}{v2,v1} 
  \fmfv{l=$V^{a}(z)$,l.a=180}{i1}
  \fmfv{l=$V_{b}(z^{\prime})$ \hspace{0.5cm} $\longrightarrow ~~~ 
        A(z;z^{\prime})$,l.a=0}{o1}
  \fmfdot{v1,v2}
  \fmffreeze
  \fmfipath{p[]}
  \fmfiset{p1}{vpath1(__v1,__v2)}
  \fmfiset{p2}{vpath2(__v2,__v1)}
  \fmfiset{p3}{vpath3(__v1,__v2)}
  \fmfiset{p4}{vpath4(__v2,__v1)}
  \fmfi{fermion}{subpath (0,length(p1)/2) of p1}
  \fmfi{fermion}{subpath (0,length(p2)/2) of p2}
  \fmfi{fermion}{subpath (length(p3)/2,length(p3)) of p3}  
  \fmfi{fermion}{subpath (length(p4)/2,length(p3)) of p4}
  \fmfiv{d.sh=cross,d.si=3mm,l=$\Phi\Phi$,
       l.a=90}{point length(p1)/2 of p1}
  \fmfiv{d.sh=cross,d.si=3mm,l=$\Phi^{\dagger}\Phi^{\dagger}$,
       l.a=-90}{point length(p3)/2 of p3}       
 \end{fmfgraph*}   
\end{fmffile} \\
\hspace*{1cm}
\begin{fmffile}{vvB}
 \begin{fmfgraph*}(150,80) 
  \fmfleft{i1} \fmfright{o1}
  \fmf{boson,tension=1.6}{i1,v1}
  \fmf{boson,tension=1.6}{v2,o1}
  \fmf{phantom,left=0.6,tension=0.3,tag=1}{v1,v2}
  \fmf{phantom,right=0.6,tension=0.3,tag=2}{v2,v1}
  \fmf{phantom,right=0.6,tension=0.3,tag=3}{v1,v2}
  \fmf{phantom,left=0.6,tension=0.3,tag=4}{v2,v1} 
  \fmfv{l=$V^{a}(z)$,l.a=180}{i1}
  \fmfv{l=$V_{b}(z^{\prime})$ \hspace{0.5cm} $\longrightarrow ~~~ 
        B(z;z^{\prime})$,l.a=0}{o1}
  \fmfdot{v1,v2}
  \fmffreeze
  \fmfipath{p[]}
  \fmfiset{p1}{vpath1(__v1,__v2)}
  \fmfiset{p2}{vpath2(__v2,__v1)}
  \fmfiset{p3}{vpath3(__v1,__v2)}
  \fmfiset{p4}{vpath4(__v2,__v1)}
  \fmfi{fermion}{subpath (0,length(p1)) of p1}
  \fmfi{fermion}{subpath (0,length(p4)) of p4}
  \fmfiv{l=$\Phi\Phi^{\dagger}$,l.a=90}{point length(p1)/2 of p1}
  \fmfiv{l=$\Phi^{\dagger}\Phi$,l.a=-90}{point length(p3)/2 of p3}       
 \end{fmfgraph*}   
\end{fmffile}

\vspace*{1.5cm}
\noindent
\hspace*{1cm}
\begin{fmffile}{vvC}
 \begin{fmfgraph*}(150,80) 
  \fmfleft{i1} \fmfright{o1}
  \fmf{boson,tension=1.4}{i1,v1}
  \fmf{boson,tension=1.4}{v1,o1}
  \fmf{phantom,tension=0.9}{v1,v1}
  \fmfv{l=$V^{a}(z)$,l.a=180}{i1}
  \fmfv{l=$V_{b}(z^{\prime})$ \hspace{0.5cm} $\longrightarrow ~~~ 
        C(z;z^{\prime})$,l.a=0}{o1}
  \fmfdot{v1}
  \fmffreeze
  \fmfipath{p[]}
  \fmfiset{p1}{vpath(__v1,__v1)}
  \fmfi{fermion}{subpath (0,length(p1)) of p1}
  \fmfiv{l=$\Phi\Phi^{\dagger}$,l.a=90}{point length(p1)/2 of p1}      
 \end{fmfgraph*}   
\end{fmffile} \\
\hspace*{1cm}
\begin{fmffile}{vvD1}
 \begin{fmfgraph*}(150,80) 
  \fmfleft{i1} \fmfright{o1}
  \fmf{boson,tension=1.6}{i1,v1}
  \fmf{boson,tension=1.6}{v2,o1}
  \fmf{phantom,left=0.6,tension=0.3,tag=1}{v1,v2}
  \fmf{phantom,right=0.6,tension=0.3,tag=2}{v2,v1}
  \fmf{phantom,right=0.6,tension=0.3,tag=3}{v1,v2}
  \fmf{phantom,left=0.6,tension=0.3,tag=4}{v2,v1} 
  \fmfv{l=$V^{a}(z)$,l.a=180}{i1}
  \fmfv{l=$V_{b}(z^{\prime})$ \hspace{0.5cm} $\longrightarrow ~~~ 
        D_{1}(z;z^{\prime})$,l.a=0}{o1}
  \fmfdot{v1,v2}
  \fmffreeze
  \fmfipath{p[]}
  \fmfiset{p1}{vpath1(__v1,__v2)}
  \fmfiset{p2}{vpath2(__v2,__v1)}
  \fmfiset{p3}{vpath3(__v1,__v2)}
  \fmfiset{p4}{vpath4(__v2,__v1)}
  \fmfi{scalar}{subpath (0,length(p1)) of p1}
  \fmfi{scalar}{subpath (0,length(p4)) of p4}
  \fmfiv{l=${\overline C}C^{\prime}$,l.a=90}{point length(p1)/2 of p1}
  \fmfiv{l=$C^{\prime}{\overline C}$,l.a=-90}{point length(p3)/2 of p3}       
 \end{fmfgraph*}   
\end{fmffile} \\
\hspace*{1cm}
\begin{fmffile}{vvD2}
 \begin{fmfgraph*}(150,80) 
  \fmfleft{i1} \fmfright{o1}
  \fmf{boson,tension=1.6}{i1,v1}
  \fmf{boson,tension=1.6}{v2,o1}
  \fmf{phantom,left=0.6,tension=0.3,tag=1}{v1,v2}
  \fmf{phantom,right=0.6,tension=0.3,tag=2}{v2,v1}
  \fmf{phantom,right=0.6,tension=0.3,tag=3}{v1,v2}
  \fmf{phantom,left=0.6,tension=0.3,tag=4}{v2,v1} 
    \fmfv{l=$V^{a}(z)$,l.a=180}{i1}
  \fmfv{l=$V_{b}(z^{\prime})$ \hspace{0.5cm} $\longrightarrow ~~~ 
        D_{2}(z;z^{\prime})$,l.a=0}{o1}
  \fmfdot{v1,v2}
  \fmffreeze
  \fmfipath{p[]}
  \fmfiset{p1}{vpath1(__v1,__v2)}
  \fmfiset{p2}{vpath2(__v2,__v1)}
  \fmfiset{p3}{vpath3(__v1,__v2)}
  \fmfiset{p4}{vpath4(__v2,__v1)}
  \fmfi{scalar}{subpath (0,length(p1)) of p1}
  \fmfi{scalar}{subpath (0,length(p4)) of p4}
  \fmfiv{l=${\overline C}^{\prime}C$,l.a=90}{point length(p1)/2 of p1}
  \fmfiv{l=$C{\overline C}^{\prime}$,l.a=-90}{point length(p3)/2 of p3}       
 \end{fmfgraph*}   
\end{fmffile} \\
\hspace*{1cm}
\begin{fmffile}{vvD3}
 \begin{fmfgraph*}(150,80) 
  \fmfleft{i1} \fmfright{o1}
  \fmf{boson,tension=1.4}{i1,v1}
  \fmf{boson,tension=1.4}{v2,o1}
  \fmf{phantom,left=0.6,tension=0.5,tag=1}{v1,v2}
  \fmf{phantom,right=0.6,tension=0.5,tag=2}{v1,v2} 
  \fmfv{l=$V^{a}(z)$,l.a=180}{i1}
  \fmfv{l=$V_{b}(z^{\prime})$ \hspace{0.5cm} $\longrightarrow ~~~ 
        D_{3}(z;z^{\prime})$,l.a=0}{o1}
  \fmfdot{v1,v2}
  \fmffreeze
  \fmfipath{p[]}
  \fmfiset{p1}{vpath1(__v1,__v2)}
  \fmfiset{p2}{vpath2(__v2,__v1)}
  \fmfi{scalar}{subpath (0,length(p1)) of p1}
  \fmfi{scalar}{subpath (0,length(p2)) of p2}
  \fmfiv{l=${\overline C}^{\prime}C$,l.a=90}{point length(p1)/2 of p1}
  \fmfiv{l=${\overline C}C^{\prime}$,l.a=-90}{point length(p2)/2 of p2}       
 \end{fmfgraph*}   
\end{fmffile} 

\vspace*{1.2cm}
\noindent
\hspace*{1cm}
\begin{fmffile}{vvE1}
 \begin{fmfgraph*}(150,80) 
  \fmfleft{i1} \fmfright{o1}
  \fmf{boson,tension=1.4}{i1,v1}
  \fmf{boson,tension=1.4}{v1,o1}
  \fmf{phantom,tension=0.9}{v1,v1}
  \fmfv{l=$V^{a}(z)$,l.a=180}{i1}
  \fmfv{l=$V_{b}(z^{\prime})$ \hspace{0.5cm} $\longrightarrow ~~~ 
        E_{1}(z;z^{\prime})$,l.a=0}{o1}
  \fmfdot{v1}
  \fmffreeze
  \fmfipath{p[]}
  \fmfiset{p1}{vpath(__v1,__v1)}
  \fmfi{scalar}{subpath (0,length(p1)) of p1}
  \fmfiv{l=${\overline C}^{\prime}C$,l.a=90}{point length(p1)/2 of p1}      
 \end{fmfgraph*}   
\end{fmffile} 

\vspace{0.2cm}
\noindent
\hspace*{1cm}
\begin{fmffile}{vvE2}
 \begin{fmfgraph*}(150,80) 
  \fmfleft{i1} \fmfright{o1}
  \fmf{boson,tension=1.4}{i1,v1}
  \fmf{boson,tension=1.4}{v1,o1}
  \fmf{phantom,tension=0.9}{v1,v1}
  \fmfv{l=$V^{a}(z)$,l.a=180}{i1}
  \fmfv{l=$V_{b}(z^{\prime})$ \hspace{0.5cm} $\longrightarrow ~~~ 
        E_{2}(z;z^{\prime})$,l.a=0}{o1}
  \fmfdot{v1}
  \fmffreeze
  \fmfipath{p[]}
  \fmfiset{p1}{vpath(__v1,__v1)}
  \fmfi{scalar}{subpath (0,length(p1)) of p1}
  \fmfiv{l=${\overline C}C^{\prime}$,l.a=90}{point length(p1)/2 of p1}      
 \end{fmfgraph*}   
\end{fmffile} \\
\hspace*{1cm}
\begin{fmffile}{vvF}
 \begin{fmfgraph*}(150,80) 
  \fmfleft{i1} \fmfright{o1}
  \fmf{boson,tension=1.4}{i1,v1}
  \fmf{boson,tension=1.4}{v2,o1}
  \fmf{phantom,left=0.6,tension=0.5,tag=1}{v1,v2}
  \fmf{phantom,right=0.6,tension=0.5,tag=2}{v1,v2} 
  \fmfv{l=$V^{a}(z)$,l.a=180}{i1}
  \fmfv{l=$V_{b}(z^{\prime})$ \hspace{0.5cm} $\longrightarrow ~~~ 
        F(z;z^{\prime})$,l.a=0}{o1}
  \fmfdot{v1,v2}
  \fmffreeze
  \fmfipath{p[]}
  \fmfiset{p1}{vpath1(__v1,__v2)}
  \fmfiset{p2}{vpath2(__v2,__v1)}
  \fmfi{boson}{subpath (0,length(p1)) of p1}
  \fmfi{boson}{subpath (0,length(p2)) of p2}
  \fmfiv{l=$VV$,l.a=90}{point length(p1)/2 of p1}
  \fmfiv{l=$VV$,l.a=-90}{point length(p2)/2 of p2}       
 \end{fmfgraph*}   
\end{fmffile} 

\vspace*{1.3cm}
\noindent
\hspace*{1cm}
\begin{fmffile}{vvG}
 \begin{fmfgraph*}(150,80) 
  \fmfleft{i1} \fmfright{o1}
  \fmf{boson,tension=1.4}{i1,v1}
  \fmf{boson,tension=1.4}{v1,o1}
  \fmf{phantom,tension=0.9}{v1,v1}
  \fmfv{l=$V^{a}(z)$,l.a=180}{i1}
  \fmfv{l=$V_{b}(z^{\prime})$ \hspace{0.5cm} $\longrightarrow ~~~ 
        G(z;z^{\prime})$,l.a=0}{o1}
  \fmfdot{v1}
  \fmffreeze
  \fmfipath{p[]}
  \fmfiset{p1}{vpath(__v1,__v1)}
  \fmfi{boson}{subpath (0,length(p1)) of p1}
  \fmfiv{l=$VV$,l.a=90}{point length(p1)/2 of p1}      
 \end{fmfgraph*}   
\end{fmffile} 

The contributions $A$ to $E$ are rather straightforward to evaluate 
much in the same way as the diagrams entering the $\langle \Phi 
\Phi^{\dagger} \rangle$ propagator. The last two graphs are more 
involved because the free propagator for the $V$ superfield is more 
complicated for generic values of the parameter $\alpha$. Moreover the
cubic and quartic vertices
\begin{eqnarray*}
&& - \frac{i}{16 \sqrt{2}} g f_{abc} \left[ {\overline D}^{2} 
\left( D^{\alpha}V^{a} \right) \right] V^{b} \left( D_{\alpha}V^{c}
 \right) \\ 
 && -\frac{1}{128} g^{2} f_{ab}{}^{e}f_{ecd} V^{a} 
\left( D^{\alpha} V^{b} \right) \left[ \left({\overline D}^{2} 
V^{c} \right) \left( D_{\alpha}V^{d} \right) \right] \, , 
\end{eqnarray*}
lead to several terms corresponding to the many ways in which the 
covariant derivatives can act on the $V$ lines. The $V^{3}$ vertex 
in particular

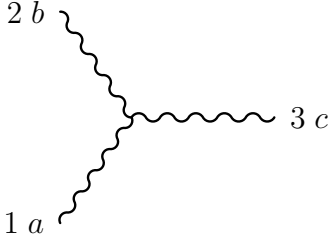
\vspace*{0.8cm}
\noindent
\hspace*{1cm}
\begin{fmffile}{vvv}
  \begin{fmfgraph*}(90,80) 
  \fmfleft{i1,i2} \fmfright{o1}
  \fmf{boson}{i1,v1}
  \fmf{boson}{i2,v1}
  \fmf{boson}{v1,o1}
  \fmfv{l=$1\;a$,l.a=180}{i1}
  \fmfv{l=$2\;b$,l.a=180}{i2}
  \fmfv{l=$3\;c$,l.a=0}{o1}
  \end{fmfgraph*}   
\end{fmffile} 

\vspace*{0.6cm}
\noindent
gives rise to six different terms.

Schematically the calculation goes as follows. 
$\tilde{A}(p)$ is a new contribution that appears because of the 
addition of the mass terms (\ref{massterm}); it is not present in the 
${\cal N}$=4 theory, \ie when $m$=0. It is useful to discuss 
separately the corrections coming from diagrams $\tilde{A}$, $\tilde{B}$ 
and $\tilde{C}$ and those obtained from $\tilde{D}_{i}$, 
$\tilde{E}_{i}$, $\tilde{F}$ and $\tilde{G}$, since the latter 
correspond to the one-loop contribution to the vector superfield 
propagator in the ${\cal N}$=1 supersymmetric Yang--Mills theory.

The first diagram, $\tilde{A}$, is logarithmically divergent and reads
\begin{equation}
	\tilde{A}(p) = \frac{3}{2} g^{2}\delta_{ab}  \int 
	\frac{d^{4}k}{(2\pi)^{4}} d^{2}\theta d^{2}{\overline\theta} \, 
	\frac{m^{2}}{(k^{2}+m^{2})[(p-k)^{2}+m^{2}]} 
	\left\{ V^{a}(p,\theta,{\overline\theta}) 
	V^{b}(-p,\theta,{\overline\theta}) \right\} ,
    \label{vvmassdiverg}
\end{equation}
For the diagram $\tilde{B}$ application of the Feynman rules gives 
rise to three different contributions, one quadratically divergent 
and two logarithmically divergent
\begin{eqnarray}
	&& \hspace{-1cm} \tilde{B}(p)= 
	\frac{3}{2} g^{2} \delta_{ab}
	\int\frac{d^{4}k}{(2\pi)^{4}} d^{2}\theta d^{2}{\overline\theta}\, 
	\frac{1}{(k^{2}+m^{2})[(p-k)^{2}+m^{2}]} \left\{ k^{2}V^{a}
	(p,\theta,{\overline\theta})V^{b}(-p,\theta,{\overline\theta})
	\rule{0pt}{18pt} - \right. \nonumber \\
	&& \hspace{-1cm} 
	\left. -\frac{i}{4} p_{\mu}\sigma^{\mu}_{\alpha{\dot\alpha}}V^{a}
	(p,\theta,{\overline\theta}) \left[
	({\overline D}^{\dot\alpha}D^{\alpha}) 
	V^{b}(-p,\theta,{\overline\theta}) \right] + \frac{1}{4}
	V^{a}(p,\theta,{\overline\theta}) \left[({\overline D}^{2}D^{2}) 
	V^{b}(-p,\theta,{\overline\theta}) \right] \right\} \, .
	\label{bcorrvv}
\end{eqnarray}
The tadpole diagram $\tilde{C}$ gives a quadratically divergent 
result of the form
\begin{equation}
	\tilde{C} = -\frac{3}{2} g^{2}\delta^{ab}
	\int \frac{d^{4}k}{(2\pi)^{4}} 
	d^{2}\theta d^{2}{\overline\theta} \, \frac{1}{(k^{2}+m^{2})} 
	\left\{ V^{a}(p,\theta,{\overline\theta}) 
	V^{b}(-p,\theta,{\overline\theta}) \right\} \, .
\end{equation}
Putting the three corrections $\tilde{A}$, 
$\tilde{B}$ and $\tilde{C}$ together gives a net result that is only 
logarithmically divergent
\begin{eqnarray}
	&& \hspace{-1.3cm} \tilde{A}(p)+\tilde{B}(p)+\tilde{C}(p) = 
	-\frac{3}{2} g^{2} \delta_{ab}
	\int\frac{d^{4}k}{(2\pi)^{4}} d^{2}\theta d^{2}{\overline\theta}\, 
	\frac{1}{(k^{2}+m^{2})[(p-k)^{2}+m^{2}]} \cdot \nonumber \\
	&& \hspace{-1.3cm} \cdot
	\left\{ \frac{i}{4} p_{\mu}\sigma^{\mu}_{\alpha{\dot\alpha}}V^{a}
	(p,\theta,{\overline\theta}) \left[
	({\overline D}^{\dot\alpha}D^{\alpha}) 
	V^{b}(-p,\theta,{\overline\theta}) \right] + \frac{1}{16}
	V^{a}(p,\theta,{\overline\theta}) \left[({\overline D}^{2}D^{2}) 
	V^{b}(-p,\theta,{\overline\theta}) \right] \right\} \, .
	\label{vvlog1}
\end{eqnarray}
Notice that in particular the logarithmically divergent contribution 
proportional to $m^{2}$ exactly cancels out. 
This is crucial because this correction would correspond to a mass 
renormalization for the vector superfield that is known to be 
excluded in any gauge theory as well as in supersymmetric theories in 
general.

The diagrams $\tilde{D}_{i}$ and $\tilde{E}_{i}$ are completely 
analogous to the previous ones, with the only difference that the mass
does not appear in the denominators and there is a minus sign 
associated with the loops. Their sum is logarithmically divergent and 
takes the form 
\begin{eqnarray}
	\tilde{D}_{1}(p)+\tilde{D}_{2}(p)+\tilde{D}_{3}(p)+
	\tilde{E}_{1}(p)+\tilde{E}_{2}(p) = \frac{1}{16} g^{2} \delta_{ab}
	\int\frac{d^{4}k}{(2\pi)^{4}} d^{2}\theta d^{2}{\overline\theta}\, 
	\frac{1}{k^{2}(p-k)^{2}} \cdot \nonumber \\
	\cdot \left\{ i p_{\mu} \sigma^{\mu}_{\alpha{\dot\alpha}}
	V^{a}(p,\theta,{\overline\theta}) \left[({\overline D}^{\dot\alpha}
	D^{\alpha}) V^{b}(-p,\theta,{\overline\theta}) \right] + 
	\frac{1}{8} V^{a}(p,\theta,{\overline\theta}) 
	\left[({\overline D}^{2}D^{2}) 
	V^{b}(-p,\theta,{\overline\theta}) \right] \right\} \, .
	\label{vvlog2} 
\end{eqnarray}

All of the above corrections are independent of the gauge parameter 
$\gamma$=$\alpha -1$ and must be summed to those coming from the last 
two diagrams. $\tilde{F}$ exactly vanishes for any $\alpha$, so that 
only $\tilde{E}$ needs to be considered. This diagram produces in 
principle 72 corrections because the Feynman rules give rise to 18 
terms (distributing the covariant derivatives associated with the two 
vertices), each of which splits into 4, since the free propagator 
itself contains two terms. Many of these contributions can be easily 
shown to vanish using the properties of the covariant derivatives. In 
particular one uses  
\begin{displaymath}
	{\overline D}^{2}D_{\alpha}{\overline D}^{2}D^{2} = 0 \, ,
\end{displaymath}
which follows from ${\overline D}^{3}=0$ and use of the (anti) 
commutation relations for the $D$'s. It is useful to separate in the 
non-vanishing part terms proportional to $\gamma$ and $\gamma^{2}$, 
from the $\gamma$-independent terms. The latter combine to give a 
logarithmically divergent correction that together with that of equation 
(\ref{vvlog2}) cancel the correction (\ref{vvlog1}) coming from the sum 
$\tilde{A}+\tilde{B}+\tilde{C}$. Actually if $m\neq 0$ this 
sum is finite and exactly vanishes at $m$=0. As a result the only 
non-vanishing corrections to the vector superfield propagator at the 
one loop level come from terms proportional to $\gamma$ and to 
$\gamma^{2}$ in $\tilde{E}(p)$. The former contain an infrared divergent 
part of the form
\begin{displaymath}
   J^{(1)}(p) = c_{1}\gamma g^{2}\delta^{a}_{b} \int \frac{d^{4}k}
   {(2\pi)^{4}} d^{2}\theta d^{2}{\overline\theta} 
   \, \frac{\sigma^{\mu}_{\alpha{\dot\alpha}}
   \sigma^{\nu}_{\beta{\dot\beta}}p_{\mu}p_{\nu}}{k^{4}(p-k)^{2}} 
   \left\{ V_{a}(p,\theta,{\overline\theta}) \left[D^{\alpha}
   {\overline D}^{\dot\alpha}{\overline D}^{\dot\beta}D^{\beta}
   V^{b}(-p,\theta,{\overline\theta}) \right] \right\} \, ,
\end{displaymath} 
but is ultraviolet finite.
Furthermore there is a correction, finite both in the ultraviolet and 
in the infrared regions, proportional to $\gamma^{2}$, that reads
\begin{eqnarray*}
   J^{(2)}(p) &=& c_{2}\gamma^{2}g^{2}\delta^{a}_{b} \int \frac{d^{4}k}
   {(2\pi)^{4}}d^{2}\theta d^{2}{\overline\theta} \, \frac{(\sigma^{\nu}
   {\overline\sigma}^{\mu}\sigma^{\lambda})_{\alpha{\dot\alpha}}
   (p-k)_{\mu}k_{\nu}p_{\lambda}}{k^{4}(p-k)^{4}} \cdot \\ 
   && \hspace{1.3cm} \cdot 
   \left\{ \left[ ({\overline D}^{2}D^{\alpha})
   V_{a}(p,\theta,{\overline\theta}) \right] 
   \left[ (D^{2}{\overline D}^{\dot\alpha}) 
   V^{b}(-p,\theta,{\overline\theta}) \right] \right\} \, .
\end{eqnarray*}
In conclusion in the ${\cal N}$=4 theory, \ie when $m$=0, the one-loop 
correction to the vector superfield propagator, just like that of the 
chiral superfield, is ultraviolet finite, but infrared singular unless 
the Fermi--Feynman gauge, $\alpha$=1, is chosen, in which case it vanishes. 
Like in the case of the chiral superfield propagator the non-vanishing 
$\gamma$-dependent corrections could be reabsorbed by a wave function 
renormalization of the superfield $V$ and are zero on-shell, 
\ie for $p^{2}$=0.

The proof of the finiteness of the theory in the presence of the mass 
terms (\ref{massterm}) given in \cite{parkeswest} is based on naive power 
counting, which gives for the superficial degree of divergence, $d$, 
of a diagram in ${\cal N}$=1 superspace 
\cite{divdegree}
\begin{displaymath}
	d = 2 - E -C \, ,
\end{displaymath}
where $E$ is the number of external (anti) chiral lines and $C$ the 
number of $\Phi \Phi$ or $\Phi^{\dagger}\Phi^{\dagger}$ propagators. 
In \cite{parkeswest} it is also argued that for corrections to the 
effective action involving only $V$ superfields the requirement of gauge 
invariance reduces the degree of divergence to
\begin{displaymath}
	d = -C \, .
\end{displaymath}
However for the purposes of \cite{japan,yoshida} it appears crucial 
that no divergences, not even corresponding to a wave function 
renormalization, be present in complete $n$-point functions for any $n$. 
The computation of the two-point functions in this section has shown 
that this actually the case at one loop. An argument for the 
generalization of this result to Green functions with an arbitrary 
number of external $V$ lines will now be briefly sketched. First note 
that diagrams involving only vector and ghost superfields are not 
modified by the inclusion of (\ref{massterm}), so that one must only 
consider graphs containing internal chiral lines. The ultraviolet 
properties of diagrams that are only logarithmically divergent in the 
original ${\cal N}$=4 theory are not modified by the presence of the 
mass in the propagators. Thus the contributions that we need to analyze 
are the quadratically divergent ones, which can acquire subleading 
logarithmic singularities, plus eventually new diagrams involving 
$\Phi\Phi$ and $\Phi^{\dagger}\Phi^{\dagger}$ propagators. The 
relevant quadratic divergences come from tadpole diagrams 

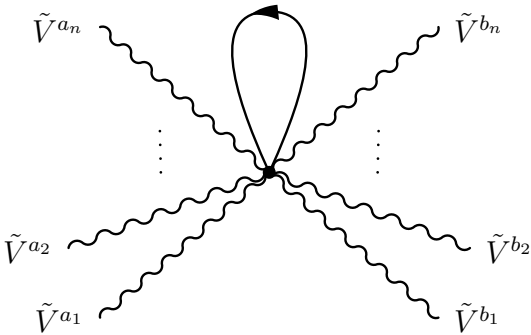
\vspace{1.5cm}
\noindent
\begin{fmffile}{vvvvC}
 \begin{fmfgraph*}(160,110) 
  \fmfleft{i1,i2,i3,i4,i5} 
  \fmfright{o1,o2,o3,o4,o5}
  \fmf{boson,tension=0.2,tag=1}{i1,v1}
  \fmf{boson,tension=0.2,tag=2}{i2,v1}
  \fmf{phantom,tension=0.2,tag=3}{i3,v1}
  \fmf{phantom,tension=0.2,tag=4}{i4,v1}
  \fmf{boson,tension=0.2,tag=5}{i5,v1}
  \fmf{boson,tension=0.2,tag=10}{v1,o1}
  \fmf{boson,tension=0.2,tag=11}{v1,o2}
  \fmf{phantom,tension=0.2,tag=12}{v1,o3}
  \fmf{phantom,tension=0.2,tag=13}{v1,o4}
  \fmf{boson,tension=0.2,tag=14}{v1,o5}
  \fmf{fermion,left=-90,tension=0.9}{v1,v1}
  \fmfv{l=$\tilde{V}^{a_{1}}$,l.a=180}{i1}
  \fmfv{l=$\tilde{V}^{a_{2}}$,l.a=180}{i2}
  \fmfv{l=$\tilde{V}^{a_{n}}$,l.a=180}{i5}
  \fmfv{l=$\tilde{V}^{b_{1}}$,l.a=0}{o1}
  \fmfv{l=$\tilde{V}^{b_{2}}$,l.a=0}{o2}
  \fmfv{l=$\tilde{V}^{b_{n}}$,l.a=0}{o5}
  \fmfdot{v1}
  \fmffreeze
  \fmfipath{p[]}
  \fmfiset{p3}{vpath3(__i3,__v1)}
  \fmfiset{p4}{vpath4(__i4,__v1)}
  \fmfiset{p12}{vpath12(__v1,__o3)}
  \fmfiset{p13}{vpath13(__v1,__o4)}
  \fmfi{dots}{point length(p3)/2 of p3
              -- point 14length(p4)/30 of p4}
  \fmfi{dots}{point length(p12)/2 of p12
              -- point 16length(p13)/30 of p13}           
 \end{fmfgraph*}   
\end{fmffile} 

\vspace{1.5cm}
\noindent
The only new diagram containing $\Phi\Phi$ and 
$\Phi^{\dagger}\Phi^{\dagger}$ propagators that must be considered 
is 

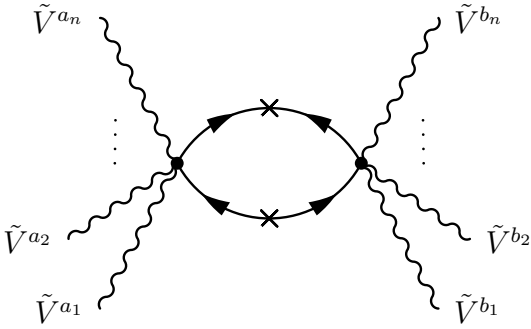
\vspace{1cm}
\noindent
\begin{fmffile}{vvvvA}
 \begin{fmfgraph*}(160,110) 
  \fmfleft{i1,i2,i3,i4,i5} 
  \fmfright{o1,o2,o3,o4,o5}
  \fmf{boson,tension=0.3,tag=1}{i1,v1}
  \fmf{boson,tension=0.3,tag=2}{i2,v1}
  \fmf{phantom,tension=0.3,tag=3}{i3,v1}
  \fmf{phantom,tension=0.3,tag=4}{i4,v1}
  \fmf{boson,tension=0.3,tag=5}{i5,v1}
  \fmf{phantom,left=0.6,tension=0.2,tag=6}{v1,v2}
  \fmf{phantom,right=0.6,tension=0.2,tag=7}{v1,v2}
  \fmf{phantom,left=0.6,tension=0.2,tag=8}{v2,v1}
  \fmf{phantom,right=0.6,tension=0.2,tag=9}{v2,v1}
  \fmf{boson,tension=0.3,tag=10}{v2,o1}
  \fmf{boson,tension=0.3,tag=11}{v2,o2}
  \fmf{phantom,tension=0.3,tag=12}{v2,o3}
  \fmf{phantom,tension=0.3,tag=13}{v2,o4}
  \fmf{boson,tension=0.3,tag=14}{v2,o5}
  \fmfv{l=$\tilde{V}^{a_{1}}$,l.a=180}{i1}
  \fmfv{l=$\tilde{V}^{a_{2}}$,l.a=180}{i2}
  \fmfv{l=$\tilde{V}^{a_{n}}$,l.a=180}{i5}
  \fmfv{l=$\tilde{V}^{b_{1}}$,l.a=0}{o1}
  \fmfv{l=$\tilde{V}^{b_{2}}$,l.a=0}{o2}
  \fmfv{l=$\tilde{V}^{b_{n}}$,l.a=0}{o5}
  \fmfdot{v1,v2}
  \fmffreeze
  \fmfipath{p[]}
  \fmfiset{p3}{vpath3(__i3,__v1)}
  \fmfiset{p4}{vpath4(__i4,__v1)}
  \fmfiset{p6}{vpath6(__v1,__v2)}
  \fmfiset{p7}{vpath7(__v1,__v2)}
  \fmfiset{p8}{vpath8(__v2,__v1)}
  \fmfiset{p9}{vpath9(__v2,__v1)}
  \fmfiset{p12}{vpath12(__v2,__o3)}
  \fmfiset{p13}{vpath13(__v2,__o4)}
  \fmfi{fermion}{subpath (0,length(p6)/2) of p6}
  \fmfi{fermion}{subpath (0,length(p9)/2) of p9}
  \fmfi{fermion}{subpath (length(p8)/2,length(p8)) of p8}
  \fmfi{fermion}{subpath (length(p7)/2,length(p7)) of p7}
  \fmfiv{d.sh=cross,d.si=3mm}{point length(p6)/2 of p6}
  \fmfiv{d.sh=cross,d.si=3mm}{point length(p7)/2 of p7}
  \fmfi{dots}{point length(p3)/2 of p3
              -- point 9length(p4)/20 of p4}
  \fmfi{dots}{point length(p12)/2 of p12
              -- point 11length(p13)/20 of p13}           
 \end{fmfgraph*}   
\end{fmffile} 

\vspace{0.5cm}
\noindent
The only covariant derivatives that are associated with the 
vertices in these graphs come from the functional 
derivatives and must act on the internal (anti) chiral lines to give 
a non-vanishing result. As a consequence the loop integral in the above 
diagrams is exactly the same as for the $\tilde{C}$ and $\tilde{A}$ 
corrections to the $VV$ propagator. Hence summing to the previous diagrams 
the contribution of

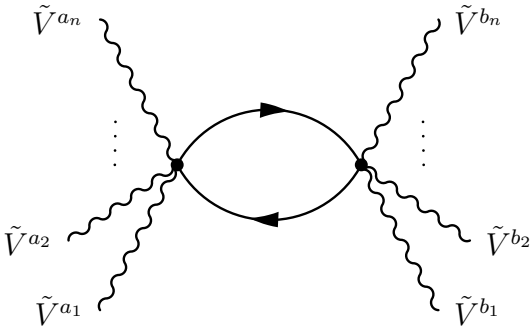
\vspace{1cm}
\noindent
\begin{fmffile}{vvvvB}
 \begin{fmfgraph*}(160,110) 
  \fmfleft{i1,i2,i3,i4,i5} 
  \fmfright{o1,o2,o3,o4,o5}
  \fmf{boson,tension=0.3,tag=1}{i1,v1}
  \fmf{boson,tension=0.3,tag=2}{i2,v1}
  \fmf{phantom,tension=0.3,tag=3}{i3,v1}
  \fmf{phantom,tension=0.3,tag=4}{i4,v1}
  \fmf{boson,tension=0.3,tag=5}{i5,v1}
  \fmf{phantom,left=0.6,tension=0.2,tag=6}{v1,v2}
  \fmf{phantom,right=0.6,tension=0.2,tag=7}{v1,v2}
  \fmf{phantom,left=0.6,tension=0.2,tag=8}{v2,v1}
  \fmf{phantom,right=0.6,tension=0.2,tag=9}{v2,v1}
  \fmf{boson,tension=0.3,tag=10}{v2,o1}
  \fmf{boson,tension=0.3,tag=11}{v2,o2}
  \fmf{phantom,tension=0.3,tag=12}{v2,o3}
  \fmf{phantom,tension=0.3,tag=13}{v2,o4}
  \fmf{boson,tension=0.3,tag=14}{v2,o5}
  \fmfv{l=$\tilde{V}^{a_{1}}$,l.a=180}{i1}
  \fmfv{l=$\tilde{V}^{a_{2}}$,l.a=180}{i2}
  \fmfv{l=$\tilde{V}^{a_{n}}$,l.a=180}{i5}
  \fmfv{l=$\tilde{V}^{b_{1}}$,l.a=0}{o1}
  \fmfv{l=$\tilde{V}^{b_{2}}$,l.a=0}{o2}
  \fmfv{l=$\tilde{V}^{b_{n}}$,l.a=0}{o5}
  \fmfdot{v1,v2}
  \fmffreeze
  \fmfipath{p[]}
  \fmfiset{p3}{vpath3(__i3,__v1)}
  \fmfiset{p4}{vpath4(__i4,__v1)}
  \fmfiset{p6}{vpath6(__v1,__v2)}
  \fmfiset{p7}{vpath7(__v1,__v2)}
  \fmfiset{p8}{vpath8(__v2,__v1)}
  \fmfiset{p9}{vpath9(__v2,__v1)}
  \fmfiset{p12}{vpath12(__v2,__o3)}
  \fmfiset{p13}{vpath13(__v2,__o4)}
  \fmfi{fermion}{subpath (0,length(p6)) of p6}
  \fmfi{fermion}{subpath (0,length(p8)) of p8}
  \fmfi{dots}{point length(p3)/2 of p3
              -- point 9length(p4)/20 of p4}
  \fmfi{dots}{point length(p12)/2 of p12
              -- point 11length(p13)/20 of p13}           
 \end{fmfgraph*}   
\end{fmffile} 

\vspace{0.5cm}
\noindent
leads to a net logarithmically divergent correction that is exactly 
the same as the one obtained in the original ${\cal N}$=4 theory. 
Like in the latter case this divergence will be cancelled by the 
contributions coming from the other one-loop diagrams involving $V$ 
and ghost internal lines. The argument given here reinforces the results 
of \cite{parkeswest}, at least at the one-loop level, and supports 
the proposal, put forward in \cite{japan,yoshida}, according to which 
the mass deformed ${\cal N}$=4 model can be considered a consistent 
supersymmetry-preserving regularization for (a class of) ${\cal N}$=1 
theories.

Notice that for the previous discussion it is not necessary to 
consider equal masses for all the (anti) chiral superfields. The same 
results can be proved giving different masses to the three superfields. 
This can be easily understood since in each diagram considered in this 
section only one chiral/anti-chiral pair is involved, because the 
propagators are diagonal in `flavor' space and the vertices containing 
vector and (anti) chiral superfields couple $\Phi^{\dagger}_{I}$ and 
$\Phi^{I}$ with the same index $I$. Basically this means that the 
above discussed cancellations apply separately to the contributions 
of each (anti) chiral superfield. From the viewpoint of the 
dimensional analysis of \cite{parkeswest} having different masses 
$m_{I}$ is irrelevant. 

As a consequence one can in particular give mass to only two of the 
(anti) chiral multiplets. This suggests the possibility of 
generalizing the approach of \cite{japan,yoshida} to the case of 
${\cal N}$=2 super Yang--Mills theories. A discussion of the effect 
of a ${\cal N}$=2 mass term in ${\cal N}$=4 supersymmetric 
Yang--Mills theory can be found in \cite{n2mass}.

\section{Three- and four-point functions of ${\cal N}$=1 superfields}
\label{threefourpointf}

The computation of Green functions with three and four external legs 
will now be considered. From now on the Fermi--Feynman gauge will be 
assumed. The three-point functions are expected to suffer from 
infrared problems of the kind encountered in the preceding section. 
This issue will not be addressed here since the calculation with 
$\alpha \neq 1$ is quite involved, even at the one-loop level, because 
of the huge number of contributions. Four-point functions on the 
contrary should be infrared finite, because they are directly related 
to physical scattering amplitudes. Notice, however, that infrared 
divergences in the Green function with four external $V$ lines were 
found in \cite{juerstorey1} with $\alpha \neq 1$. In 
that paper beyond the infrared singularity, it was shown that the 
adimensionality of the superfield $V$ implies that it requires a 
non-linear renormalization, in the sense that the renormalized field 
$V_{R}$ will be a non-linear function of the bare field $V$
\begin{displaymath}
	V_{R}=f(V) \, .
\end{displaymath}

An example of three-point function will be briefly discussed here and 
then the more interesting case of a four-point function will be studied 
in greater detail.
 
\subsection{Three-point functions}

The simplest three point function that one can consider corresponds to 
the correction to the vertex $\varepsilon_{IJK}\tr\left(\Phi^{I}\Phi^{J}
\Phi^{K}\right)$ (or $\varepsilon^{IJK}\tr\left(\Phi^{\dagger}_{I}
\Phi^{\dagger}_{J}\Phi^{\dagger}_{K}\right)$), which is determined by one 
single diagram at the one loop level. However, as a less trivial example 
of computation of three-point functions, the one-loop correction to the 
vertex 

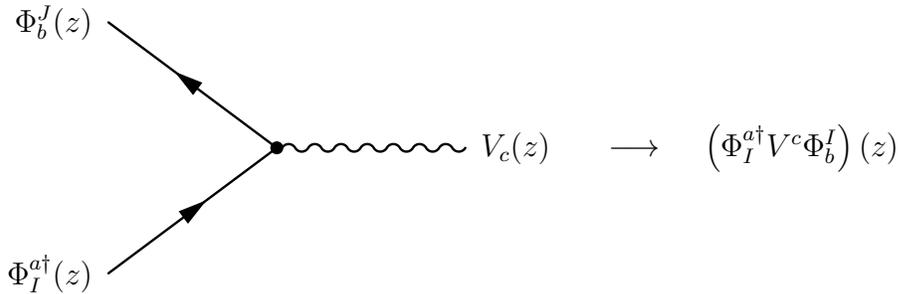
\vspace{1cm}
\noindent
\hspace*{0.7cm}
\begin{fmffile}{phi+vphi}
 \begin{fmfgraph*}(150,95) 
  \fmfleft{i1,i2} \fmfright{o1}
  \fmf{phantom,tension=0.7,tag=1}{i1,v1}
  \fmf{phantom,tension=0.7,tag=2}{i2,v1}
  \fmf{phantom,tension=0.7,tag=3}{v1,i1}
  \fmf{phantom,tension=0.7,tag=4}{v1,i2} 
  \fmf{boson,tension=2.5}{v1,o1}
  \fmfv{l=$\Phi^{a\dagger}_{I}(z)$,l.a=180}{i1}
  \fmfv{l=$\Phi^{J}_{b}(z)$,l.a=180}{i2}
  \fmfv{l=$V_{c}(z)$ \hspace{0.5cm} 
          $\longrightarrow ~~~ \left(\Phi^{a\dagger}_{I}
		  V^{c}\Phi_{b}^{I}\right)(z)$,l.a=0}{o1}
  \fmfdot{v1}
  \fmffreeze
  \fmfipath{p[]}
  \fmfiset{p1}{vpath1(__i1,__v1)}
  \fmfiset{p2}{vpath2(__i2,__v1)}
  \fmfiset{p3}{vpath3(__v1,__i1)}
  \fmfiset{p4}{vpath4(__v1,__i2)}
  \fmfi{fermion}{subpath (0,length(p1)) of p1}
  \fmfi{fermion}{subpath (0,length(p4)) of p4}
 \end{fmfgraph*}   
\end{fmffile} 

\vspace{1cm}
\noindent
will be considered here. The correction at the one-loop level comes 
from the following diagrams (the notation for the propagators is the 
same as in the previous section)

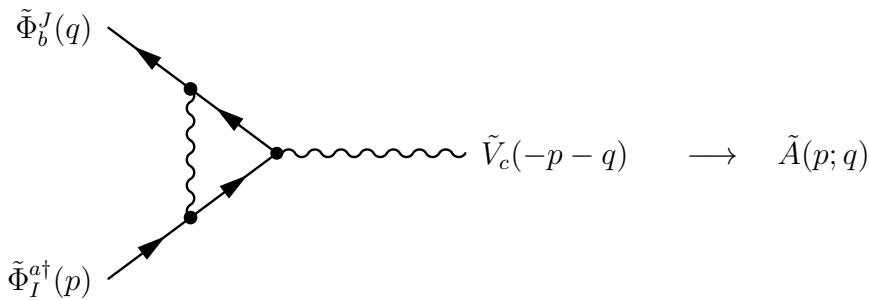
\vspace{1cm}
\noindent
\hspace*{0.7cm}
\begin{fmffile}{phi+vphiA}
 \begin{fmfgraph*}(150,95) 
  \fmfleft{i1,i2} \fmfright{o1}
  \fmf{phantom,tension=0.7,tag=1}{i1,v1}
  \fmf{phantom,tension=0.7,tag=2}{i2,v1}
  \fmf{phantom,tension=0.7,tag=3}{v1,i1}
  \fmf{phantom,tension=0.7,tag=4}{v1,i2} 
  \fmf{boson,tension=2.5}{v1,o1}
  \fmfv{l=$\tilde{\Phi}^{a\dagger}_{I}(p)$,l.a=180}{i1}
  \fmfv{l=$\tilde{\Phi}^{J}_{b}(q)$,l.a=180}{i2}
  \fmfv{l=$\tilde{V}_{c}(-p-q)$ \hspace{0.5cm} 
          $\longrightarrow ~~~ \tilde{A}(p;q)$,l.a=0}{o1}
  \fmfdot{v1}
  \fmffreeze
  \fmfipath{p[]}
  \fmfiset{p1}{vpath1(__i1,__v1)}
  \fmfiset{p2}{vpath2(__i2,__v1)}
  \fmfiset{p3}{vpath3(__v1,__i1)}
  \fmfiset{p4}{vpath4(__v1,__i2)}
  \fmfi{fermion}{subpath (0,length(p1)/2) of p1}
  \fmfi{fermion}{subpath (length(p1)/2,length(p1)) of p1}
  \fmfi{fermion}{subpath (0,length(p4)/2) of p4}
  \fmfi{fermion}{subpath (length(p4)/2,length(p4)) of p4}
  \fmfi{boson}{point length(p1)/2 of p1 -- point
               length(p2)/2 of p2}
  \fmfiv{d.sh=circle,d.fill=full,d.si=1.5mm}{point length(p1)/2 of p1}      
  \fmfiv{d.sh=circle,d.fill=full,d.si=1.5mm}{point length(p2)/2 of p2}
 \end{fmfgraph*}   
\end{fmffile} 

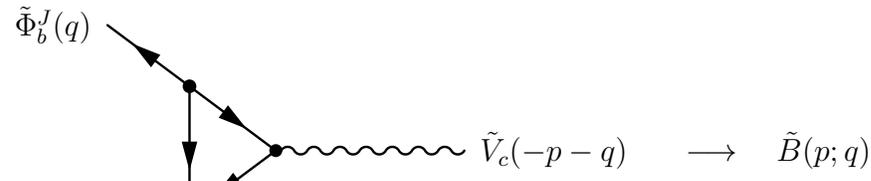
\vspace{1cm}
\noindent
\hspace*{0.7cm}
\begin{fmffile}{phi+vphiB}
 \begin{fmfgraph*}(150,95) 
  \fmfleft{i1,i2} \fmfright{o1}
  \fmf{phantom,tension=0.7,tag=1}{i1,v1}
  \fmf{phantom,tension=0.7,tag=2}{i2,v1}
  \fmf{phantom,tension=0.7,tag=3}{v1,i1}
  \fmf{phantom,tension=0.7,tag=4}{v1,i2} 
  \fmf{boson,tension=2.5}{v1,o1}
  \fmfv{l=$\tilde{\Phi}^{a\dagger}_{I}(p)$,l.a=180}{i1}
  \fmfv{l=$\tilde{\Phi}^{J}_{b}(q)$,l.a=180}{i2}
  \fmfv{l=$\tilde{V}_{c}(-p-q)$ \hspace{0.5cm} 
          $\longrightarrow ~~~ \tilde{B}(p;q)$,l.a=0}{o1}
  \fmfdot{v1}
  \fmffreeze
  \fmfipath{p[]}
  \fmfiset{p1}{vpath1(__i1,__v1)}
  \fmfiset{p2}{vpath2(__i2,__v1)}
  \fmfiset{p3}{vpath3(__v1,__i1)}
  \fmfiset{p4}{vpath4(__v1,__i2)}
  \fmfi{fermion}{subpath (0,length(p1)/2) of p1}
  \fmfi{fermion}{subpath (0,length(p3)/2) of p3}
  \fmfi{fermion}{subpath (length(p2)/2,length(p2)) of p2}
  \fmfi{fermion}{subpath (length(p4)/2,length(p4)) of p4}
  \fmfi{fermion}{point length(p2)/2 of p2 -- point length(p1)/2 of p1}
  \fmfiv{d.sh=circle,d.fill=full,d.si=1.5mm}{point length(p1)/2 of p1}      
  \fmfiv{d.sh=circle,d.fill=full,d.si=1.5mm}{point length(p2)/2 of p2}
 \end{fmfgraph*}   
\end{fmffile} 

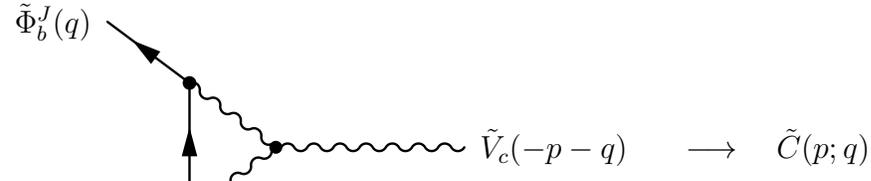
\vspace{1cm}
\noindent
\hspace*{0.7cm}
\begin{fmffile}{phi+vphiC}
 \begin{fmfgraph*}(150,95) 
  \fmfleft{i1,i2} \fmfright{o1}
  \fmf{phantom,tension=0.7,tag=1}{i1,v1}
  \fmf{phantom,tension=0.7,tag=2}{i2,v1}
  \fmf{phantom,tension=0.7,tag=3}{v1,i1}
  \fmf{phantom,tension=0.7,tag=4}{v1,i2} 
  \fmf{boson,tension=2.5}{v1,o1}
  \fmfv{l=$\tilde{\Phi}^{a\dagger}_{I}(p)$,l.a=180}{i1}
  \fmfv{l=$\tilde{\Phi}^{J}_{b}(q)$,l.a=180}{i2}
  \fmfv{l=$\tilde{V}_{c}(-p-q)$ \hspace{0.5cm} 
          $\longrightarrow ~~~ \tilde{C}(p;q)$,l.a=0}{o1}
  \fmfdot{v1}
  \fmffreeze
  \fmfipath{p[]}
  \fmfiset{p1}{vpath1(__i1,__v1)}
  \fmfiset{p2}{vpath2(__i2,__v1)}
  \fmfiset{p3}{vpath3(__v1,__i1)}
  \fmfiset{p4}{vpath4(__v1,__i2)}
  \fmfi{fermion}{subpath (0,length(p1)/2) of p1}
  \fmfi{boson}{subpath (0,length(p3)/2) of p3}
  \fmfi{boson}{subpath (0,length(p4)/2) of p4}
  \fmfi{fermion}{point length(p1)/2 of p1 -- point length(p2)/2 of p2}
  \fmfi{fermion}{subpath (length(p4)/2,length(p4)) of p4}
  \fmfiv{d.sh=circle,d.fill=full,d.si=1.5mm}{point length(p1)/2 of p1}      
  \fmfiv{d.sh=circle,d.fill=full,d.si=1.5mm}{point length(p2)/2 of p2}
 \end{fmfgraph*}   
\end{fmffile} 

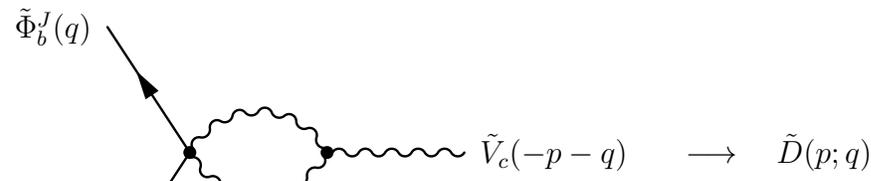
\vspace{1.5cm}
\noindent
\hspace*{0.7cm}
\begin{fmffile}{phi+vphiD}
 \begin{fmfgraph*}(150,95) 
  \fmfleft{i1,i2} \fmfright{o1}
  \fmf{fermion}{i1,v1}
  \fmf{fermion}{v1,i2}
  \fmf{boson,left=0.6,tension=0.6}{v1,v2}
  \fmf{boson,right=0.6,tension=0.6}{v1,v2}
  \fmf{boson,tension=1.2}{v2,o1}
  \fmfv{l=$\tilde{\Phi}^{a\dagger}_{I}(p)$,l.a=180}{i1}
  \fmfv{l=$\tilde{\Phi}^{J}_{b}(q)$,l.a=180}{i2}
  \fmfv{l=$\tilde{V}_{c}(-p-q)$ \hspace{0.5cm} 
          $\longrightarrow ~~~ \tilde{D}(p;q)$,l.a=0}{o1}	  
  \fmfdot{v1,v2}
 \end{fmfgraph*}   
\end{fmffile} 

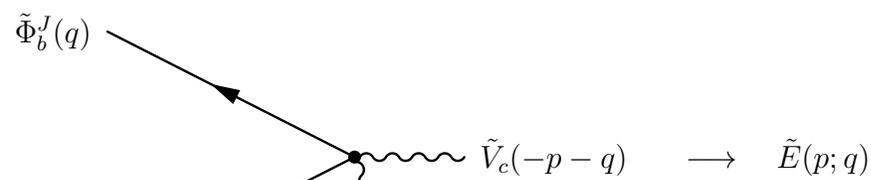
\vspace{1.5cm}
\noindent
\hspace*{0.7cm}
\begin{fmffile}{phi+vphiE}
 \begin{fmfgraph*}(150,95) 
  \fmfleft{i1,i2} \fmfright{o1}
  \fmf{phantom}{i1,v1,v2,v3}
  \fmf{phantom}{v3,v4,v5,i2}
  \fmf{boson,tension=1.5}{v3,o1}
  \fmffreeze
  \fmf{fermion}{i1,v1}
  \fmf{fermion}{v1,v3}
  \fmf{fermion}{v3,i2}
  \fmf{boson,right,tension=0}{v1,v3} 
  \fmfv{l=$\tilde{\Phi}^{a\dagger}_{I}(p)$,l.a=180}{i1}
  \fmfv{l=$\tilde{\Phi}^{J}_{b}(q)$,l.a=180}{i2}
  \fmfv{l=$\tilde{V}_{c}(-p-q)$ \hspace{0.5cm} 
          $\longrightarrow ~~~ \tilde{E}(p;q)$,l.a=0}{o1}
  \fmfdot{v1,v3}
 \end{fmfgraph*}   
\end{fmffile} 

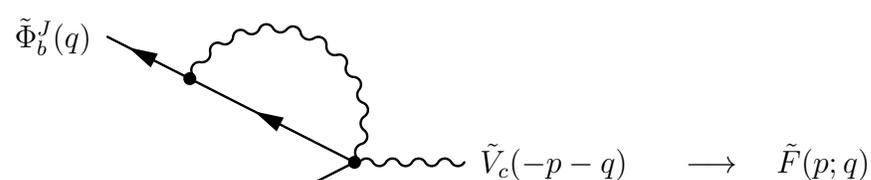
\vspace{1.5cm}
\noindent
\hspace*{0.7cm}
\begin{fmffile}{phi+vphiF}
 \begin{fmfgraph*}(150,95) 
  \fmfleft{i1,i2} \fmfright{o1}
  \fmf{phantom}{i1,v1,v2,v3}
  \fmf{phantom}{v3,v4,v5,i2}
  \fmf{boson,tension=1.5}{v3,o1}
  \fmffreeze
  \fmf{fermion}{i1,v3}
  \fmf{fermion}{v3,v5}
  \fmf{fermion}{v5,i2}
  \fmf{boson,right,tension=0}{v3,v5} 
  \fmfv{l=$\tilde{\Phi}^{a\dagger}_{I}(p)$,l.a=180}{i1}
  \fmfv{l=$\tilde{\Phi}^{J}_{b}(q)$,l.a=180}{i2}
  \fmfv{l=$\tilde{V}_{c}(-p-q)$ \hspace{0.5cm} 
          $\longrightarrow ~~~ \tilde{F}(p;q)$,l.a=0}{o1}
  \fmfdot{v3,v5}
 \end{fmfgraph*}   
\end{fmffile}

\vspace{1cm}
\noindent
In the presence of mass terms for the chiral and antichiral 
superfields there are two additional contributions

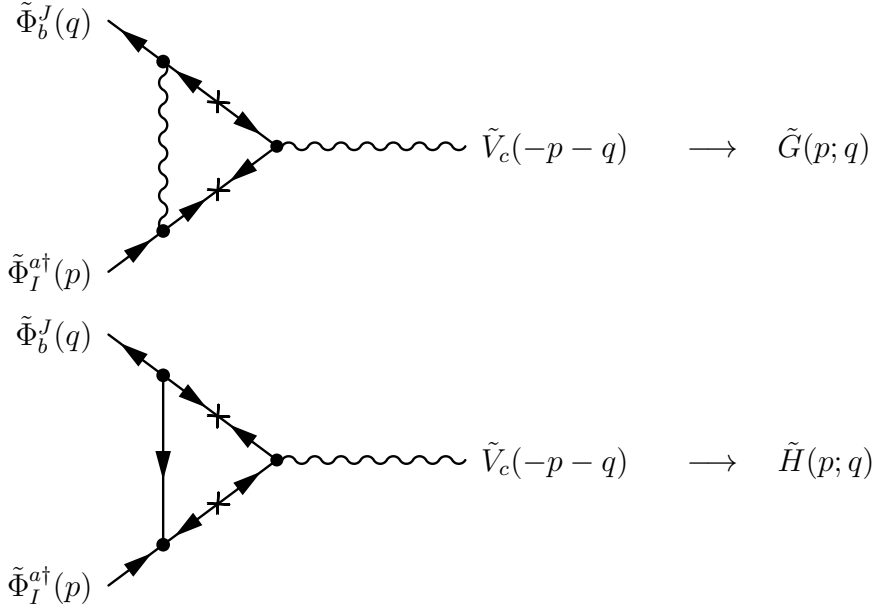
\vspace*{0.8cm}
\noindent
\hspace*{0.7cm}
\begin{fmffile}{phi+vphiG}
 \begin{fmfgraph*}(150,95) 
  \fmfleft{i1,i2} \fmfright{o1}
  \fmf{phantom,tension=0.7,tag=1}{i1,v1}
  \fmf{phantom,tension=0.7,tag=2}{i2,v1}
  \fmf{phantom,tension=0.7,tag=3}{v1,i1}
  \fmf{phantom,tension=0.7,tag=4}{v1,i2} 
  \fmf{boson,tension=2.5}{v1,o1}
  \fmfv{l=$\tilde{\Phi}^{a\dagger}_{I}(p)$,l.a=180}{i1}
  \fmfv{l=$\tilde{\Phi}^{J}_{b}(q)$,l.a=180}{i2}
  \fmfv{l=$\tilde{V}_{c}(-p-q)$ \hspace{0.5cm} 
          $\longrightarrow ~~~ \tilde{G}(p;q)$,l.a=0}{o1}
  \fmfdot{v1}
  \fmffreeze
  \fmfipath{p[]}
  \fmfiset{p1}{vpath1(__i1,__v1)}
  \fmfiset{p2}{vpath2(__i2,__v1)}
  \fmfiset{p3}{vpath3(__v1,__i1)}
  \fmfiset{p4}{vpath4(__v1,__i2)}
  \fmfi{fermion}{subpath (0,length(p1)/3) of p1}
  \fmfi{fermion}{subpath (length(p1)/3,2length(p1)/3) of p1}
  \fmfi{fermion}{subpath (0,length(p3)/3) of p3}
  \fmfi{fermion}{subpath (2length(p2)/3,length(p2)) of p2}
  \fmfi{fermion}{subpath (length(p4)/3,2length(p4)/3) of p4}
  \fmfi{fermion}{subpath (2length(p4)/3,length(p4)) of p4}
  \fmfi{boson}{point length(p1)/3 of p1 -- point
               length(p2)/3 of p2}
  \fmfiv{d.sh=circle,d.fill=full,d.si=1.5mm}{point length(p1)/3 of p1}      
  \fmfiv{d.sh=circle,d.fill=full,d.si=1.5mm}{point length(p2)/3 of p2}
  \fmfiv{d.sh=cross,d.si=3mm,d.a=42}{point 2length(p1)/3 of p1}      
  \fmfiv{d.sh=cross,d.si=3mm,d.a=42}{point 2length(p2)/3 of p2}
 \end{fmfgraph*}   
\end{fmffile} 

\vspace*{0.8cm}
\noindent
\hspace*{0.7cm}
\begin{fmffile}{phi+vphiH}
 \begin{fmfgraph*}(150,95) 
  \fmfleft{i1,i2} \fmfright{o1}
  \fmf{phantom,tension=0.7,tag=1}{i1,v1}
  \fmf{phantom,tension=0.7,tag=2}{i2,v1}
  \fmf{phantom,tension=0.7,tag=3}{v1,i1}
  \fmf{phantom,tension=0.7,tag=4}{v1,i2} 
  \fmf{boson,tension=2.5}{v1,o1}
  \fmfv{l=$\tilde{\Phi}^{a\dagger}_{I}(p)$,l.a=180}{i1}
  \fmfv{l=$\tilde{\Phi}^{J}_{b}(q)$,l.a=180}{i2}
  \fmfv{l=$\tilde{V}_{c}(-p-q)$ \hspace{0.5cm} 
          $\longrightarrow ~~~ \tilde{H}(p;q)$,l.a=0}{o1}
  \fmfdot{v1}
  \fmffreeze
  \fmfipath{p[]}
  \fmfiset{p1}{vpath1(__i1,__v1)}
  \fmfiset{p2}{vpath2(__i2,__v1)}
  \fmfiset{p3}{vpath3(__v1,__i1)}
  \fmfiset{p4}{vpath4(__v1,__i2)}
  \fmfi{fermion}{subpath (0,length(p1)/3) of p1}
  \fmfi{fermion}{subpath (length(p3)/3,2length(p3)/3) of p3}
  \fmfi{fermion}{subpath (2length(p1)/3,length(p1)) of p1}
  \fmfi{fermion}{subpath (0,length(p4)/3) of p4}
  \fmfi{fermion}{subpath (length(p2)/3,2length(p2)/3) of p2}
  \fmfi{fermion}{subpath (2length(p4)/3,length(p4)) of p4}
  \fmfi{fermion}{point length(p2)/3 of p2 -- point
               length(p1)/3 of p1}
  \fmfiv{d.sh=circle,d.fill=full,d.si=1.5mm}{point length(p1)/3 of p1}      
  \fmfiv{d.sh=circle,d.fill=full,d.si=1.5mm}{point length(p2)/3 of p2}
  \fmfiv{d.sh=cross,d.si=3mm,d.a=42}{point 2length(p1)/3 of p1}      
  \fmfiv{d.sh=cross,d.si=3mm,d.a=42}{point 2length(p2)/3 of p2}
 \end{fmfgraph*}   
\end{fmffile}

\vspace{0.5cm}
\noindent
The calculation of these diagrams is completely analogous to those 
that were presented in the previous section. The last two 
diagrams are finite and vanish in the ${\cal N}$=4 theory, \ie at  
$m$=0, as they are proportional to $m^{2}$; they will not be considered 
here. The other contributions will be briefly studied in the limit 
$m$=0.

The diagram $\tilde{D}(p,q)$ is zero as a consequence of the 
contraction among the colour indices, so that there are five 
contributions to be calculated. Dimensional analysis gives a 
vanishing superficial degree of divergence $d$, corresponding to a 
logarithmic divergence, for the Green function under consideration. 
Single diagrams actually contain a logarithmically divergent term 
plus finite terms. A straightforward but rather lengthy computation, 
based on elementary $D$-algebra, allows to prove that the divergent 
part of all the diagrams is of the form
\begin{displaymath}
	I_{{\rm log}} = c \, \delta^{I}_{J} f^{abc}
	\int \frac{d^{4}k}{(2\pi)^{4}}d^{2}\theta d^{2}{\overline\theta} 
	\, \frac{1}{k^{2}(p-k)^{2}} \left\{ 
	\Phi^{\dagger}_{aI}(p,\theta,{\overline \theta}) V_{b}
	(-p-q,\theta,{\overline \theta}) \Phi^{J}_{c}(q,\theta,
	{\overline \theta}) \right\} \, ,
\end{displaymath}
where $c$ is a constant. Moreover one finds a finite contribution of 
the form
\begin{eqnarray*}
	I_{{\rm finite}} &=& \delta^{I}_{J} f^{abc}
	\int \frac{d^{4}k}{(2\pi)^{4}}d^{2}\theta d^{2}{\overline\theta} 
	\, \frac{\Phi^{\dagger}_{aI}
	(p,\theta,{\overline \theta})}{k^{2}(q+k)^{2}(p-k)^{2}}  
	\left\{ c_{1} \left[ D^{\alpha}{\overline D}^{2}D_{\alpha} 
	V_{b}(-p-q,\theta,{\overline \theta}) \right] \right. \cdot \\ 
	&& \hspace{1cm} \left. \cdot 
        \Phi^{J}_{c}(q,\theta,{\overline \theta})  +
	c_{2}\sigma^{\mu}_{\alpha{\dot\alpha}}(q+k)_{\mu} \left[
	D^{\alpha}{\overline D}^{\dot\alpha}V_{b}
	(-p-q,\theta,{\overline \theta}) \right] 
	\Phi^{J}_{c}(q,\theta,{\overline \theta}) \right\} \, ,
\end{eqnarray*}
where $c_{1}$ and $c_{2}$ are numerical constants.
The sum of all the logarithmically divergent terms contained in the 
diagrams $\tilde{A}$, $\tilde{B}$, $\tilde{C}$, $\tilde{E}$ 
and $\tilde{F}$ vanishes. The residual finite part can be shown to be 
zero as well. 

In conclusion the one-loop correction to the three-point function 
$\langle \Phi^{\dagger} V \Phi \rangle$ exactly vanishes in the 
Fermi--Feynman gauge. The same result can be shown to hold for the 
other three-point functions.

The superfield formalism does not lead to significant simplifications 
in the calculation of three-point functions with respect to the same 
computation in components (in the Wess--Zumino gauge!). Actually for 
the Green function considered here the number of diagrams to be evaluated 
is approximately the same as in the component formulation. However the 
power of superspace techniques becomes clear in the computation of 
four-point functions that will be considered in next subsection.

\subsection{Four-point functions}

The computation of four-point functions in components is extremely 
complicated even at the one-loop level and in the Wess--Zumino gauge. 
In this case the choice of the WZ gauge should not lead to extra 
divergences because the complete four-point function must finally give 
the physical scattering amplitude. In the ${\cal N}$=1 superfield 
formulation the calculation of four-point functions, though rather 
lengthy, is much more simple. 

In this section the computation of the one-particle irreducible 
one-loop correction to the Green function 
$\langle\Phi^{\dagger}\Phi\Phi^{\dagger}\Phi\rangle$ in the 
Fermi--Feynman gauge will be presented. There are several diagrams to 
be considered: each of them is free of ultraviolet divergences, as 
immediately follows from dimensional analysis. Moreover in the 
Fermi--Feynman gauge each single diagram is infrared finite. In 
conclusion one finds a finite and non-vanishing result.

The first subset of contributions corresponds to the diagram

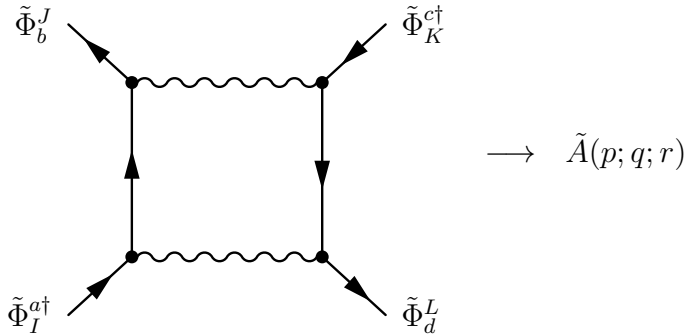
\vspace{1cm}
\noindent
\begin{fmffile}{phi+phiphi+phiA}
 \begin{fmfgraph*}(150,110) 
  \fmfleft{i1,i2} 
  \fmfright{o1,o2}
  \fmf{fermion,tension=0.6}{i1,v1}
  \fmf{fermion,tension=0.6}{v2,i2}
  \fmf{fermion,tension=0.6}{v3,o1}
  \fmf{fermion,tension=0.6}{o2,v4}
  \fmf{phantom,tension=0.1,tag=1}{v1,v2}
  \fmf{phantom,tension=0.1,tag=2}{v2,v1}
  \fmf{phantom,tension=0.1,tag=3}{v2,v4}
  \fmf{phantom,tension=0.1,tag=4}{v4,v2}
  \fmf{phantom,tension=0.1,tag=5}{v4,v3}
  \fmf{phantom,tension=0.1,tag=6}{v3,v4}
  \fmf{phantom,tension=0.1,tag=7}{v3,v1}
  \fmf{phantom,tension=0.1,tag=8}{v1,v3}
  \fmfv{l=$\tilde{\Phi}^{a\dagger}_{I}$,l.a=180}{i1}
  \fmfv{l=$\tilde{\Phi}^{J}_{b}$,l.a=180}{i2}
  \fmfv{l=$\tilde{\Phi}^{L}_{d}$,l.a=0}{o1}
  \fmfv{l=$\tilde{\Phi}^{c\dagger}_{K}$,l.a=0}{o2}
  \fmfdot{v1,v2,v3,v4}
  \fmffreeze
  \fmfipath{p[]}
  \fmfiset{p1}{vpath1(__v1,__v2)}
  \fmfiset{p2}{vpath2(__v2,__v1)}
  \fmfiset{p3}{vpath3(__v2,__v4)}
  \fmfiset{p4}{vpath4(__v4,__v2)}
  \fmfiset{p5}{vpath5(__v4,__v3)}
  \fmfiset{p6}{vpath6(__v3,__v4)}
  \fmfiset{p7}{vpath7(__v3,__v1)}
  \fmfiset{p8}{vpath8(__v1,__v3)}
  \fmfi{fermion}{subpath (0,length(p1)) of p1}
  \fmfi{boson}{subpath (0,length(p7)) of p7}
  \fmfi{boson}{subpath (0,length(p3)) of p3}
  \fmfi{fermion}{subpath (0,length(p5)) of p5}
  \fmfiv{l=\hspace{7cm}$\longrightarrow ~~
            \tilde{A}(p;q;r)$,l.a=90}
            {point 3length(p5)/5 of p5}
 \end{fmfgraph*}   
\end{fmffile} 

\vspace{0.5cm}
\noindent
together with those obtained from this one by crossing, for a total 
of four terms of this kind. Steps completely analogous to those 
entering the evaluation of the propagators allow one to find the 
following results.
\begin{displaymath}
	\tilde{A}_{1}(p,q,r) = \left(\frac{1}{4}\right)^{2}
	\frac{g^{4}}{4} \,\kappa (\delta_{a}^{b}\delta^{d}_{c} + 
	\delta_{ac}\delta^{bd})\delta^{I}_{J}\delta^{K}_{L} 
	I_{1\:bdIK}^{(A)acJL}(p,q,r)
\end{displaymath}
where
\begin{eqnarray}
	I_{1\:bdIK}^{(A)acJL}(p,q,r) & = & \int \frac{d^{4}k}{(2\pi)^{4}}
	d^{2}\theta d^{2}{\overline\theta}
	\frac{1}{k^{2}(p-k)^{2}(q+k)^{2}(q+k-r)^{2}} \cdot  
	\nonumber \\
	&& \hspace{-2.5cm} \cdot \left\{ \left[(D^{2}{\overline D}^{2}) 
	\Phi^{a\dagger}_{I}(p,\theta,{\overline \theta}) \right]
	\Phi^{c\dagger}_{K}(p+q-r,\theta,{\overline \theta})
	\Phi_{d}^{L}(r,\theta,{\overline\theta}) \Phi_{b}^{J}(q,\theta,
	{\overline\theta}) + \Phi^{a\dagger}_{I}(p,\theta,{\overline \theta})
	\cdot \right. \nonumber \\
	&& \hspace{-2.5cm} \left. \cdot\left[(D^{2}{\overline D}^{2}) 
	\Phi^{c\dagger}_{K}(p+q-r,\theta,{\overline \theta})\right] 
	\Phi_{d}^{L}(r,\theta,{\overline\theta}) \Phi_{b}^{J}
	(q,\theta,{\overline\theta}) 
	+ 2\left[(D^{2}{\overline D}_{\dot\alpha})\Phi^{a\dagger}_{I}(p,
	\theta,{\overline \theta})\right] \cdot \right. \nonumber \\ 
	&&\hspace{-2.5cm} \left.\cdot\left[{\overline D}^{\dot\alpha}
	\Phi^{c\dagger}(p+q-r,\theta,{\overline \theta})\right] 
	\Phi_{d}^{L}(r,\theta,{\overline\theta}) \Phi_{b}^{J}(q,\theta,
	{\overline\theta}) + 2\left[{\overline D}_{\dot\alpha}
	\Phi^{a\dagger}_{I}(p,\theta,{\overline \theta})\right]
	\cdot \right. \nonumber \\
	&& \hspace{-2.5cm} \left. \cdot 
	\left[(D^{2}{\overline D}^{\dot\alpha})\Phi^{c\dagger}
	(p+q-r,\theta,{\overline \theta})\right] \Phi_{d}^{L}(r,\theta,
	{\overline\theta}) \Phi_{b}^{J}(q,\theta,{\overline\theta}) + 
	4\left[(D^{\alpha}{\overline D}_{\dot\alpha})\Phi^{a\dagger}_{I}
	(p,\theta,{\overline \theta})\right] \cdot \right. \nonumber \\
	&& \hspace{-2.5cm} \left. \cdot \left[D_{\alpha}
	{\overline D}^{\dot\alpha}\Phi^{c\dagger}(p+q-r,\theta,{\overline 
	\theta})\right] \Phi_{d}^{L}(r,\theta,
	{\overline\theta}) \Phi_{b}^{J}(q,\theta,{\overline\theta}) 
	\right\} \equiv \nonumber \\
	&& \hspace{-2.5cm}\equiv \int \frac{d^{4}k}{(2\pi)^{4}}
	d^{2}\theta d^{2}{\overline\theta}
	\frac{1}{k^{2}(p-k)^{2}(q+k)^{2}(q+k-r)^{2}} K_{bdIK}^{acJL}
	(p,q,r;\theta;{\overline\theta}) 
	\label{4pointa1}
\end{eqnarray}
and the constant $\kappa$ is a group theory factor defined by
\begin{displaymath}
	f_{aef}f^{bfg}f_{cgh}f^{dhe} = \kappa 
	(\delta_{a}^{b}\delta_{c}^{d} + \delta_{a}^{d}\delta_{c}^{b}) \, ;
\end{displaymath}
\vspace{1cm}
\begin{displaymath}
	\tilde{A}_{2}(p,q,r) = \left(\frac{1}{4}\right)^{2}
	\frac{g^{4}}{4}\, \kappa (\delta_{a}^{d}\delta^{b}_{c} + 
	\delta_{ac}\delta^{bd})\delta^{I}_{L}\delta^{K}_{J} 
	I_{2\:bdIK}^{(A)acJL}(p,q,r) \, ,
\end{displaymath}
where
\begin{eqnarray}
	I_{2\:bdIK}^{(A)acJL}(p,q,r) &=& \int \frac{d^{4}k}{(2\pi)^{4}}
	d^{2}\theta d^{2}{\overline\theta}
	\frac{1}{k^{2}(p-k)^{2}(p-k-r)^{2}(q+k)^{2}(p+q-k-r)^{2}} 
	\cdot \nonumber \\
	&& \hspace{2cm} \rule{0pt}{20pt}
	\cdot K_{bdIK}^{acJL} (p,q,r;\theta;{\overline\theta}) \, ;
	\label{4pointa2}
\end{eqnarray}
\vspace{1cm}
\begin{displaymath}
	\tilde{A}_{3}(p,q,r) = \left(\frac{1}{4}\right)^{2}
	\frac{g^{4}}{4}\, \kappa (\delta_{a}^{b}\delta^{d}_{c} + 
	\delta_{a}^{d}\delta^{b}_{c})\delta^{I}_{J}\delta^{K}_{L} 
	I_{3\:bdIK}^{(A)acJL}(p,q,r) \, ,
\end{displaymath}
where
\begin{eqnarray}
	I_{3\:bdIK}^{(A)acJL}(p,q,r) & = & \int \frac{d^{4}k}{(2\pi)^{4}}
	d^{2}\theta d^{2}{\overline\theta} 
	\frac{1}{k^{2}(p-k)^{2}(q+k)^{2}(q+k-r)^{2}} \cdot  
	\nonumber \\ 
	&& \hspace{-3cm} \cdot \left\{ \left[{\overline 
	D}^{2}\Phi^{a\dagger}_{I}(p,\theta,{\overline \theta})\right] \left[ 
	D^{2} \Phi^{J}_{b}(q,\theta,{\overline \theta})\right] 
	\Phi^{c\dagger}_{K}(p+q-r,\theta,{\overline \theta}) \Phi^{L}_{d}
	(r,\theta,{\overline \theta}) + 
	\right.  \label{4pointa3}\\ 
	&& \hspace{-3cm} \left. + 8i\sigma^{\mu}_{\alpha{\dot\alpha}}
	(p-k-r)_{\mu} \left[ {\overline D}^{\dot\alpha}\Phi^{a\dagger}_{I}
	(p,\theta,{\overline \theta})\right] \left[ 
	D^{\alpha} \Phi^{J}_{b}(q,\theta,{\overline \theta})\right] 
	\Phi^{c\dagger}_{K}(p+q-r,\theta,{\overline \theta}) 
	\cdot \right. \nonumber \\
	&& \hspace{-3cm}\cdot\left.\Phi^{L}_{d}(r,\theta,{\overline\theta}) 
	- 16 (p-k-r)^{2} \Phi^{a\dagger}_{I}
	(p,\theta,{\overline \theta})\Phi^{J}_{b}(q,\theta,{\overline\theta})
	\Phi^{c\dagger}_{K}(p+q-r,\theta,{\overline \theta}) \Phi^{L}_{d}
	(r,\theta,{\overline \theta}) \right\} \, ;
	\nonumber
\end{eqnarray}
\vspace{1cm}
\begin{displaymath}
	\tilde{A}_{4}(p,q,r) = \left(\frac{1}{4}\right)^{2}
	\frac{g^{4}}{4}\, \kappa (\delta_{a}^{b}\delta^{d}_{c} + 
	\delta_{a}^{d}\delta^{b}_{c})\delta^{I}_{J}\delta^{K}_{L} 
	I_{4\:bdIK}^{(A)acJL}(p,q,r) \, ,
\end{displaymath}
where
\begin{eqnarray}
	I_{4\:bdIK}^{(A)acJL}(p,q,r) & = & \int \frac{d^{4}k}{(2\pi)^{4}}
	d^{2}\theta d^{2}{\overline\theta}
	\frac{1}{k^{2}(p-k)^{2}(p-k-r)^{2}(q+k)^{2}} \cdot \nonumber \\ 
	&& \hspace{-3cm} \cdot \left\{ \left[{\overline 
	D}^{2}\Phi^{a\dagger}_{I}(p,\theta,{\overline \theta})\right] 
	\Phi^{J}_{b}(q,\theta,{\overline \theta}) 
	\Phi^{c\dagger}_{K}(p+q-r,\theta,{\overline \theta}) 
	\left[ D^{2} \Phi^{L}_{d}(r,\theta,{\overline \theta}) \right]+ 
	\right. \nonumber \\ 
	&& \hspace{-3cm} \left. + 8i\sigma^{\mu}_{\alpha{\dot\alpha}}
	(q+k)_{\mu} \left[ {\overline D}^{\dot\alpha}\Phi^{a\dagger}_{I}
	(p,\theta,{\overline \theta})\right] \Phi^{J}_{b}(q,\theta,
	{\overline \theta}) \Phi^{c\dagger}_{K}
	(p+q-r,\theta,{\overline \theta}) \left[ 
	D^{\alpha} \Phi^{L}_{d}(r,\theta,{\overline\theta}) \right] -
	\right. \nonumber \\
	&& \hspace{-3cm} \left.
	- 16 (q+k)^{2} \Phi^{a\dagger}_{I}
	(p,\theta,{\overline \theta})\Phi^{J}_{b}(q,\theta,{\overline\theta})
	\Phi^{c\dagger}_{K}(p+q-r,\theta,{\overline \theta}) \Phi^{L}_{d}
	(r,\theta,{\overline \theta}) \right\} \, .
	\label{4pointa4}
\end{eqnarray}

The second kind of contributions correspond to the diagram

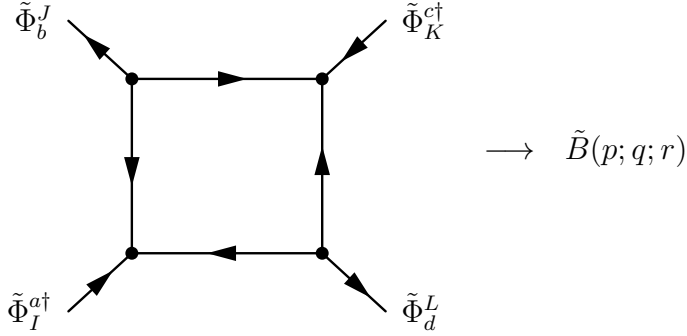
\vspace{1cm}
\noindent
\begin{fmffile}{phi+phiphi+phiB}
 \begin{fmfgraph*}(150,110) 
  \fmfleft{i1,i2} 
  \fmfright{o1,o2}
  \fmf{fermion,tension=0.6}{i1,v1}
  \fmf{fermion,tension=0.6}{v2,i2}
  \fmf{fermion,tension=0.6}{v3,o1}
  \fmf{fermion,tension=0.6}{o2,v4}
  \fmf{phantom,tension=0.1,tag=1}{v1,v2}
  \fmf{phantom,tension=0.1,tag=2}{v2,v1}
  \fmf{phantom,tension=0.1,tag=3}{v2,v4}
  \fmf{phantom,tension=0.1,tag=4}{v4,v2}
  \fmf{phantom,tension=0.1,tag=5}{v4,v3}
  \fmf{phantom,tension=0.1,tag=6}{v3,v4}
  \fmf{phantom,tension=0.1,tag=7}{v3,v1}
  \fmf{phantom,tension=0.1,tag=8}{v1,v3}
  \fmfv{l=$\tilde{\Phi}^{a\dagger}_{I}$,l.a=180}{i1}
  \fmfv{l=$\tilde{\Phi}^{J}_{b}$,l.a=180}{i2}
  \fmfv{l=$\tilde{\Phi}^{L}_{d}$,l.a=0}{o1}
  \fmfv{l=$\tilde{\Phi}^{c\dagger}_{K}$,l.a=0}{o2}
  \fmfdot{v1,v2,v3,v4}
  \fmffreeze
  \fmfipath{p[]}
  \fmfiset{p1}{vpath1(__v1,__v2)}
  \fmfiset{p2}{vpath2(__v2,__v1)}
  \fmfiset{p3}{vpath3(__v2,__v4)}
  \fmfiset{p4}{vpath4(__v4,__v2)}
  \fmfiset{p5}{vpath5(__v4,__v3)}
  \fmfiset{p6}{vpath6(__v3,__v4)}
  \fmfiset{p7}{vpath7(__v3,__v1)}
  \fmfiset{p8}{vpath8(__v1,__v3)}
  \fmfi{fermion}{subpath (0,length(p2)) of p2}
  \fmfi{fermion}{subpath (0,length(p7)) of p7}
  \fmfi{fermion}{subpath (0,length(p3)) of p3}
  \fmfi{fermion}{subpath (0,length(p6)) of p6}
  \fmfiv{l=\hspace{7cm}$\longrightarrow ~~
            \tilde{B}(p;q;r)$,l.a=90}
            {point 3length(p5)/5 of p5}
 \end{fmfgraph*}   
\end{fmffile} 

\vspace{0.5cm}
\noindent
In this case the diagrams obtained by crossing are identical to the 
one depicted here, so that they are accounted for by giving the correct 
weight to $\tilde{B}(p,q,r)$. One finds
\begin{displaymath}
	\tilde{B}(p,q,r) = \frac{g^{4}}{24}
	\, \kappa (\delta_{a}^{b}\delta^{d}_{c} + 
	\delta_{a}^{d}\delta^{b}_{c})(\delta^{I}_{J}\delta^{K}_{L}
	+\delta^{I}_{L}\delta^{K}_{J}) I_{bdIK}^{(B)acJL}(p,q,r) \, ,
\end{displaymath}
where
\begin{eqnarray}
	I_{bdIK}^{(B)acJL}(p,q,r) & = & \int \frac{d^{4}k}{(2\pi)^{4}}
	d^{2}\theta d^{2}{\overline\theta} 
	\frac{1}{k^{2}(p-k)^{2}(p+q-r)^{2}(q+k)^{2}} \cdot  
	\nonumber \\ 
	&& \hspace{-3cm} \cdot \left\{ \left[{\overline 
	D}^{2}\Phi^{a\dagger}_{I}(p,\theta,{\overline \theta})\right] 
	\Phi^{J}_{b}(q,\theta,{\overline \theta}) 
	\Phi^{c\dagger}_{K}(p+q-r,\theta,{\overline \theta}) 
	\left[ D^{2} \Phi^{L}_{d}(r,\theta,{\overline \theta}) \right]+ 
	\right. \nonumber \\ 
	&& \hspace{-3cm} \left. + 8i\sigma^{\mu}_{\alpha{\dot\alpha}}
	(p-k)_{\mu} \left[ {\overline D}^{\dot\alpha}\Phi^{a\dagger}_{I}
	(p,\theta,{\overline \theta})\right] \Phi^{J}_{b}(q,\theta,
	{\overline \theta}) \Phi^{c\dagger}_{K}
	(p+q-r,\theta,{\overline \theta}) \left[ 
	D^{\alpha} \Phi^{L}_{d}(r,\theta,{\overline\theta}) \right] - 
	\right. \nonumber \\
	&& \hspace{-3cm} \left.
	- 16 (p-k)^{2} \Phi^{a\dagger}_{I} (p,\theta,{\overline \theta})
	\Phi^{J}_{b}(q,\theta,{\overline\theta})
	\Phi^{c\dagger}_{K}(p+q-r,\theta,{\overline \theta}) \Phi^{L}_{d}
	(r,\theta,{\overline \theta}) \right\} \, .
	\label{4pointb}
\end{eqnarray}

The next one-loop correction is 

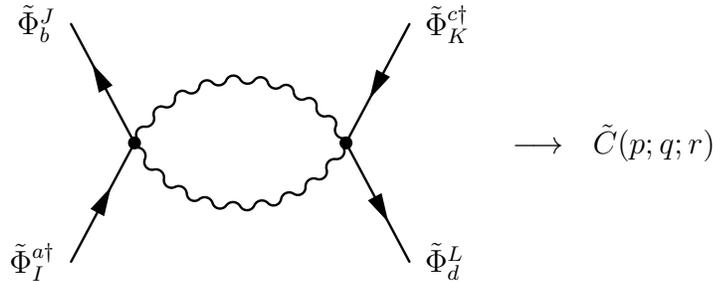
\vspace{1cm}
\noindent
\begin{fmffile}{phi+phiphi+phiC}
 \begin{fmfgraph*}(160,90) 
  \fmfleft{i1,i2} 
  \fmfright{o1,o2}
  \fmf{fermion,tension=1}{i1,v1}
  \fmf{fermion,tension=1}{v1,i2}
  \fmf{boson,left=0.6,tension=0.3}{v1,v2}
  \fmf{boson,right=0.6,tension=0.3}{v1,v2}
  \fmf{fermion,tension=1}{v2,o1}
  \fmf{fermion,tension=1}{o2,v2}
  \fmfv{l=$\tilde{\Phi}^{a\dagger}_{I}$,l.a=180}{i1}
  \fmfv{l=$\tilde{\Phi}^{J}_{b}$,l.a=180}{i2}
  \fmfv{l=$\tilde{\Phi}^{L}_{d}$,l.a=0}{o1}
  \fmfv{l=$\tilde{\Phi}^{c\dagger}_{K}$,l.a=0}{o2}
  \fmfv{l=\hspace{2cm}$\longrightarrow ~~
            \tilde{C}(p;q;r)$,l.a=0}{v2}
  \fmfdot{v1,v2}
 \end{fmfgraph*}   
\end{fmffile} 

\vspace{0.5cm}
\noindent
This contribution is trivially zero because it contains the product 
\begin{displaymath}
	\delta_{4}(\theta_{1}-\theta_{2})
	\delta_{4}(\theta_{1}-\theta_{2}) \equiv 0 \, .
\end{displaymath}
This would not be true in a gauge different from the 
Fermi--Feynman gauge, \ie with $\alpha \neq 1$, because in that case 
there would be projectors acting on the $\delta$'s. The vanishing 
of $\tilde{C}(p,q,r)$, which is completely trivial in the superfield 
formulation, corresponds, in the component formulation, to a 
complicated cancellation among various terms coming from graphs with 
the same topology. 

Another subset of diagrams includes

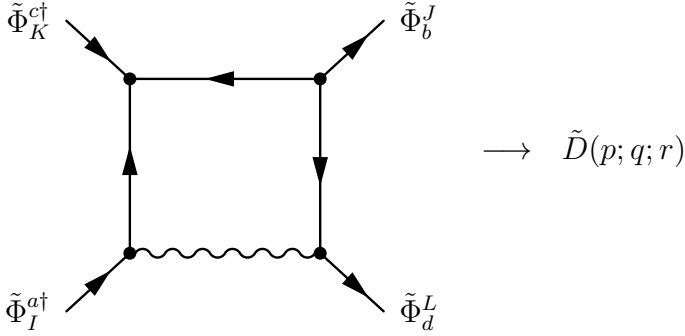
\vspace{1cm}
\noindent
\begin{fmffile}{phi+phiphi+phiD}
 \begin{fmfgraph*}(150,110) 
  \fmfleft{i1,i2} 
  \fmfright{o1,o2}
  \fmf{fermion,tension=0.6}{i1,v1}
  \fmf{fermion,tension=0.6}{i2,v2}
  \fmf{fermion,tension=0.6}{v3,o1}
  \fmf{fermion,tension=0.6}{v4,o2}
  \fmf{phantom,tension=0.1,tag=1}{v1,v2}
  \fmf{phantom,tension=0.1,tag=2}{v2,v1}
  \fmf{phantom,tension=0.1,tag=3}{v2,v4}
  \fmf{phantom,tension=0.1,tag=4}{v4,v2}
  \fmf{phantom,tension=0.1,tag=5}{v4,v3}
  \fmf{phantom,tension=0.1,tag=6}{v3,v4}
  \fmf{phantom,tension=0.1,tag=7}{v3,v1}
  \fmf{phantom,tension=0.1,tag=8}{v1,v3}
  \fmfv{l=$\tilde{\Phi}^{a\dagger}_{I}$,l.a=180}{i1}
  \fmfv{l=$\tilde{\Phi}^{J}_{b}$,l.a=0}{o2}
  \fmfv{l=$\tilde{\Phi}^{L}_{d}$,l.a=0}{o1}
  \fmfv{l=$\tilde{\Phi}^{c\dagger}_{K}$,l.a=180}{i2}
  \fmfdot{v1,v2,v3,v4}
  \fmffreeze
  \fmfipath{p[]}
  \fmfiset{p1}{vpath1(__v1,__v2)}
  \fmfiset{p2}{vpath2(__v2,__v1)}
  \fmfiset{p3}{vpath3(__v2,__v4)}
  \fmfiset{p4}{vpath4(__v4,__v2)}
  \fmfiset{p5}{vpath5(__v4,__v3)}
  \fmfiset{p6}{vpath6(__v3,__v4)}
  \fmfiset{p7}{vpath7(__v3,__v1)}
  \fmfiset{p8}{vpath8(__v1,__v3)}
  \fmfi{fermion}{subpath (0,length(p1)) of p1}
  \fmfi{fermion}{subpath (0,length(p4)) of p4}
  \fmfi{fermion}{subpath (0,length(p5)) of p5}
  \fmfi{boson}{subpath (0,length(p7)) of p7}
  \fmfiv{l=\hspace{7cm}$\longrightarrow ~~
            \tilde{D}(p;q;r)$,l.a=90}
            {point 3length(p5)/5 of p5}
 \end{fmfgraph*}   
\end{fmffile} 

\vspace{0.5cm}
\noindent
as well as the crossed ones. There are three inequivalent 
crossed diagrams. The result of the calculation consists of the 
following four terms ($\tilde{D}_{1}$, $\tilde{D}_{2}$, 
$\tilde{D}_{3}$, $\tilde{D}_{4}$)
\begin{displaymath}
	\tilde{D}_{1}(p,q,r) = \left(\frac{1}{4}\right)^{2}
	\frac{g^{4}}{2}\, \kappa (\delta_{ac}\delta^{bd} + 
	\delta_{a}^{d}\delta^{b}_{c})(\delta^{I}_{J}\delta^{K}_{L}
	-\delta^{I}_{L}\delta^{K}_{J}) I_{1\:bdIK}^{(D)acJL}(p,q,r) \, ,
\end{displaymath}
where
\begin{eqnarray}
	I_{1\:bdIK}^{(D)acJL}(p,q,r) & = & \int \frac{d^{4}k}{(2\pi)^{4}}
	d^{2}\theta d^{2}{\overline\theta} 
	\frac{1}{k^{2}(p-k)^{2}(k+r-p-q)^{2}(p-k-r)^{2}} \cdot  
	\nonumber \\ 
	&& \hspace{-3cm} \cdot \left\{ \left[{\overline 
	D}^{2}\Phi^{a\dagger}_{I}(p,\theta,{\overline \theta})\right] 
	\left[ D^{2} \Phi^{J}_{b}(q,\theta,{\overline \theta})\right] 
	\Phi^{c\dagger}_{K}(p+q-r,\theta,{\overline \theta}) 
	\Phi^{L}_{d}(r,\theta,{\overline \theta}) + 
	\right. \label{4pointd1} \\ 
	&& \hspace{-3cm} \left. + 8i\sigma^{\mu}_{\alpha{\dot\alpha}}
	(k+r-p)_{\mu} \left[ {\overline D}^{\dot\alpha}\Phi^{a\dagger}_{I}
	(p,\theta,{\overline \theta})\right] \left[ D^{\alpha} 
	\Phi^{J}_{b}(q,\theta,{\overline \theta}) \right] 
	\Phi^{c\dagger}_{K}(p+q-r,\theta,{\overline \theta}) 
	\cdot \right. \nonumber \\
	&& \hspace{-3cm}\cdot\left.
	\Phi^{L}_{d}(r,\theta,{\overline\theta}) 
	- 16 (k+r-p)^{2} \Phi^{a\dagger}_{I}
	(p,\theta,{\overline \theta})\Phi^{J}_{b}(q,\theta,{\overline\theta})
	\Phi^{c\dagger}_{K}(p+q-r,\theta,{\overline \theta}) \Phi^{L}_{d}
	(r,\theta,{\overline \theta}) \right\} \nonumber  \, ;
\end{eqnarray}
\vspace{1cm}
\begin{displaymath}
	\tilde{D}_{2}(p,q,r) = \left(\frac{1}{4}\right)^{2}
	\frac{g^{4}}{2}\, \kappa (\delta_{ac}\delta^{bd} + 
	\delta_{a}^{b}\delta^{d}_{c})(\delta^{I}_{L}\delta^{K}_{J}
	-\delta^{I}_{J}\delta^{K}_{L}) I_{2\:bdIK}^{(D)acJL}(p,q,r) \, ,
\end{displaymath}
where
\begin{eqnarray}
	I_{2\:bdIK}^{(D)acJL}(p,q,r) & = & \int \frac{d^{4}k}{(2\pi)^{4}}
	d^{2}\theta d^{2}{\overline\theta} 
	\frac{1}{k^{2}(p-k)^{2}(k+r-p-q)^{2}(p+q-k)^{2}} \cdot  
	\nonumber \\ 
	&& \hspace{-3cm} \cdot \left\{ \left[{\overline 
	D}^{2}\Phi^{a\dagger}_{I}(p,\theta,{\overline \theta})\right] 
	\Phi^{J}_{b}(q,\theta,{\overline \theta}) 
	\Phi^{c\dagger}_{K}(p+q-r,\theta,{\overline \theta}) 
	\left[ D^{2} \Phi^{L}_{d}(r,\theta,{\overline \theta}) \right] + 
	\right. \nonumber \\ 
	&& \hspace{-3cm} \left. + 8i\sigma^{\mu}_{\alpha{\dot\alpha}}
	(k-p-q)_{\mu} \left[ {\overline D}^{\dot\alpha}\Phi^{a\dagger}_{I}
	(p,\theta,{\overline \theta})\right] 
	\Phi^{J}_{b}(q,\theta,{\overline \theta}) 
	\Phi^{c\dagger}_{K}(p+q-r,\theta,{\overline \theta}) 
	\left[ D^{\alpha}\Phi^{L}_{d}(r,\theta,{\overline\theta})\right]
	- \right. \nonumber \\
	&& \hspace{-3cm} \left.
	- 16 (k-p-q)^{2} \Phi^{a\dagger}_{I}
	(p,\theta,{\overline \theta})\Phi^{J}_{b}(q,\theta,{\overline\theta})
	\Phi^{c\dagger}_{K}(p+q-r,\theta,{\overline \theta}) \Phi^{L}_{d}
	(r,\theta,{\overline \theta}) \right\} \label{4pointd2} \, ;
\end{eqnarray}
\vspace{1cm}
\begin{displaymath}
	\tilde{D}_{3}(p,q,r) = \left(\frac{1}{4}\right)^{2}
	\frac{g^{4}}{2}\, \kappa ( \delta_{a}^{d}\delta^{b}_{c}
	+ \delta_{ac}\delta^{bd})(\delta^{I}_{J}\delta^{K}_{L}
	-\delta^{I}_{L}\delta^{K}_{J}) I_{3\:bdIK}^{(D)acJL}(p,q,r) \, ,
\end{displaymath}
where
\begin{eqnarray}
	I_{3\:bdIK}^{(D)acJL}(p,q,r) & = & \int \frac{d^{4}k}{(2\pi)^{4}}
	d^{2}\theta d^{2}{\overline\theta} 
	\frac{1}{k^{2}(p-k)^{2}(k+r-p-q)^{2}(p+q-k)^{2}} \cdot  
	\nonumber \\ 
	&& \hspace{-3cm} \cdot \left\{ \Phi^{a\dagger}_{I}
	(p,\theta,{\overline \theta})
	\Phi^{J}_{b}(q,\theta,{\overline \theta}) 
	\left[{\overline D}^{2}\Phi^{c\dagger}_{K}
	(p+q-r,\theta,{\overline \theta}) \right] 
	\left[ D^{2} \Phi^{L}_{d}(r,\theta,{\overline \theta}) \right] - 
	\right. \nonumber \\ 
	&& \hspace{-3cm} \left. - 8i\sigma^{\mu}_{\alpha{\dot\alpha}}
	(p-k-r)_{\mu} \Phi^{a\dagger}_{I}(p,\theta,{\overline \theta}) 
	\Phi^{J}_{b}(q,\theta,{\overline \theta}) 
	\left[ {\overline D}^{\dot\alpha}\Phi^{c\dagger}_{K}
	(p+q-r,\theta,{\overline \theta}) \right]
	\left[ D^{\alpha}\Phi^{L}_{d}(r,\theta,{\overline\theta})\right]
	- \right. \nonumber \\
	&& \hspace{-3cm} \left.
	- 16 (p-k-r)^{2} \Phi^{a\dagger}_{I}
	(p,\theta,{\overline \theta})\Phi^{J}_{b}(q,\theta,{\overline\theta})
	\Phi^{c\dagger}_{K}(p+q-r,\theta,{\overline \theta}) \Phi^{L}_{d}
	(r,\theta,{\overline \theta}) \right\} \label{4pointd3} \, ;
\end{eqnarray}
\vspace{1cm}
\begin{displaymath}
	\tilde{D}_{4}(p,q,r) = \left(\frac{1}{4}\right)^{2}
	\frac{g^{4}}{2}\, \kappa ( \delta_{a}^{b}\delta^{d}_{c}
	+ \delta_{ac}\delta^{bd})(\delta^{I}_{L}\delta^{K}_{J}
	-\delta^{I}_{J}\delta^{K}_{L}) I_{4\:bdIK}^{(D)acJL}(p,q,r) \, ,
\end{displaymath}
where
\begin{eqnarray}
	I_{4\:bdIK}^{(D)acJL}(p,q,r) & = & \int \frac{d^{4}k}{(2\pi)^{4}}
	d^{2}\theta d^{2}{\overline\theta} 
	\frac{1}{k^{2}(p-k)^{2}(k+r-p-q)^{2}(p+q-k)^{2}} \cdot  
	\nonumber \\ 
	&& \hspace{-3cm} \cdot \left\{ \Phi^{a\dagger}_{I}
	(p,\theta,{\overline \theta})
	\left[ D^{2} \Phi^{J}_{b}(q,\theta,{\overline \theta})\right] 
	\left[{\overline D}^{2}\Phi^{c\dagger}_{K}
	(p+q-r,\theta,{\overline \theta}) \right] 
	\Phi^{L}_{d}(r,\theta,{\overline \theta}) - 
	\right. \label{4pointd4} \\ 
	&& \hspace{-3cm} \left. - 8i\sigma^{\mu}_{\alpha{\dot\alpha}}
	(p+q-k)_{\mu} \Phi^{a\dagger}_{I}(p,\theta,{\overline \theta}) 
	\left[ D^{\alpha}\Phi^{J}_{b}(q,\theta,{\overline \theta})\right] 
	\left[ {\overline D}^{\dot\alpha}\Phi^{c\dagger}_{K}
	(p+q-r,\theta,{\overline \theta}) \right]
	\cdot \right. \nonumber \\
	&& \hspace{-3cm}\cdot\left.
	\Phi^{L}_{d}(r,\theta,{\overline\theta}) 
	- 16 (p+q-k)^{2} \Phi^{a\dagger}_{I}
	(p,\theta,{\overline \theta})\Phi^{J}_{b}(q,\theta,{\overline\theta})
	\Phi^{c\dagger}_{K}(p+q-r,\theta,{\overline \theta}) \Phi^{L}_{d}
	(r,\theta,{\overline \theta}) \right\} \nonumber \, .
\end{eqnarray}

The last family of one-loop diagrams consists of the one below plus 
again those obtained by crossing.

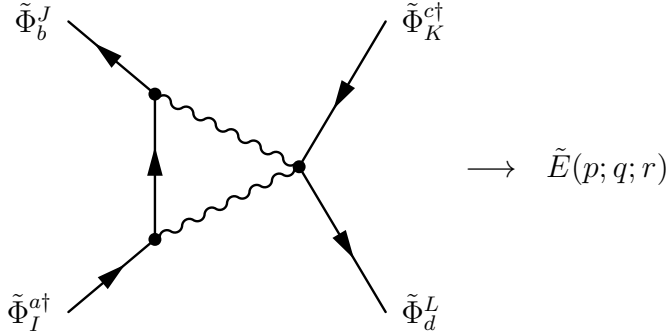
\vspace{1cm}
\noindent
\begin{fmffile}{phi+phiphi+phiE}
 \begin{fmfgraph*}(150,110) 
  \fmfleft{i1,i2} 
  \fmfright{o1,o2}
  \fmf{fermion,tension=1}{i1,v1}
  \fmf{fermion,tension=1}{v2,i2}
  \fmf{fermion,tension=0.2}{v1,v2}
  \fmf{boson,tension=0.6}{v1,v3}
  \fmf{boson,tension=0.6}{v2,v3}
  \fmf{fermion,tension=1}{v3,o1}
  \fmf{fermion,tension=1}{o2,v3}
  \fmfv{l=$\tilde{\Phi}^{a\dagger}_{I}$,l.a=180}{i1}
  \fmfv{l=$\tilde{\Phi}^{J}_{b}$,l.a=180}{i2}
  \fmfv{l=$\tilde{\Phi}^{L}_{d}$,l.a=0}{o1}
  \fmfv{l=$\tilde{\Phi}^{c\dagger}_{K}$,l.a=0}{o2}
  \fmfv{l=\hspace{2cm}$\longrightarrow ~~
            \tilde{E}(p;q;r)$,l.a=0}{v3}
  \fmfdot{v1,v2,v3}
 \end{fmfgraph*}   
\end{fmffile} 

\vspace{0.5cm}
\noindent
The resulting contributions to the four-point Green function are the 
following four terms ($\tilde{E}_{1}$, $\tilde{E}_{2}$, 
$\tilde{E}_{3}$, $\tilde{E}_{4}$)
\begin{displaymath}
	\tilde{E}_{1}(p,q,r) = \left(\frac{1}{4}\right)^{2}
	\frac{g^{4}}{4}\, \kappa ( \delta_{a}^{b}\delta^{d}_{c}
	+ \delta_{ac}\delta^{bd})\delta^{I}_{J}\delta^{K}_{L}
	I_{1\:bdIK}^{(E)acJL}(p,q,r) \, ,
\end{displaymath}
where
\begin{eqnarray}
	I_{1\:bdIK}^{(E)acJL}(p,q,r) & = & \int \frac{d^{4}k}{(2\pi)^{4}}
	d^{2}\theta d^{2}{\overline\theta} 
	\frac{1}{k^{2}(p-k)^{2}(p+q-k)^{2}} \cdot  
	\nonumber \\ 
	&& \hspace{-2.5cm} \cdot \left\{ \Phi^{a\dagger}_{I}
	(p,\theta,{\overline \theta})\Phi^{J}_{b}(q,\theta,{\overline\theta})
	\Phi^{c\dagger}_{K}(p+q-r,\theta,{\overline \theta}) 
	\Phi^{L}_{d}(r,\theta,{\overline \theta}) \right\} \, ;
\end{eqnarray}
% \vspace{1cm}
\begin{displaymath}
	\tilde{E}_{2}(p,q,r) = \left(\frac{1}{4}\right)^{2}
	\frac{g^{4}}{4}\, \kappa ( \delta_{a}^{d}\delta^{b}_{c}
	+ \delta_{ac}\delta^{bd})\delta^{I}_{L}\delta^{K}_{J}
	I_{2\:bdIK}^{(E)acJL}(p,q,r) \, ,
\end{displaymath}
where
\begin{eqnarray}
	I_{2\:bdIK}^{(E)acJL}(p,q,r) & = & \int \frac{d^{4}k}{(2\pi)^{4}}
	d^{2}\theta  d^{2}{\overline\theta}
	\frac{1}{k^{2}(p-k)^{2}(r-k)^{2}} \cdot  
	\nonumber \\ 
	&& \hspace{-2.5cm} \cdot \left\{ \Phi^{a\dagger}_{I}
	(p,\theta,{\overline \theta})\Phi^{J}_{b}(q,\theta,{\overline\theta})
	\Phi^{c\dagger}_{K}(p+q-r,\theta,{\overline \theta}) 
	\Phi^{L}_{d}(r,\theta,{\overline \theta}) \right\} \, ;
\end{eqnarray}
\vspace{1cm}
\begin{displaymath}
	\tilde{E}_{3}(p,q,r) = \left(\frac{1}{4}\right)^{2}
	\frac{g^{4}}{4}\, \kappa ( \delta_{a}^{b}\delta^{d}_{c}
	+ \delta_{ac}\delta^{bd})\delta^{I}_{L}\delta^{K}_{J}
	I_{3\:bdIK}^{(E)acJL}(p,q,r) \, ,
\end{displaymath}
where
\begin{eqnarray}
	I_{3\:bdIK}^{(E)acJL}(p,q,r) & = & \int \frac{d^{4}k}{(2\pi)^{4}}
	d^{2}\theta d^{2}{\overline\theta} 
	\frac{1}{k^{2}(p+q-k)^{2}(k-r)^{2}} \cdot  
	\nonumber \\ 
	&& \hspace{-2.5cm} \cdot \left\{ \Phi^{a\dagger}_{I}
	(p,\theta,{\overline \theta})\Phi^{J}_{b}(q,\theta,{\overline\theta})
	\Phi^{c\dagger}_{K}(p+q-r,\theta,{\overline \theta}) 
	\Phi^{L}_{d}(r,\theta,{\overline \theta}) \right\} \, ;
\end{eqnarray}
\vspace{1cm}
\begin{displaymath}
	\tilde{E}_{4}(p,q,r) = \left(\frac{1}{4}\right)^{2}
	\frac{g^{4}}{4}\, \kappa ( \delta_{a}^{d}\delta^{b}_{c}
	+ \delta_{ac}\delta^{bd})\delta^{I}_{L}\delta^{K}_{J}
	I_{4\:bdIK}^{(E)acJL}(p,q,r) \, ,
\end{displaymath}
where
\begin{eqnarray}
	I_{4\:bdIK}^{(E)acJL}(p,q,r) & = & \int \frac{d^{4}k}{(2\pi)^{4}}
	d^{2}\theta d^{2}{\overline\theta} 
	\frac{1}{k^{2}(q+k)^{2}(k+r-p)^{2}} \cdot  
	\nonumber \\ 
	&& \hspace{-2.5cm} \cdot \left\{ \Phi^{a\dagger}_{I}
	(p,\theta,{\overline \theta})\Phi^{J}_{b}(q,\theta,{\overline\theta})
	\Phi^{c\dagger}_{K}(p+q-r,\theta,{\overline \theta}) 
	\Phi^{L}_{d}(r,\theta,{\overline \theta}) \right\} \, .
\end{eqnarray}

The sum of all the preceding terms results in a finite and non 
vanishing total one-loop correction to the four point function. 
The final expression contains terms with six different tensorial 
structures
\begin{displaymath}
	\langle \Phi^{\dagger}\Phi\Phi^{\dagger}\Phi \rangle = 
	\sum_{i=1}^{6} G^{(i)} \, ,
\end{displaymath}
where
\begin{eqnarray*}
	G^{(1)} &=& \kappa \left(\frac{1}{4}\right)^{2}g^{4}\, 
	\delta_{a}^{b}\delta_{c}^{d}\delta_{J}^{I}\delta_{L}^{K}\left(
	\frac{1}{4}I_{1\:bdIK}^{(A)acJL} + \frac{1}{4}I_{3\:bdIK}^{(A)acJL}
	+\frac{1}{6}I_{bdIK}^{(B)acJL}-\frac{1}{2}I_{2\:bdIK}^{(D)acJL}- 
	\right. \\
	&-& \left.
	\frac{1}{2}I_{4\:bdIK}^{(D)acJL}+\frac{1}{2}I_{1\:bdIK}^{(E)acJL}+
	\frac{1}{2}I_{3\:bdIK}^{(E)acJL} \right) \\
	G^{(2)} &=& \kappa \left(\frac{1}{4}\right)^{2}g^{4}\, 
	\delta_{ac}\delta^{bd}\delta_{J}^{I}\delta_{L}^{K}\left(
	\frac{1}{4}I_{1\:bdIK}^{(A)acJL} + \frac{1}{2}I_{1\:bdIK}^{(D)acJL}-
	\frac{1}{2}I_{2\:bdIK}^{(D)acJL}+\frac{1}{2}I_{3\:bdIK}^{(D)acJL}- 
	\right. \\ 
	&-& \left.
	\frac{1}{2}I_{4\:bdIK}^{(D)acJL}+\frac{1}{2}I_{1\:bdIK}^{(E)acJL}+
	\frac{1}{2}I_{3\:bdIK}^{(E)acJL} \right) \\
	G^{(3)} &=& \kappa \left(\frac{1}{4}\right)^{2}g^{4}\, 
	\delta_{a}^{b}\delta_{c}^{d}\delta_{L}^{I}\delta_{J}^{K}\left(
	\frac{1}{4}I_{2\:bdIK}^{(A)acJL} + \frac{1}{6}I_{bdIK}^{(B)acJL}+
	\frac{1}{2}I_{2\:bdIK}^{(D)acJL}+\frac{1}{2}I_{4\:bdIK}^{(D)acJL}
	\right) \\
	G^{(4)} &=& \kappa \left(\frac{1}{4}\right)^{2}g^{4}\, 
	\delta_{a}^{d}\delta_{c}^{b}\delta_{L}^{I}\delta_{J}^{K}\left(
	\frac{1}{4}I_{2\:bdIK}^{(A)acJL} + \frac{1}{4}I_{4\:bdIK}^{(A)acJL}
	+\frac{1}{6}I_{bdIK}^{(B)acJL}-\frac{1}{2}I_{1\:bdIK}^{(D)acJL}- 
	\right. \\
	&-& \left.
	\frac{1}{2}I_{3\:bdIK}^{(D)acJL}+\frac{1}{2}I_{2\:bdIK}^{(E)acJL}+
	\frac{1}{2}I_{4\:bdIK}^{(E)acJL} \right) \\
	G^{(5)} &=& \kappa \left(\frac{1}{4}\right)^{2}g^{4}\, 
	\delta_{a}^{d}\delta_{c}^{b}\delta_{J}^{I}\delta_{L}^{K}\left(
	\frac{1}{4}I_{3\:bdIK}^{(A)acJL}+\frac{1}{6}I_{bdIK}^{(B)acJL}
	+\frac{1}{2}I_{1\:bdIK}^{(D)acJL}+\frac{1}{2}I_{3\:bdIK}^{(E)acJL}
	\right) \\
	G^{(6)} &=& \kappa \left(\frac{1}{4}\right)^{2}g^{4}\, 
	\delta_{ac}\delta^{bd}\delta_{L}^{I}\delta_{J}^{K}\left(
	\frac{1}{4}I_{4\:bdIK}^{(A)acJL} - \frac{1}{2}I_{1\:bdIK}^{(D)acJL}+
	\frac{1}{2}I_{2\:bdIK}^{(D)acJL}-\frac{1}{2}I_{3\:bdIK}^{(D)acJL}+ 
	\right. \\ 
	&+& \left. 
	\frac{1}{2}I_{4\:bdIK}^{(D)acJL}+\frac{1}{2}I_{2\:bdIK}^{(E)acJL}+
	\frac{1}{2}I_{4\:bdIK}^{(E)acJL} \right)
\end{eqnarray*}
Notice that, since in the Fermi--Feynman gauge the one-loop corrections 
to the propagators and vertices are zero, the total cross section is 
completely determined by the sum of the above contributions, which is non 
vanishing for on-shell external momenta, \ie $p^{2}$=$q^{2}$=$r^{2}$=0.

In the presence of a mass term for the (anti) chiral superfields the 
expressions given above are modified by the presence of the 
mass in the free propagators and furthermore there are two additional 
sets of contributions corresponding to the diagrams

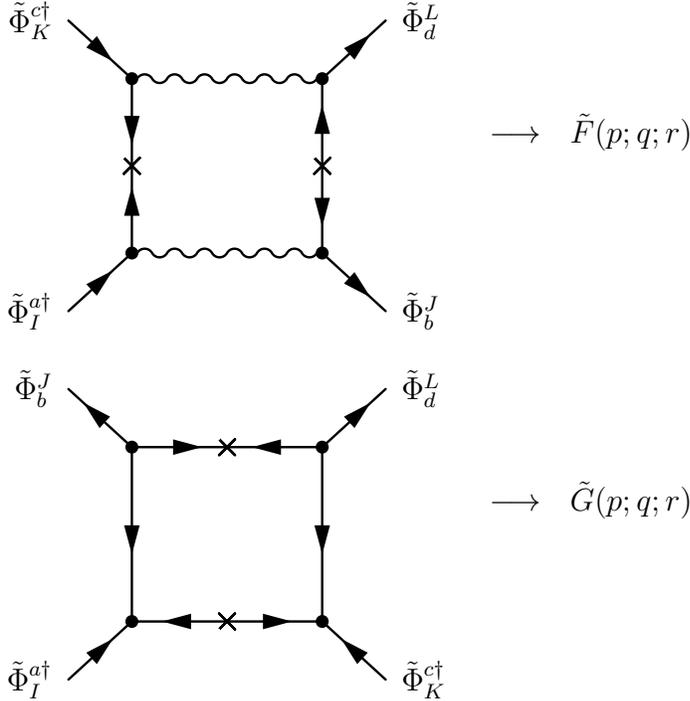
\vspace{1cm}
\noindent
\begin{fmffile}{phi+phiphi+phiF}
 \begin{fmfgraph*}(150,110) 
  \fmfleft{i1,i2} 
  \fmfright{o1,o2}
  \fmf{fermion,tension=0.6}{i1,v1}
  \fmf{fermion,tension=0.6}{i2,v2}
  \fmf{fermion,tension=0.6}{v3,o1}
  \fmf{fermion,tension=0.6}{v4,o2}
  \fmf{phantom,tension=0.1,tag=1}{v1,v2}
  \fmf{phantom,tension=0.1,tag=2}{v2,v1}
  \fmf{phantom,tension=0.1,tag=3}{v2,v4}
  \fmf{phantom,tension=0.1,tag=4}{v4,v2}
  \fmf{phantom,tension=0.1,tag=5}{v4,v3}
  \fmf{phantom,tension=0.1,tag=6}{v3,v4}
  \fmf{phantom,tension=0.1,tag=7}{v3,v1}
  \fmf{phantom,tension=0.1,tag=8}{v1,v3}
  \fmfv{l=$\tilde{\Phi}^{a\dagger}_{I}$,l.a=180}{i1}
  \fmfv{l=$\tilde{\Phi}^{J}_{b}$,l.a=0}{o1}
  \fmfv{l=$\tilde{\Phi}^{L}_{d}$,l.a=0}{o2}
  \fmfv{l=$\tilde{\Phi}^{c\dagger}_{K}$,l.a=180}{i2}
  \fmfdot{v1,v2,v3,v4}
  \fmffreeze
  \fmfipath{p[]}
  \fmfiset{p1}{vpath1(__v1,__v2)}
  \fmfiset{p2}{vpath2(__v2,__v1)}
  \fmfiset{p3}{vpath3(__v2,__v4)}
  \fmfiset{p4}{vpath4(__v4,__v2)}
  \fmfiset{p5}{vpath5(__v4,__v3)}
  \fmfiset{p6}{vpath6(__v3,__v4)}
  \fmfiset{p7}{vpath7(__v3,__v1)}
  \fmfiset{p8}{vpath8(__v1,__v3)}
  \fmfi{fermion}{subpath (0,length(p1)/2) of p1}
  \fmfi{fermion}{subpath (0,length(p2)/2) of p2}
  \fmfi{fermion}{subpath (length(p6)/2,length(p6)) of p6}
  \fmfi{fermion}{subpath (length(p5)/2,length(p5)) of p5}
  \fmfi{boson}{subpath (0,length(p3)) of p3}
  \fmfi{boson}{subpath (0,length(p7)) of p7}
  \fmfiv{d.sh=cross,d.si=3mm}{point length(p1)/2 of p1}
  \fmfiv{d.sh=cross,d.si=3mm,l=\raisebox{-20pt}{\hspace{7cm}
        $\longrightarrow ~~\tilde{F}(p;q;r)$},l.a=90}
        {point length(p5)/2 of p5}
 \end{fmfgraph*}   
\end{fmffile} 

\vspace{1cm}
\noindent
\begin{fmffile}{phi+phiphi+phiG}
 \begin{fmfgraph*}(150,110) 
  \fmfleft{i1,i2} 
  \fmfright{o1,o2}
  \fmf{fermion,tension=0.6}{i1,v1}
  \fmf{fermion,tension=0.6}{v2,i2}
  \fmf{fermion,tension=0.6}{o1,v3}
  \fmf{fermion,tension=0.6}{v4,o2}
  \fmf{phantom,tension=0.1,tag=1}{v1,v2}
  \fmf{phantom,tension=0.1,tag=2}{v2,v1}
  \fmf{phantom,tension=0.1,tag=3}{v2,v4}
  \fmf{phantom,tension=0.1,tag=4}{v4,v2}
  \fmf{phantom,tension=0.1,tag=5}{v4,v3}
  \fmf{phantom,tension=0.1,tag=6}{v3,v4}
  \fmf{phantom,tension=0.1,tag=7}{v3,v1}
  \fmf{phantom,tension=0.1,tag=8}{v1,v3}
  \fmfv{l=$\tilde{\Phi}^{a\dagger}_{I}$,l.a=180}{i1}
  \fmfv{l=$\tilde{\Phi}^{J}_{b}$,l.a=180}{i2}
  \fmfv{l=$\tilde{\Phi}^{L}_{d}$,l.a=0}{o2}
  \fmfv{l=$\tilde{\Phi}^{c\dagger}_{K}$,l.a=0}{o1}
  \fmfdot{v1,v2,v3,v4}
  \fmffreeze
  \fmfipath{p[]}
  \fmfiset{p1}{vpath1(__v1,__v2)}
  \fmfiset{p2}{vpath2(__v2,__v1)}
  \fmfiset{p3}{vpath3(__v2,__v4)}
  \fmfiset{p4}{vpath4(__v4,__v2)}
  \fmfiset{p5}{vpath5(__v4,__v3)}
  \fmfiset{p6}{vpath6(__v3,__v4)}
  \fmfiset{p7}{vpath7(__v3,__v1)}
  \fmfiset{p8}{vpath8(__v1,__v3)}
  \fmfi{fermion}{subpath (0,length(p2)) of p2}
  \fmfi{fermion}{subpath (0,length(p3)/2) of p3}
  \fmfi{fermion}{subpath (0,length(p4)/2) of p4}
  \fmfi{fermion}{subpath (0,length(p5)) of p5}
  \fmfi{fermion}{subpath (length(p8)/2,length(p8)) of p8}
  \fmfi{fermion}{subpath (length(p7)/2,length(p7)) of p7}
  \fmfiv{d.sh=cross,d.si=3mm}{point length(p3)/2 of p3}
  \fmfiv{d.sh=cross,d.si=3mm}{point length(p7)/2 of p7}
  \fmfiv{l=\raisebox{-20pt}{\hspace{7cm}
        $\longrightarrow ~~\tilde{G}(p;q;r)$},l.a=90}
        {point length(p5)/2 of p5}
 \end{fmfgraph*}   
\end{fmffile}

\vspace{0.5cm}
\noindent
Both of these graphs give corrections proportional to $m^{2}$, that can 
be calculated much in the same way as the previous ones.

\section{Discussion}
\label{discussion}

The infrared divergences found in the calculation of the propagators 
of the chiral and the vector superfields are due to the fact that the 
vector superfield is dimensionless, so 
that it contains in particular, as its lowest component, the scalar 
$C$ that is itself dimensionless and hence has a propagator which 
behaves, in momentum space, like
\begin{displaymath}
	\langle (CC)(k) \rangle \sim \frac{1}{k^{4}} \, .
\end{displaymath}
The contribution of the scalar $C$ to the $\langle VV \rangle$ 
propagator leads to an infrared divergence whenever a diagram contains 
a loop involving a $VV$ line. In the Fermi--Feynman gauge the problem 
is not present because the choice $\alpha=1$ gives a kinetic matrix 
of the form of equation (\ref{kineticmatrices}) in the component 
expansion. The corresponding inverse matrix $M^{-1}$ in  
(\ref{inverskinmatr}) shows that no $\langle CC \rangle$ free 
propagator is present. On the contrary any choice $\alpha \neq 1$ 
produces such a propagator, \ie gives a matrix $M^{-1}$ with a 
non-vanishing $(M^{-1})_{11}$ entry, leading to the previously 
discussed infrared problem. An explicit calculation in components with 
$\alpha \neq 1$ and without fixing the Wess--Zumino gauge should 
display problematic infrared divergences analogous to those encountered 
here. These problems with infrared divergences are not peculiar of 
the ${\cal N}$=4 super Yang--Mills theory, but appear in any  
supersymmetric gauge theory. Analogous infrared divergences in Green 
functions have been observed in \cite{juerstorey2,juerstorey1}. The 
conclusion proposed in these papers is that there exist no way to 
remove the infrared divergences, so that the choice $\alpha$=1 is 
somehow necessary, at least for the computation of gauge-dependent 
quantities.

There are two general theorems concerning infrared divergences in 
quantum field theory. The first is the Kinoshita--Lee--Nauenberg 
theorem \cite{kinoleenauen}, which deals with infrared divergences 
in cross sections. It states that no infrared problem is present 
in physical cross sections of a renormalizable quantum field theory, 
if the appropriate sums over degenerate initial and final states, 
associated with soft and collinear particles, are considered. 
The second theorem was proved by Kinoshita, Poggio and Quinn 
\cite{kinopoggioquinn} and concerns Green functions. The statement of 
the theorem is that the proper (one-particle irreducible) Euclidean 
Green functions with non-exceptional external momenta are free of 
infrared divergences in a renormalizable quantum field theory. A set 
of momenta $p_{i}$, $i=1,2,\ldots,n$ is said to be exceptional if any 
of the partial sums
\begin{displaymath}
	\sum_{i\in I} p_{i}\, , \qquad {\rm with}~~I~~{\rm a~subset~of}~~ 
	\{1,2,\ldots,n\} \, ,
\end{displaymath}
vanishes. The reason for the absence of divergences is that the 
external momenta, if non-exceptional, play the r\^ole of an infrared 
regulator in the Green functions. 

The proof of the theorem is based on dimensional analysis which does 
not work in the presence of a $\frac{1}{k^{4}}$ propagator. 
In particular it fails in the case of 
supersymmetric gauge theories in general gauges, because the 
adimensionality of the $V$ superfield implies that a propagator 
$\frac{1}{k^{4}}$ can appear in loop integrals. The apparent violation 
of the Kinoshita--Poggio--Quinn theorem can be understood from the 
viewpoint of the component  formulation: supersymmetric gauge theories, 
being non-polynomial and thus containing an infinite number of 
interaction terms, are not {\em formally} renormalizable. The choice 
of the Wess--Zumino gauge makes the theory polynomial. In fact in 
this case no infrared divergence is found in Green functions and the 
general theorems are satisfied. However in the case of the ${\cal 
N}$=4 theory the choice of the WZ gauge results in a change of the 
ultraviolet properties of the model, at least for what concerns gauge 
dependent quantities, {\em e.g.} the propagators.

In theories with less supersymmetry, in which ultraviolet divergences 
are present the problem might be less severe, since the subtraction 
of the ultraviolet infinities could also cure infrared singularities. 

The infrared problems discussed here are however gauge artifacts and 
cannot affect gauge invariant quantities. The mechanism leading to 
the cancellation of infrared divergences in gauge independent Green 
functions has not been verified in detail yet, but can be understood 
starting from the super Feynman rules. When the propagator $\langle 
VV \rangle$ is inserted into a Feynman graph it is connected to 
vertices which carry covariant derivatives. These derivatives come 
directly from the form of the action for vertices involving only $V$ 
fields and from the definition of the functional derivatives for 
vertices involving (anti) chiral superfields. 
The situation in diagrams for gauge invariant 
quantities is such that at least one covariant 
derivative can always be brought to act on the $V$ propagator by 
integrations by parts. In this way an additional factor of the momentum 
$k$ is generated in the numerator. More precisely for the infrared 
problematic term $-\frac{1}{k^{4}}({\overline 
D}^{2}D^{2}+D^{2}{\overline D}^{2})\delta_{4}(\theta-\theta^{\prime})$ 
one gets
\begin{displaymath}
	D_{\alpha}\left[-\frac{1}{k^{4}}({\overline D}^{2}
	D^{2}+D^{2}{\overline D}^{2})\delta_{4}(\theta-\theta^{\prime})
	\right] = -4i\frac{k_{\mu}\sigma^{\mu}_{\alpha{\dot\alpha}}}{k^{4}}
	{\overline D}^{\dot\alpha}D^{2}
\end{displaymath}
and
\begin{displaymath}
	{\overline D}^{\dot\alpha}\left[-\frac{1}{k^{4}}({\overline D}^{2}
	D^{2}+D^{2}{\overline D}^{2})\delta_{4}(\theta-\theta^{\prime})
	\right] = 4i\frac{k_{\mu}{\overline \sigma}^{\mu{\dot\alpha}\alpha}}
	{k^{4}} D_{\alpha}{\overline D}^{2} \, .
\end{displaymath}
A rigorous and general check of this mechanism in gauge invariant 
Green functions has not been achieved yet.

With a gauge choice different from the Fermi--Feynman gauge the 
computation of four-point Green 
functions is more complicated. Single diagrams involving vector 
superfield propagators contain new contributions, some of which are 
infrared divergent. The correction $\tilde{C}$ is not zero anymore, 
because there are projection operators acting on the $\delta$-functions. 
Moreover further diagrams must be included in the calculation of 
scattering amplitudes at the same order as a consequence of the 
non-vanishing of the one-loop correction to the propagators and vertices. 
For example one must consider diagrams such as

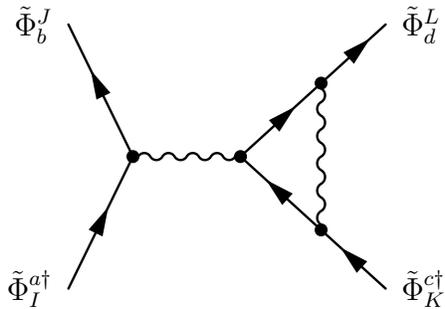
\vspace{1cm}
\noindent
\begin{fmffile}{phi+phiphi+phiH}
 \begin{fmfgraph*}(150,100) 
  \fmfleft{i1,i2} 
  \fmfright{o1,o2}
  \fmf{fermion,tension=1}{i1,v1}
  \fmf{fermion,tension=1}{v1,i2}
  \fmf{boson,tension=1.2}{v1,v2}
  \fmf{fermion,tension=0.8}{v2,v3}
  \fmf{fermion,tension=0.8}{v4,v2}
  \fmf{fermion}{o1,v4}
  \fmf{fermion}{v3,o2}
  \fmffreeze
  \fmf{boson}{v3,v4}
  \fmfv{l=$\tilde{\Phi}^{J}_{b}$,l.a=180}{i2}
  \fmfv{l=$\tilde{\Phi}^{L}_{d}$,l.a=0}{o2}
  \fmfv{l=$\tilde{\Phi}^{c\dagger}_{K}$,l.a=0}{o1}
  \fmfv{l=$\tilde{\Phi}^{a\dagger}_{I}$,l.a=180}{i1}
  \fmfdot{v1,v2,v3,v4}
 \end{fmfgraph*}   
\end{fmffile} 

\vspace{1cm}
\noindent
as well as diagrams obtained by the insertion of one-loop corrections 
to the propagators in tree graphs. Notice that the inclusion of these 
contributions, each one separately infrared divergent, is fundamental 
in order to get a finite total cross section.

The techniques illustrated in this paper for the calculation of 
Green functions in the ${\cal N}$=1 formalism can be applied to 
correlation functions of composite operators as well. Green functions 
of gauge invariant composite operators such as those that form 
the multiplet of currents (see \cite{currents} for the explicit form of the 
multiplet in the Abelian case) play  a crucial r\^ole in the correspondence 
with type IIB superstring theory on AdS space. The application of 
${\cal N}$=1 superspace to this problem has not been considered here, 
but is under active investigation \cite{bkrs}. 
It can be shown that the extension of the formalism to the case of 
composite operators is in principle rather straightforward, the fundamental 
difference being that the complete Green functions and not the proper 
parts must be considered.

\vspace{1cm}
 
{\bf Acknowledgments}

\vspace{0.3cm}
 
\noindent
I am very grateful to Massimo Bianchi and Giancarlo Rossi for many useful 
discussions and for their comments on the manuscript. I also wish to 
thank Claudio Fusi, Yassen Stanev and Kensuke Yoshida for stimulating 
discussions.


\begin{thebibliography}{123}
% 
\bibitem{brinkscherkschwarz}{L.~Brink, J.~Scherk and J.H.~Schwarz, 
``Supersymmetric Yang-Mills theories'', 
{\it Nucl. Phys.} {\bf B121} (1977) 77}. 
\bibitem{gso}{F.~Gliozzi, D.I.~Olive and J.~Scherk, 
``Supersymmetry, supergravity and the dual spinor model'', 
{\it Nucl. Phys.} {\bf B122} (1977) 253.} 
\bibitem{sohniuswest}{M.F.~Sohnius and P.C.~West, ``Conformal invariance 
in $N=4$ supersymmetric Yang--Mills theory'', {\it Phys. Lett.} 
{\bf 100B} (1981) 245.}
\bibitem{white}{P.L.~White, ``Analysis of the superconformal 
cohomology structure of $N$=4 super Yang--Mills'', {\it Class. Quant. 
Grav.} {\bf 9} (1992) 413.}
\bibitem{hagsohlop}{R.~Haag, J.T.~{\L}opusza\'{n}ski and M.~Sohnius, 
``All possible generators of supersymmetries of the $S$--matrix'', 
{\it Nucl. Phys.} {\bf B88} (1975) 257.}
\bibitem{olivemont}{K.~Montonen and D.I.~Olive, ``Magnetic monopoles 
as gauge particles'', {\it Phys. Lett.} {\bf 66B} (1977) 61.} 
\bibitem{osborn}{H.~Osborn, ``Topological charges for $N$=4 
supersymmetric gauge theories and monopoles of spin one'', {\it Phys. 
Lett.} {\bf 83B} 321.}
\bibitem{n4qm}{J.P.~Gauntlett, ``Low energy dynamics of $N$=2 
supersymmetric monopoles'', {\it Nucl. Phys.} {\bf 
B411} (1994) 443, {\tt hep-th/9305068}; J.D.~Blum, ``Supersymmetric 
quantum mechanics of monopoles in $N$=4 Yang--Mills theory'', 
{\it Phys. Lett.} {\bf B333} (1994) 92, {\tt hep-th/9401133}.}
\bibitem{sen}{A.~Sen, ``Dyon-monopole bound states, self dual harmonic 
forms on the multimonopole moduli space'', {\it Phys. Lett.} {\bf 
329B} (1994) 217, {\tt hep-th/9402032}.}
\bibitem{sym1}{S.~Ferrara and B.~Zumino, ``Supergauge invariant 
Yang--Mills theories'', {\it Nucl. Phys.} {\bf B79} (1974) 413.}
\bibitem{2loop}{D.T.R.~Jones, ``Charge renormalization in a 
supersymmetric Yang--Mills theory'', {\it Phys Lett.} {\bf 72B} 
(1977) 199; E.~Poggio and H.~Pendleton, ``Vanishing of charge 
renormalization and anomalies in a supersymmetric gauge theory'', 
{\it Phys Lett.} {\bf 72B} (1977) 200.}
\bibitem{tarasov}{A.A.~Vladimirov and O.V.~Tarasov, ``Vanishing of the 
three-loop charge renormalization function in a supersymmetric gauge 
theory'', {\it Phys. Lett.} {\bf 96B} (1980) 94.}
\bibitem{grirocsieg2}{M.~Grisaru, M.~Ro\v{c}ek and W.~Siegel, 
``Zero value of the three-loop $\beta$ function in $N=4$ 
supersymmetric Yang--Mills theory'', {\it Phys. Rev. Lett.} 
{\bf 45} (1980) 1063.}
\bibitem{caswzanon}{W.E.~Caswell and D.~Zanon, ``Zero three-loop beta 
function in the $N=4$ supersymmetric Yang--Mills theory'', {\it Nucl. 
Phys.} {\bf B182} (1981) 125.}
\bibitem{n2finite}{P.S.~Howe, K.S.~Stelle and P.K.~Townsend, ``The 
relaxed hypermultiplet: an unconstrained $N=2$ superfield theory'', 
{\it Nucl. Phys.} {\bf B214} (1983) 519; P.S.~Howe, K.S.~Stelle and 
P.C.~West, ``A class of finite four-dimensional supersymmetric field 
theories'', {\it Phys. Lett.} {\bf 124B} (1983) 55; P.S.~Howe, 
K.S.~Stelle and P.K.~Townsend, ``Miraculous ultraviolet cancellations in 
supersymmetry made manifest'', {\it Nucl. Phys.} {\bf B236} (1984) 125.}
\bibitem{mandelstam}{S.~Mandelstam, ``Light-cone superspace and the 
ultraviolet finiteness of the $N=4$ model'', {\it Nucl. Phys.} 
{\bf B213} (1983) 149.}
\bibitem{dinesei}{M.~Dine and N.~Seiberg, ``Comments on higher 
derivative operators in some SUSY field theories'', {\it Phys. 
Lett.} {\bf B409} (1997) 239, {\tt hep-th/9705057}.}
\bibitem{maldacena}{J. Maldacena, ``The large $N$ limit of   
superconformal field theories and supergravity'', {\it Adv. Theor. 
Math. Phys.} {\bf 2} (1998) 231, {\tt hep-th/9711200}.} 
\bibitem{adspert}{E. Witten, ``Anti de Sitter space and  
holography'', {\it Adv. Theor. Math. Phys.} {\bf 2} (1998) 253, 
{\tt hep-th/9802150}; W. M\"uck and K.S. Viswanathan, 
``Conformal field theory correlators from classical field theory 
on anti-de Sitter space, I and II'', 
{\tt hep-th/9804035} and {\tt hep-th/9805145}; 
D.Z.~Freedman, S.D.~Mathur, A.~Matusis and L.~Rastelli, 
``Correlation functions in the CFT$_{d}$/AdS$_{d+1}$ correspondence'', 
{\tt hep-th/9804058}; G. Chalmers, H. Nastase, K. Schalm and R. Siebelink, 
``R-current correlators in N=4 Super Yang--Mills theory from 
anti-de Sitter supergravity'', {\tt hep-th/9805105}; S. Lee, S. Minwalla, M. 
Rangamani and N. Seiberg, ``Three-point functions of chiral operators in 
$D$=4 $N$=4 SYM at large N'', {\tt hep-th/9806074}; D.Z.~Freedman, 
E.~D'Hoker and W.~Skiba, ``Field theory tests for correlators in the 
AdS/CFT correspondence'', {\tt hep-th/9807098}; H.~Liu and 
A.A.~Tseytlin, ``On four point functions in the CFT/AdS correspondence'', 
{\tt hep-th/9807097}; D.Z.~Freedman, S.D.~Mathur, 
A.~Matusis and L.~Rastelli, ``Comments on 4 point functions 
in the CFT/AdS correspondence'', {\tt hep-th/9808006}; G.~Chalmers 
and K.~Schalm, ``The large $N_{c}$ limit of four point functions in $N$=4 
super Yang--Mills theory from anti-de Sitter supergravity'', 
{\tt hep-th/9810051}; H.~Liu, ``Scattering in anti-de Sitter space and 
operator product expansion'', {\tt hep-th/9811152}; B.~Eden, P.S.~Howe, 
C.~Schubert, E. Sokatchev and P.C.~West, ``Four point functions in $N$=4 
supersymmetric Yang--Mills theory at two loops'', {\tt hep-th/9811172}}
\bibitem{adsnonpert}{M.~Bianchi, M.B.~Green, S.~Kovacs and G.C.~Rossi, 
``Instantons in supersymmetric Yang--Mills and D-instantons in IIB 
superstring theory'', {\it JHEP} {\bf 08} (1998) 013, {\tt 
hep-th/9807033}; N.~Dorey, V.V.~Khoze, M.P.~Mattis and S.~Vandoren,  
``Yang--Mills Instantons in the large-$N$ limit and the AdS/CFT 
correspondence'', {\tt hep-th/9808157}; N.~Dorey, T.J.~Hollowood, 
V.V.~Khoze, M.P.~Mattis and S.~Vandoren, ``Multi-instantons and Maldacena's 
conjecture'', {\tt hep-th/9810243}; J.H.~Brodie and M.~Gutperle, ``String 
corrections to four point functions in the AdS/CFT correspondence'', 
{\tt hep-th/9809067}.}
\bibitem{piguetrouet}{O.~Piguet and A.~Rouet, ``Supersymmetric BPHZ 
renormalization 2: supersymmetric extension of pure Yang--Mills model'', 
{\it Nucl. Phys.} {\bf B108} (1976) 265.}
\bibitem{wessbagger}{J.~Wess and J.~Bagger, {\it Supersymmetry and 
supergravity}, 2nd Edition, Princeton University Press (1992).}
\bibitem{storey}{D.~Storey, ``General gauge calculations in $N$=4 
super Yang--Mills theory'', {\it Phys. Lett.} {\bf 105B} (1981) 171.}
\bibitem{parkeswest}{A.J.~Parkes and P.C.~West, ``$N$=1 supersymmetric 
mass terms in the $N$=4 supersymmetric Yang--Mills theory'', {\it 
Phys. Lett.} {\bf 122B} (1983) 365.}
\bibitem{japan}{N.~Arkani-Hamed and H.~Murayama, ``Holomorphy, 
Rescaling Anomalies and exact $\beta$ functions in supersymmetric 
gauge theories'', {\tt hep-th/9707133}.}
\bibitem{yoshida}{S.~Arnone, C.~Fusi and K.~Yoshida, ``Exact 
renormalization group equation in presence of rescaling anomaly'', 
{\tt hep-th/9812022}.}
\bibitem{juerstorey2}{J.W.~Juer and D.~Storey, ``One-loop 
renormalization of superfield Yang--Mills theories'', {\it Nucl. 
Phys.} {\bf B216} (1983) 185.}
\bibitem{grisaruroceksiegel}{M.T.~Grisaru, M.~Ro\v{c}ek and 
W.~Siegel, ``Improved methods for supergraphs'', {\it Nucl. Phys.} 
{\bf B159} (1979)  429.}
\bibitem{n1nonrenormal}{J.~Wess and B.~Zumino, ``A Lagrangian model 
invariant under supergauge transformations'', {\it Phys. Lett.} 
{\bf 49B} (1974) 52; J.~Iliopoulos and B.~Zumino, ``Broken supergauge 
symmetry and renormalization'', {\it Nucl. Phys.} {\bf B76} (1974) 310; 
S.~Ferrara, J.~Iliopoulos and B.~Zumino, ``Supergauge invariance and 
the Gell-Mann Low eigenvalue'', {\it Nucl. Phys.} {\bf B77} (1974) 41.}
\bibitem{divdegree}{S.~Ferrara and O.~Piguet, ``Perturbation 
theory and renormalization of supersymmetric Yang--Mills theories'', 
{\it Nucl. Phys.} {\bf B93} (1975) 261; D.~Capper and G.~Leibbrandt, 
``On the degree of divergence of Feynman diagrams in superfield 
theories'', {\it Nucl. Phys.} {\bf B85} (1975) 492; K.~Fujiikawa and 
W.~Lang, ``Perturbation calculations for the scalar multiplet in a 
superfield formulation'', {\it Nucl. Phys.} {\bf B88} (1975) 61.}
\bibitem{n2mass}{M.A.~Namazie, A.~Salam and J.~Strathdee, ``Finiteness 
of broken $N$=4 super Yang--Mills theory'', {\it Phys.Rev.} {\bf D28} 
(1983) 1481.}
\bibitem{juerstorey1}{J.W.~Juer and D.~Storey, ``Nonlinear 
renormalization in superfield gauge theories'', {\it Phys. Lett.} 
{\bf 119B} (1982) 125.}
\bibitem{kinoleenauen}{T.~Kinoshita, ``Mass singularities of Feynman 
amplitudes'', {\it J. Math. Phys.} {\bf 3} (1962) 650; T.D.~Lee and 
M.~Nauenberg, ``Degenerate systems and mass singularities'', {\it 
Phys. Rev.} {\bf 133} (1964) B1562.}
\bibitem{kinopoggioquinn}{T.~Kinoshita and A.~Ukawa, ``New approach to 
the singularities of Feynman amplitudes in the zero-mass limit'', {\it 
Phys. Rev.} {\bf D13} (1976) 1573; E.C.~Poggio and H.R.~Quinn, ``The 
infrared behavior of zero-mass Green's functions and the absence of 
quark confinement in perturbation theory'', {\it Phys. Rev.} {\bf 
D14} (1976) 578; G.~Sterman, ``Kinoshita's theorem in Yang--Mills 
theories'', {\it Phys. Rev.} {\bf D14} (1976) 2123.}
\bibitem{currents}{E.~Bergshoeff, M.~de Roo and B.~de Wit, 
``Extended conformal supergravity'', {\it Nucl. Phys.} {\bf B182} 
(1981) 173.} 
\bibitem{bkrs}{M.~Bianchi, S.~Kovacs, G.C.~Rossi and Ya.~Stanev, in 
preparation.}

\end{thebibliography}
\end{document}